\documentclass[11pt]{article}
\usepackage{amssymb}
\usepackage{amsmath}
\usepackage{natbib}
\usepackage{graphicx}
\usepackage[titletoc,title]{appendix}
\usepackage[linesnumbered,lined,ruled,longend]{algorithm2e}
\usepackage{subfigure}
\usepackage{multirow}
\usepackage{dcolumn}
\usepackage{setspace}
\usepackage{amsthm}
\usepackage{wasysym}
\usepackage{bbm}
\usepackage{enumerate}
\usepackage{blkarray}
\usepackage{authblk}
\usepackage{bm}
\usepackage{arydshln}
\usepackage{hyperref}

\usepackage[margin=3cm]{geometry}

\numberwithin{equation}{section}

\newcommand{\vect}[1]{\boldsymbol{#1}}
\newcommand{\vectx}{\boldsymbol{x}}
\newcommand{\vecty}{\boldsymbol{y}}
\newcommand{\vectd}{\boldsymbol{d}}
\newcommand{\vectt}{\boldsymbol{\theta}}
\newcommand{\vectS}{\boldsymbol{S}}

\newcommand{\training}{\mathcal{T}}

\begin{document}

\setlength{\parindent}{0pc}
\setlength{\parskip}{1ex}

\date{6 April 2022}

\title{Optimal Bayesian Design for Model Discrimination via Classification\footnote{The original article available under \url{https://doi.org/10.1007/s11222-022-10078-2} has been published under the CC--BY $4.0$ licence (\url{https://creativecommons.org/licenses/by/4.0/}). This document combines the main article and the supplementary document and contains a few minor corrections.}}

\author[1,2]{Markus Hainy \footnote{e-mail: \texttt{markus.hainy@jku.at}}}
\author[3,4,5]{David J. Price}
\author[5]{Olivier Restif}
\author[2,6,7]{Christopher Drovandi}
\affil[1]{Department of Applied Statistics, Johannes Kepler University, 4040~Linz, Austria}
\affil[2]{School of Mathematical Sciences, Queensland University of Technology,\newline Brisbane QLD 4000, Australia}
\affil[3]{Centre for Epidemiology and Biostatistics, Melbourne School of Population and Global Health, The University of Melbourne, VIC 3010, Australia}
\affil[4]{The Department of Infectious Diseases at The Peter Doherty Institute for Infection and Immunity, The University of Melbourne and Royal Melbourne Hospital, \newline VIC 3000, Australia}
\affil[5]{Department of Veterinary Medicine, University of Cambridge,\newline Cambridgeshire CB3~0ES, United Kingdom}
\affil[6]{ARC Centre of Excellence for Mathematical \& Statistical Frontiers}
\affil[7]{QUT Centre for Data Science}

\maketitle

\begin{center}\textbf{Abstract}\end{center}

Performing optimal Bayesian design for discriminating between competing models is computationally intensive as it involves estimating posterior model probabilities for thousands of simulated datasets.  This issue is compounded further when the likelihood functions for the rival models are computationally expensive.  A new approach using supervised classification methods is developed to perform Bayesian optimal model discrimination design. This approach requires considerably fewer simulations from the candidate models than previous approaches using approximate Bayesian computation. Further, it is easy to assess the performance of the optimal design through the misclassification error rate.  The approach is particularly useful in the presence of models with intractable likelihoods but can also provide computational advantages when the likelihoods are manageable.


\section{Introduction} \label{sec:intro}

In many applications, finding the most appropriate model among a class of possible models is an important goal of statistical inference. In the classical literature, these decisions are commonly based on model selection criteria such as the Akaike information criterion or related criteria \citep{konishi2008}. The Bayesian approach, where the model indicator is regarded as an additional unknown random variable, offers a coherent decision-theoretic framework for inference and model discrimination \citep*{key1999}. Common options to carry out model selection in a Bayesian context are Bayes factors \citep{kass_raftery_1995}, the deviance information criterion \citep{Spiegelhalter2002}, or the computation of the marginal likelihoods or evidence \citep{Friel}. Given the prior model probabilities, the marginal likelihoods can be turned into posterior model probabilities. Classical model selection criteria only provide a ranking of the models, whereas posterior model probabilities contain useful information about the relative likeliness of the various models as well. In addition, the posterior model probabilities permit model-averaged predictions.

Prior to conducting an experiment, it is pertinent to determine the optimal combination of the controllable factors so as to maximise the (expected) information gain of the experiment with respect to the desired statistical objective (e.g., parameter inference, model discrimination, prediction). This is achieved by applying the principles and methods of optimal experimental design \citep*[see, e.g.,][]{atkinson_donev_tobias_2007}.
In optimal experimental design, one seeks to find the optimal combination of the controllable factors in order to maximise the (expected) information gain of the experiment \citep{atkinson_donev_tobias_2007}. If the main goal of statistical inference is to determine which statistical process is the most suitable representation of the phenomenon of interest, it is pertinent to employ design criteria specifically developed for the purpose of model discrimination. For example, in epidemiology it is paramount to understand the transmission dynamics of a disease in order to be able to implement effective countermeasures \citep*[see, e.g.,][]{Dehideniya2018synth}. The most commonly used classical design criterion is T-optimality \citep{AtkinsonFedorov1975a,AtkinsonFedorov1975b,dette_titoff_2009}, with extensions to Bayesian T-optimality \citep{ponce_atkinson_1992} to incorporate prior information. Except for robust T-optimal designs \citep{vajjah_duffull_2012}, one model has to be selected as the assumed true model. For classical T-optimality, one seeks to maximise the $L_2$-norm of the difference between the assumed true model's predictor and the other model's predictor with respect to the design measure, where for each design the most unfavourable parameter setting with respect to the predictor difference is chosen for the second model. Therefore, T-optimal designs are generally computationally expensive. \citet{Harman2020} propose a symmetric criterion based on the linearised distance between the mean-value surfaces of the models, which can be computed quickly. Their designs depend on the set of parameters over which the criterion is optimised, so they suggest to consider different parameter set sizes and to choose the size of the set based on ones ``confidence'' about the true parameter value.

Fully Bayesian experimental design provides a consistent framework to handle parameter and model uncertainty when planning the experiment \citep{ChalonerVerdinelli,ryan_drovandi_mcgree_pettitt_2015}. For model discrimination, the most popular design criterion is the mutual information between the model indicator and the data, which is measured by the Kullback-Leibler divergence between the joint and marginal distributions of those two random variables \citep[see][]{BoxHill1967}. This criterion requires the computation of the evidence of each model for many potential observations, so its use has been confined to a limited set of applications such as simple models with conjugate priors \citep{ng_chick_2004}, cases where numerical quadrature is feasible \citep{cavagnaro_myung_pitt_kujala_2010}, or sequential design settings \citep*{drovandi_mcgree_pettitt_2014}. \citet*{Overstall2018} employ normal-based approximations to find optimal designs for several criteria including mutual information and misclassification error. For the case of intractable likelihoods, \citet*{Dehideniya2018} use approximate Bayesian computation (ABC) to estimate these criteria. The ABC approach only requires the ability to simulate from all the candidate models. However, their approach is simulation- and memory-intensive and is thus limited to low-dimensional designs. \citet{Overstall2019} propose an approach based on auxiliary models, whereas \citet{Dehideniya2018synth} employ synthetic likelihoods. An extension of \citet{Overstall2018} for models with intractable likelihoods is developed by \citet*{Dehideniya2019}. \citet{kleinegesse2019} develop a design approach based on likelihood-free inference by ratio estimation \citep[see][]{thomas2022}, which is suitable for the commonly used mutual information-based design criteria, with an extension to sequential designs in \citet*{kleinegesse2020}. Another approach for mutual information-based criteria which can also be applied to intractable likelihood models is presented by \citet{foster2019}, who use amortised variational inference to find an approximation to the posterior distribution that is part of the criterion.

Like \citet{Dehideniya2018}, we suggest a simulation-based approach.  However, we use the outputs of standard supervised classification procedures from machine learning \citep*[see][]{Hastie2009} to estimate the design criteria. In particular, we employ classification trees \citep{Breiman1984} and random forests \citep{Breiman2001}. We demonstrate that this approach considerably reduces the required number of simulations compared to ABC. In order to keep the computational burden manageable, \citet{Dehideniya2018} pre-simulate a large sample from the prior predictive distribution at a grid of possible design points and re-use these simulations for all the designs they consider during the optimisation process, refining the grid over time. However, as we require fewer simulations for the classification approach, it is not necessary to pre-simulate the data. As a consequence, the classification approach is much more flexible and suitable for much higher-dimensional designs. Furthermore, the classification approach does not require direct approximations of posterior quantities such as the posterior model probabilities, which may only be reliably estimated with great computational effort, making it a viable alternative for many models with tractable likelihoods. Another advantage of the classification approach is that one can readily use the output from the classification procedures to assess the designs by estimating misclassification error rates or misclassification matrices. Our method represents a novel approach using supervised learning methods for optimal Bayesian design for model discrimination.

Section~\ref{sec:design} reviews Bayesian experimental design and the associated expected utility and loss functions. Our classification approach is presented in Section~\ref{sec:classification} along with a discussion of classification and regression trees (CART) and random forests. In Section~\ref{sec:examples}, we provide three examples to demonstrate the utility of the classification approach: discriminating between the epidemiological Markov process models of the same type as considered by \citet{Dehideniya2018} (Section~\ref{subsec:epi}), a two-model variation of the previous example to be able to make comparisons with likelihood-based designs and apply our method to higher-dimensional settings (Section~\ref{subsec:epi2}), and discriminating between three Markov process models describing the dynamics of bacteria within phagocytic cells (Section~\ref{subsec:macro}). The appendix contains further details on CART and random forests, a description of the variant of the coordinate exchange algorithm that we employ for all our examples, a comparative investigation of the computational performances of the different methods for the three examples in Section~\ref{sec:examples}, detailed model descriptions and further results for the three examples in Section~\ref{sec:examples}, and two additional examples. The first additional example is a  logistic regression example which has been considered for Bayesian experimental design before \cite[e.g.,][]{Overstall2018}, the second is about discriminating between three spatial extremes models for which \citet{lee2018} perform ABC model discrimination for a given design.

\section{Optimal Bayesian Design for Model Discrimination} \label{sec:design}

We assume there are $K$ candidate statistical models for a process of interest, one of them being the true underlying model. The models are indexed by the model indicator random variable $m \in \{1,2,\ldots,K\}$. Each model $m$ has a likelihood function $p(\vecty|\vectt_m,m,\vectd)$, with data $\vecty \in \mathcal{Y}$, and parameter vector $\vectt_m \in \Theta_m$.  In the experimental design context, the likelihood depends on the design vector $\vectd \in \mathcal{D}$, which is a vector of controllable variables of the experiment that might influence the informativeness of the data $\vecty$.  In the Bayesian framework, a prior distribution $p(\vectt_m|m)$ is assigned to the parameters of each model $m$.  Furthermore, we assign a prior probability $p(m)$ to each model such that $\sum_{m=1}^K p(m) = 1$. One can then derive the following important quantities from these elements: $p(\vecty|m,\vectd) = \int_{\vectt_m} p(\vecty|\vectt_m,m,\vectd) \, p(\vectt_m|m) \, \mathrm{d} \vectt_m$ is the \emph{marginal likelihood}, \emph{evidence}, or \emph{prior predictive distribution} for model $m$; $p(\vecty|\vectd) = \sum_{m=1}^K p(\vecty|m,\vectd) \, p(m)$ is the overall or model-averaged \emph{marginal likelihood} or \emph{prior predictive distribution}; and $p(m|\vecty,\vectd) =  p(\vecty|m,\vectd) \, p(m) \bigl/ p(\vecty|\vectd)$ is the \emph{posterior model probability} of model $m$.

Optimal experimental design requires the specification of a design criterion that encodes the goal of the experiment. In Bayesian design, a function $l$ that quantifies the loss of an experiment needs to be specified, see, e.g., \citet{Overstall2018}. Apart from the design $\vectd$, this loss function usually also depends on the model indicator $m$ and the data $\vecty$ observed at the experiment. It may also depend on the parameters $\vectt_m$ at each of the models. For experimental design, the expected or integrated loss,
\begin{equation}
l(\vectd) = \mathrm{E}_{\vectt_m,\vecty,m|\vectd}[\, l(\vectd,\vectt_m,\vecty,m)],
\end{equation}	
is of interest, where the expectation is taken with respect to all the unknown variables. The optimal design is given by $\vectd^* = \arg \min_{\vectd \in \mathcal{D}}l(\vectd)$, where $\mathcal{D}$ is the set of admissible designs, which in general is a challenging optimisation problem. Alternatively, the design problem may be formulated in terms of a utility function instead of a loss function. Then the goal is to maximise the expected utility function.

In Bayesian model discrimination, we are interested in finding a design $\vectd$ that is likely to produce data $\vecty$ from which we can infer the posterior distribution of the model indicator $m$ with minimal uncertainty. The most popular measure of uncertainty of a distribution is its \emph{Shannon entropy} \citep[see, e.g.,][]{Lindley1956}. For a given dataset $\vecty$, the conditional entropy of the model indicator is given by 
\[
l_{MD}(\vectd,\vecty) = - \sum_{m=1}^K p(m|\vecty,\vectd)\log p(m|\vecty,\vectd).
\]
The conditional entropy features the loss function 
\[
l_{MD}(\vectd,\vecty,m) = - \log p(m|\vecty,\vectd),
\]
which is called the \emph{multinomial deviance loss} \citep{Hastie2009}.

Since $\vecty$ is not known in advance, we take the average over the marginal distribution of $\vecty$, $p(\vecty|\vectd)$.
For discrete data $\vecty$, the expected multinomial deviance loss is
\begin{equation}
l_{MD}(\vectd) = - \sum_{\vecty \in \mathcal{Y}} p(\vecty|\vectd) \sum_{m=1}^K p(m|\vecty,\vectd)\log p(m|\vecty,\vectd). \label{eq:eMDL2}
\end{equation}
The negative of the expected multinomial deviance loss is also known as the \emph{mutual information utility} \citep[see, e.g.,][]{drovandi_mcgree_pettitt_2014}.

Another common loss function for model discrimination is the \emph{0--1 loss} \citep[see, e.g.,][]{Overstall2018}. Let $\hat{m}(\vecty|\vectd)$ be a classifier function that assigns one of the class labels $1,\ldots,K$ to the data $\vecty$. The 0--1 loss function is defined as
\[
l_{01}(\vectd,\vecty,m) = \mathrm{I}[\hat{m}(\vecty|\vectd) \neq m] = 1 - \mathrm{I}[\hat{m}(\vecty|\vectd) = m],
\]
where $\mathrm{I}[\cdot]$ is the indicator function, which takes the value $1$ if the argument is true and 0 otherwise. Therefore, the 0--1 loss is 1 if the data is misclassified and 0 if it is classified correctly. A generalisation of this loss function would be a loss matrix that assigns different loss values to all the combinations of true and selected models. Averaging the 0--1 loss function over the prior predictive distribution of the data and the model indicators yields the \emph{misclassification error rate} or \emph{prior error rate} \citep{Pudlo2016}, which for discrete data $\vecty$ is given by
\begin{equation}
l_{01}(\vectd)  =  \sum_{\vecty \in \mathcal{Y}} p(\vecty|\vectd) \sum_{m=1}^K p(m|\vecty,\vectd) \{1 - \mathrm{I}[\hat{m}(\vecty|\vectd) = m]\}. \label{eq:MER}
\end{equation} 
The classifier $\hat{m}(\vecty|\vectd) = \arg \max_{m \in \{1,\ldots,K\}} p(m|\vecty,\vectd)$ -- also known as the \emph{Bayes classifier} -- classifies the data according to the posterior modal model. It can be shown that the Bayes classifier minimises the expected 0--1 loss (\ref{eq:MER}). The misclassification error rate for the Bayes classifier is called the \emph{Bayes error rate}  \citep[see][]{Hastie2009}.

In the continuous case, the sums over $\vecty \in \mathcal{Y}$ in the expected loss functions (\ref{eq:eMDL2}) and (\ref{eq:MER}) have to be replaced by integrals. The integrals and sums involved in (\ref{eq:eMDL2}) and (\ref{eq:MER}) can be high-dimensional, analytically intractable and computationally intensive to approximate accurately.  One approach is to estimate the expected loss functions using Monte Carlo integration. Let $\vecty^{m,j} \sim p(\vecty|m,\vectd)$ for $j=1,\ldots,J_m$ and $m=1,\ldots,K$. That is, $J_m$ draws $\vecty^{m,j}$ from the prior predictive distribution under model $m$ are generated, for each of the models in turn. Then we can estimate the expected loss (\ref{eq:eMDL2}) by
\begin{equation}
\hat{l}_{MD}(\vectd) = - \sum_{m=1}^K p(m) \frac{1}{J_m} \sum_{j=1}^{J_m} \log p(m|\vecty^{m,j},\vectd), \label{eq:eMDL_est}
\end{equation}
and the expected loss (\ref{eq:MER}) by 
\begin{equation}
\hat{l}_{01}(\vectd) = 1 - \sum_{m=1}^K p(m) \frac{1}{J_m} \sum_{j=1}^{J_m} \mathrm{I}[\hat{m}(\vecty^{m,j}|\vectd) = m], \label{eq:MER_est}
\end{equation}
respectively, where $\hat{m}(\vecty^{m,j}|\vectd) = \arg \max_{m \in \{1,\ldots,K\}} p(m|\vecty^{m,j},\vectd)$.

The first issue with these approximations is that the $J_m$ may need to be large to estimate the expected loss with low variance.  The second issue is that the posterior model probability, $p(m|\vecty,\vectd)$, is generally not available analytically and is difficult to approximate accurately.  In fact, estimating this quantity is a research problem in its own right in the Bayesian community \citep{Friel2012}. For an efficient recent approach using Gaussian quadrature, see \citet{chai2019}. In the Bayesian optimal design setting, an estimate of the expected loss requires $J = \sum_{m=1}^K J_m$ evaluations/approximations of $p(m|\vecty,\vectd)$, one for each dataset $\vecty$ drawn from the prior predictive distribution. Then, the expected loss must be optimised over a potentially large design space $\mathcal{D}$, and therefore often many thousands of posterior model probabilities must be calculated to arrive at an optimal design.  This is why only relatively simple models and experimental settings have been considered in the Bayesian design literature for model discrimination in comparison to the elaborate models that can be analysed in Bayesian inference \citep[see, e.g., the application in][]{Drovandi2014}. 

Further complications arise for estimating $p(m|\vecty,\vectd)$ when the likelihood function $p(\vecty|\vectt_m,m,\vectd)$ for the models of interest is computationally intractable. \citet{Dehideniya2018} present a rather general ABC approach to tackle the problem of Bayesian design for model discrimination for models with intractable likelihoods. However, their approach is very simulation-intensive and therefore only suitable for low-dimensional designs. The approach of \linebreak \citet{Overstall2019} relies on finding suitable auxiliary models for the intractable models of interest and uses Gaussian processes to model the relationship between the parameters of the true model and the corresponding auxiliary model parameters. The marginal likelihood is modelled by a copula, which aims to capture the dependence induced by marginalising out the parameters. \citet{Dehideniya2018synth}, on the other hand, use a synthetic likelihood approach to approximate the true likelihood function. This approach works best if the likelihood function depends on summary statistics whose distribution is close to normal. A more computationally efficient approach is presented in \citet{Dehideniya2019}, where Laplace-based approximations are used to estimate the design criteria instead of performing Monte Carlo integration. In order to find the posterior mode and curvature required for the Laplace approximation, synthetic likelihoods are used.

The ultimate goal of this paper is to expand the set of models and design settings for which it is possible to obtain optimal Bayesian designs for the purpose of model discrimination without having to rely on the availability of suitable parametric likelihood approximations.

\section{The Classification Approach} \label{sec:classification}

\subsection{Methodology}

In this paper we take a classification perspective on the Bayesian model discrimination problem to greatly reduce the computational burden highlighted in the previous section.  As a by-product, we also obtain several other advantages over the standard Bayesian approach.  The only requirement to apply our methodology is that it is computationally efficient to simulate from each of the $K$ models. Therefore, the class of models that can be considered in optimal design for model selection increases dramatically. In addition, the generality of the proposed approach allows for implementations that are less application-specific.  Furthermore, we find that the performance of the optimal design can be assessed easily via the misclassification error rate, as opposed to performing more posterior calculations at the optimal and sub-optimal designs. 

For each design $\vectd$ proposed in the design optimisation algorithm, our approach involves simulating $J$ samples from the joint distribution of data and model indicators, 
\[
p(\vecty,m|\vectd) = \int_{\vectt_m} p(\vecty|\vectt_m,m,\vectd) \, p(\vectt_m|m) \, p(m) \, \mathrm{d} \vectt_m,
\]
to generate the training sample $\training = \{ (m^j,\vecty^j): \: j = 1,\ldots,J\}$.

We can use this training sample to train a supervised classification algorithm, where we consider the model indicator $m$ as a categorical response or `target' variable and the simulated data $\vecty$ as the features. As a result, we obtain a classifier function $\hat{m}_{C}(\vecty|\vectd,\training)$ that we can use in Equation~(\ref{eq:MER_est}) instead of the Bayes classifier to estimate the misclassification error rate.

Alternatively, we can write the sample $\training$ as 
$$\training = \left\{ (m, \, \vecty^{m,j}): \: j=1,\ldots,J_m; \: m = 1,\ldots,K \right\}, $$
where $J_m$ is the number of samples from model $m$ in $\training$. Given $m$, the data are sampled from $\vecty^{m,j} \sim p(\vecty|m,\vectd)$. The numbers $J_m$ may be fixed in advance, usually selected to be proportional to the prior model probabilities. However, if the prior model probabilities are highly imbalanced, there may only be a few observations from the models with small prior model probabilities in $\training$. For training the classifier, it may then be advantageous to have a more balanced training sample. If the sample proportions do not reflect the prior model probabilities, it is necessary to adjust the classifier accordingly, for example by weighting the observations.

Due to overfitting, it is not advisable to use the same sample $\training$ for training the classifier as well as for evaluating the expected 0--1 loss in (\ref{eq:MER_est}). To deal with this problem, one possibility to estimate the expected loss in practice is to use $L$-fold cross-validation \citep[see, e.g.,][]{Hastie2009}, where the full sample $\training$ is randomly split into $L$ folds of approximately equal size: $\training = \{\training^{1},\ldots,\training^L\}$.  Let $\training^{-i} = \{\training^{1},\ldots,\training^{i-1},\training^{i+1},\ldots,\training^L\}$ denote the full sample without the $i$th fold and let $\training^i_m$ be defined as $\training^i_m = \{\vecty_*: (m_*,\vecty_*) \in \training^i \: \wedge \: m_*=m \}$. The procedure is repeated $L$ times. At each step $i$ ($i = 1,\ldots,L$), the classifier is trained on $\training^{-i}$ and validated on the subsample $\training^i$. Thus, at step $i$ the expected 0--1 loss is computed as
\begin{equation}
\hat{l}_{01,i}^{\mathrm{cv}}(\vectd) = 1 - \sum_{m=1}^K p(m) \frac{1}{J_m^i} \sum_{\vecty \in \training_m^i}  \mathrm{I}[\hat{m}_{C}(\vecty|\vectd,\training^{-i}) = m], \label{eq:MER_est2_comp}
\end{equation}
where $J_m^i = \mathrm{card}(\training_m^i)$.

The final estimate of the expected 0--1 loss is then obtained by averaging over the $L$ expected loss estimates:

\begin{equation}
\hat{l}_{01}^{\mathrm{cv}}(\vectd) = \frac{1}{L} \sum_{i=1}^L \hat{l}_{01,i}^{\mathrm{cv}}(\vectd). \label{eq:MER_est2}
\end{equation}

In our examples, we always perform stratified sampling of the fold indicators. That is, first we divide the total sample $\training$ into $m$ subsamples according to the model indicators. Then we randomly split the subsample for each model into $L$ equal-sized folds. Finally, we combine all the subsample folds with the same fold indicator $i$ across all model subsets into fold $\training^i$. In this way, we guarantee that the model proportions are the same in all folds.

An alternative to cross-validation would be to generate an independent test or validation sample and evaluate the expected loss function on that sample. Depending on how cheap it is to simulate the data and how expensive it is to run the classifier, this approach might be preferable to cross-validation. In our examples, we only report the results for cross-validation since both approaches are qualitatively very similar.

Larger values of $J$ allow for a more accurate estimate of the misclassification error rate, and therefore lead to a less noisy objective function to optimise over, although the time to estimate the error rate increases. However, for intractable likelihood models the sample size $J$ needed for the classification approach to obtain a reasonably precise approximation of the expected loss function is several orders of magnitude less than the sample size required for ABC \citep{Pudlo2016}. Moreover, for many other models the classification approach may be more time-efficient than estimating $p(m|\vecty,\vectd)$ in a conventional way.

Many classification methods also provide estimates of the posterior model probabilities,  \linebreak $\hat{p}_C(m|\vecty,\vectd,\training)$, which can be used to estimate the expected multinomial deviance loss~(\ref{eq:eMDL2}) in a similar way as the misclassification error rate is estimated by Equations~(\ref{eq:MER_est2_comp}) and (\ref{eq:MER_est2}). However, the estimates for the posterior model probabilities provided by many computationally efficient methods such as classification trees or linear discriminant analysis are rather crude, noisy and biased \citep[see, e.g.,][]{Breiman1984,Hastie2009}.

Even if the posterior model probabilities are estimated poorly, the classification method can perform quite well at the task of assigning the correct class labels to the observations. All that matters is that the posterior modal model is identified correctly. If a classifier assigns the posterior modal model $\arg \max_m p(m|\vecty,\vectd)$ to each dataset $\vecty \in \mathcal{Y}$, it is called an \emph{order-correct classifier} \citep{Breiman1996}. For an order-correct classifier, the misclassification error rate corresponds to the Bayes error rate and is therefore minimal. The misclassification error rate of a classifier that is order-correct everywhere except for a small subset of the sample space $\mathcal{Y}$ will still be very close to the Bayes error rate. 
Therefore, the misclassification error rate is relatively robust to inaccurate estimates of the posterior model probabilities. For this reason, we focus mainly on finding designs which are optimal with respect to the misclassification error rate. However, the misclassification error rate is not estimated very well if the posterior modal model is hard to identify among several highly probable models in a non-negligible subset of the sample space $\mathcal{Y}$, which may happen, for example, if the data is generally not very informative.

\subsection{CARTs and Random Forests}

\label{subsec:CART_RF}

There are a plethora of supervised classification algorithms that are suitable candidates for the task of estimating the expected loss. As the optimal design procedure estimates the expected loss many times, we require a fast classification method. As a generic and fast nonparametric classification approach, we adopt \emph{classification and regression trees} \citep[CART, see][]{Breiman1984} to estimate the expected loss at each design visited during the design procedure.

One disadvantage of trees is their high variance. Slight changes in the data might lead to widely different trees. To reduce the variance, \citet{Breiman2001} proposes \emph{random forests}, which  consist of an ensemble of trees. For classification, the class prediction of a random forest is obtained by majority vote among the individual trees of the forest. More information about the structure, properties, and estimation of CARTs and random forests can be found in Appendix~\ref{app:CART_RF}.

Random forests have been used successfully in many applications and compare favourably to many other more computationally intensive classification methods such as boosting or neural networks, see  \citet{Hastie2009}. Their nonparametric nature allows for capturing complex dependencies between the model indicator and the features and so they are more flexible than many parametric methods such as logistic regression. Another advantage of trees and random forests is that the scaling of the features does not matter, so there is no need to standardise or transform the features. For our purpose it is also important that random forests do not require any tuning for each new dataset and design because the standard settings work reasonably well in most situations. A further advantage of random forests is that the misclassification error rate can be estimated using \emph{out-of-bag} class predictions \citep{Breiman2001}, so there is no need to perform cross-validation or to generate a test set.

\citet{Pudlo2016} note that random forests can easily cope with many noisy, weakly informative and correlated input features. Nevertheless, if the dimension of the raw data is very high, summary statistics may need to be used to improve the classification performance. However, random forests make it possible to include a relatively large amount of informative summary statistics. This may alleviate the loss of information regarding model discrimination when using non-sufficient summary statistics reported by \citet{robert2011}. The standard kernel-based ABC approaches for intractable likelihood problems suffer from the curse of dimensionality much more strongly and require low-dimensional summary statistics to work efficiently \citep[see, e.g.,][]{blum2010}.

It is possible to obtain estimates for the posterior model probabilities $p(m|\vecty,\vectd)$ from trees and random forests. However, these estimates are not smooth and very rough, in particular for trees. It might happen that the estimated posterior model probabilities for some observations are 0, which causes problems when estimating the expected multinomial deviance loss. Appendix~\ref{app:CART_RF} discusses this issue in more detail and explains how we deal with it.

\subsection{Assessing the Performance of a Design}

Once we have found optimal or close-to-optimal designs using a variety of our design search methods for some design dimensions, we are also able to assess the performance of those designs with the classification method. For example, it may be of interest to assess the ability to discriminate between models as the sample size or design dimension is increased, or to investigate which design search methods lead to more efficient designs. 
We want this assessment to be as accurate as possible. Given that only a relatively small number of designs need to be assessed, we suggest that more effort can be placed in the classification procedure. For example, we can simulate both a large training and a large test set and fit an elaborate classifier such as a random forest with a large number of trees.
Then, the classification performance in terms of the misclassification error rate can be estimated by applying the fitted model to the test dataset.

\section{Examples} \label{sec:examples}

In this section, we consider several examples to highlight the utility of our proposed method.  To perform the design optimisation, we use a modification of the coordinate exchange (CE) algorithm \citep{meyer_nachtsheim_1995}, which involves cycling through each of the design variables iteratively, trialling a set of candidate replacements and updating the value of the design variable if the objective/loss function is reduced. This is continued until no updates to the design are made in a given cycle. To guard against possible local optima, we run the algorithm in parallel 20 times with random starts. We acknowledge the stochastic nature of our objective function by considering the (up to) six last designs visited in each of the 20 runs as candidates for the overall optimal design. For each of the candidates, we compute the loss function ten times to reduce the noise. The best design found through this algorithm is the one with the lowest average loss among the candidate designs across all runs. As an additional post-processing step, we combine all the candidate designs and estimated loss function values from all the runs. Then we employ Gaussian process regression \citep{Rasmussen2006} on them to obtain a smooth estimate for the expected loss surface, which we seek to minimise with respect to the design. Finally, we compare the expected loss at this new design to the expected loss found previously by the coordinate exchange algorithm. This is done by estimating the expected loss 100 times at each of the two designs and selecting the design with the lower average expected loss as the optimal design. A detailed description of the optimisation algorithm that we employ is provided in Appendix~\ref{app:mce_algo}. We do not expend any effort on finding the best optimisation algorithm for each of the examples as this is not the focus of the paper. We find that the CE algorithm performs adequately to illustrate the findings of the paper.

The first example in Section~\ref{subsec:epi} compares the results of our supervised classification approach to ABC for different loss functions for an infectious disease application. It demonstrates that ABC and the computationally much more tractable classification approaches lead to designs with similar efficiency. The second example in Section~\ref{subsec:epi2} is a modification of the first example. It only considers the first two models of the first example, which have reasonably tractable likelihoods. This makes it possible to obtain likelihood-based loss estimates and find likelihood-based designs at least for lower dimensions, which we can use for comparisons with our classification approach. In addition, we demonstrate how we are also able to apply our approach successfully to higher-dimensional design settings. The third example is a practically important application in the field of experimental biology. The goal is to obtain good designs for discriminating between different hypotheses about unobserved heterogeneity with respect to the reproduction of bacteria within phagocytic cells. We apply our classification-based design method to two further examples in Appendices~\ref{app:logreg} and \ref{app:spatextr}. The first is a fairly high-dimensional logistic regression example with fixed and random effects, for which previous attempts on finding Bayesian optimal designs were only possible by making some additional approximations \citep{Overstall2018}. The second example is an application to intractable max-stable spatial extremes models, for which designs were previously only found on a very limited number of candidate design points using the ABC approach \citep*{hainy2016}.

Listings of computational runtime performance statistics for the different methods and design settings for all the examples in this section can be found in Appendix~\ref{app:performance}.

\subsection{Stochastic Models in Epidemiology} \label{subsec:epi}

\subsubsection{Problem Formulation}

An example involving four competing continuous-time Markov process models for the spread of an infectious disease is considered in \citet{Dehideniya2018}.  Let $S(t)$, $E(t)$ and $I(t)$ denote the number of susceptible, exposed and infected individuals at time $t$ in a closed population of size $N=50$ such that $S(t) + E(t) + I(t) = N$ for all $t$.  The possible transitions in an infinitesimal time $\delta_t$ for each of the four models are shown in Table \ref{tab:epi_models}.  Models 1 -- 4 are referred to as the death, SI, SEI and SEI2 models, respectively.  Models 1 and 2 do not have an exposed population. The algorithm of \citet{gillespie1977} can be used to efficiently generate samples from all the models. The prior distributions for all the parameters of each model are provided in Table~\ref{tab:epi_priors} in Appendix~\ref{app:epi_prior}.  All models are assumed equally likely \emph{a priori}.

\begin{table}
	\centering
	\caption{Four competing models considered in the infectious disease example of Section \ref{subsec:epi}\label{tab:epi_models}}
	\begin{tabular}{cccc}
		\hline\noalign{\smallskip}
		Model & Event type & Update & Rate \\
		\noalign{\smallskip}\hline\noalign{\smallskip}
		1 & Infected & $S(t)-1$, $I(t)+1$ & $b_1^{(1)}S(t)$ \\
		\noalign{\smallskip}\hline\noalign{\smallskip}
		2 & Infected & $S(t)-1$, $I(t)+1$ & $[b_1^{(2)} + b_2^{(2)}I(t)] \, S(t)$ \\
		\noalign{\smallskip}\hline\noalign{\smallskip}
		3 & Exposed & $S(t)-1$, $E(t)+1$ & $b_1^{(3)}S(t)$ \\
		& Infected & $E(t)-1$, $I(t)+1$ & $\gamma^{(3)}E(t)$ \\
		\noalign{\smallskip}\hline\noalign{\smallskip}
		4 & Exposed & $S(t)-1$, $E(t)+1$ & $[b_1^{(4)} + b_2^{(4)}I(t)] \, S(t)$ \\
		& Infected & $E(t)-1$, $I(t)+1$ & $\gamma^{(4)}E(t)$ \\
		\noalign{\smallskip}\hline
	\end{tabular}
\end{table}

We consider the design problem of determining the optimal times (in days) $\vectd = (d_1,d_2,\ldots,d_n)$, where $d_1 < d_2 < \cdots < d_n \leq 10$, to observe the stochastic process in order to best discriminate between the four models under the available prior information.  Only the infected population can be observed.  Unfortunately, the likelihood functions for all but the simplest model are computationally cumbersome as they require computing the matrix exponential \citep[see, e.g.,][]{Drovandi2008}.  Whilst computing a single posterior distribution is feasible, as in a typical data analysis, computing the posterior distribution or posterior model probabilities for thousands of prior predictive simulations, as in a standard optimal Bayesian design approach, is computationally intractable. 

\subsubsection{Approximate Bayesian Computation}
\label{subsec:epi_ABC}

\citet{Dehideniya2018} develop a likelihood-free approach based on approximate Bayesian computation (ABC) to solve this model discrimination design problem. Given a particular level of discretisation of the design space (time in this case), the ABC approach involves generating a large number of prior predictive simulations at all discrete time points and storing them in the so-called reference table.  
Then, for a particular `outer' draw from the prior predictive distribution, $\vecty$, at some proposed design, $\vectd$, the ABC rejection algorithm of \citet{Grelaud2009} is used to estimate the posterior model probabilities and in further consequence the loss functions. This means that the posterior model probability $p(m|\vecty,\vectd)$ is estimated by computing the proportion of model $m$ simulations in the retained sample, where the retained sample is composed of those simulations from the reference table which are `closest' to the process realisation $\vecty$ with respect to some distance such as Euclidean or Manhattan distance. The size of the retained sample is only a very small fraction of the size of the reference table. The estimated posterior model probability is used to compute the estimated loss for process realisation $\vecty$. Finally, the estimated expected loss is obtained by averaging the loss estimates for all the `outer' draws.
The reader is referred to \citet{Dehideniya2018} for more details. \citet{Price2016} improve the efficiency for these models by making use of the discrete nature of the data to efficiently estimate the expected loss. 

\subsubsection{Simulation Settings}

For each of the classification methods from machine learning, we use a sample of 5K simulations from each model to train the classifier and to estimate the expected loss at each new design. For the classification trees, we use tenfold cross-validation to estimate the expected loss functions. When using random forests, we employ out-of-bag class predictions. As a criterion, we use expected 0--1 loss as well as expected multinomial deviance loss. When computing the expected multinomial deviance loss, we set the posterior model probability of the correct model to 0.001 whenever it is estimated to be 0, see Appendix~\ref{app:CART_RF} for more information. We could follow the ABC method and draw the simulations from a large bank of prior predictive process realisations simulated at the whole design grid to reduce the computing time. However, since the machine learning classification method requires significantly fewer simulations, we find that it is still fast to draw a fresh process realisation for each proposed design. For the ABC approach, the reference table contains 100K stored prior predictive simulations for each model. To compute the expected loss, we average the estimated loss over 500 `outer' draws from $p(\vecty|m,\vectd)$ for each model and retain a sample of size 2K from the reference table for each draw. For all the methods, the optimal design search was conducted over a grid of time points from 0.25 to 10 with a spacing of 0.25.

\subsubsection{One-dimensional Estimated Expected Loss Curves} \label{subsubsec:onedim_losscurve}

\begin{figure}
	\centering
	\includegraphics[height=0.3\textheight]{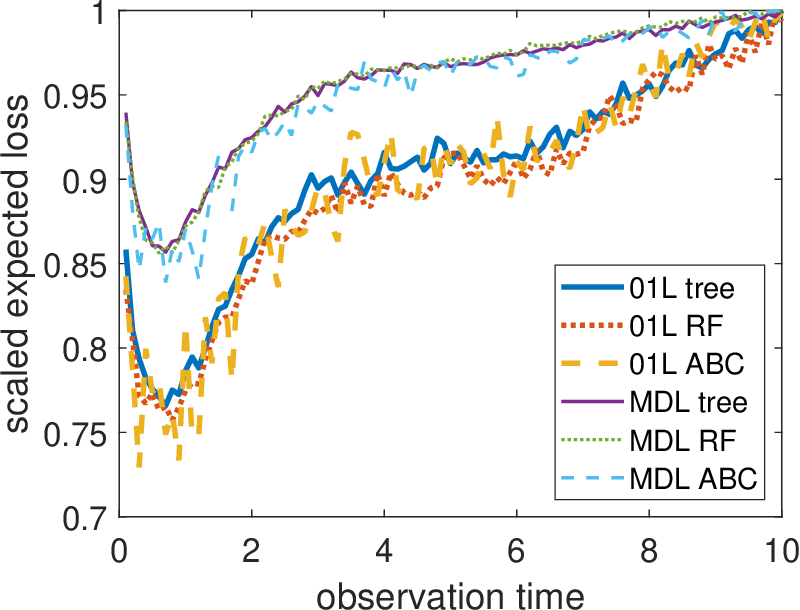}
	\caption{Plots of the approximated expected loss functions produced by the tree classification approach with cross-validation (solid), the random forest classification approach using out-of-bag class predictions (dotted), and the ABC approach (dashed) under the 0--1 loss (thick lines) and multinomial deviance loss (thin lines) for the infectious disease example. The expected losses have been scaled by dividing through the maximum loss for an easier comparison}
	\label{fig:1d_results}
\end{figure}

Figure~\ref{fig:1d_results} shows the approximate expected loss functions for 1 design observation under several estimation approaches and loss functions over a grid of design points with spacing 0.1. 
It is evident that all the functions are qualitatively similar and produce the same optimal design around $0.5 - 0.7$ days. In particular, one can see that the expected loss curves for both the 0--1 loss and the multinomial deviance loss seem to be minimised at around the same observation time. However, the times needed to construct the curves are vastly different between the different approaches. On our workstation, it took less than half a minute for the cross-validated tree classification approach (single core), between $4$ and $5$ minutes for the random forest classification approach (single core), and between $9.5$ and $10$ minutes using $8$ parallel cores for the ABC approach to generate the respective graphs. Creating the reference table with 400K simulations required only between $3.5$ and $4$ seconds in this example, since sampling via the Gillespie algorithm is very efficient. In our example, what is causing the computational inefficiency of ABC is having to sort the large reference table for each outer draw to obtain the retained ABC sample. Despite the much higher computational effort needed for the ABC approach, its estimates of the expected loss functions are still considerably noisier than the estimates of the classification approaches, which is mostly due to the relatively small outer sample size of 2000.

\subsubsection{Optimal Designs}

The optimal designs obtained by the machine learning and ABC approaches are shown in Table~\ref{tab:epi_designs1} for $n=1$ to $n=3$ time points and Table~\ref{tab:epi_designs2} in Appendix~\ref{app:epi_designs} for $n=4$ and $n=5$ time points. The machine learning methods lead to designs with a general preference for later sampling times. The designs obtained by trees and random forests are very similar. The ABC approach produces designs with notably lower sampling times. However, the results obtained by the ABC approach should be taken with caution, since the high noise of the expected loss estimates makes it harder to optimise over the design space, especially for higher dimensions. Moreover, the approximation of the posterior gets worse the higher the dimension. It is also interesting to note that there are hardly any differences between the two loss functions for any given method. This reaffirms our decision to consider only the 0--1 loss in the other examples.

\begin{table}[htbp!]
	\centering
	\caption{Optimal designs obtained by tree classification (cross-validated), random forest classification (using out-of-bag class predictions), and ABC approaches under the 0--1 loss (01L) or multinomial deviance loss (MDL) ($n = 1$, $2$, and $3$) for the infectious disease example. The equidistant designs are also shown \label{tab:epi_designs1}}
	\begin{tabular}{lD{.}{.}{3}@{\hskip 2.5em}*{2}{D{.}{.}{3}}@{\hskip 2.5em}*{3}{D{.}{.}{3}}}
		\hline
		\multicolumn{1}{l}{Method/Loss} & \multicolumn{1}{c@{\hskip 2.5em}}{$n = 1$} & \multicolumn{2}{c@{\hskip 2.5em}}{$n = 2$} & \multicolumn{3}{c}{$n = 3$}\\
		\hline
		Tree 01L & 0.598 & 0.787 & 4.437 & 0.818 & 4.568 & 9.493 \\
		RF 01L   & 0.611 & 0.823 & 4.433 & 0.750 & 4.000 & 10.000 \\
		ABC 01L  & 0.597 & 0.877 & 4.357 & 0.750 & 2.250 & 5.750 \\
		\hline
		Tree MDL & 0.621 & 0.750 & 4.750 & 0.750 & 4.750 & 10.000 \\
		RF MDL & 0.633 & 0.750 & 4.750 & 0.750 & 4.500 & 9.000 \\
		ABC MDL  & 0.556 & 0.750 & 3.500 & 0.500 & 1.750 & 4.750 \\
		\hline 
		Equidistant & 5.000 & 3.333 & 6.667 & 2.500 & 5.000 & 7.500 \\
		\hline		  
		
	\end{tabular}
\end{table}

\subsubsection{Classification Performance Evaluations of Optimal Designs}

As our next step, we compare the optimal designs found under the different approaches using a random forest classifier.  For each of the optimal designs, we train a random forest with 100 trees based on 10K simulations from each model. The misclassification error rates and the misclassification matrices are estimated from a fresh set of 10K simulations from each model. This is repeated 100 times to be able to quantify the random error in estimating the misclassification error rates. The results for all the optimal designs as well as for the equispaced designs are shown in Table~\ref{tab:epi_classvalidation}. For more than two observations, the designs that clearly perform best are those found under the machine learning classification approaches. However, also the ABC optimal designs generally perform well except for $n = 5$ design times. We can also observe that the loss function used for optimisation has little effect on the performance of the optimal design, only for the designs found using ABC there is a notable difference for $n = 4$. The equispaced designs perform substantially worse than all the optimal designs up until $n = 4$ observations. Table~\ref{tab:epi_classvalidation} also shows that there is almost no gain in the classification performance by increasing the number of observations beyond 2. Any additional observation will only add a negligible amount of information regarding model discrimination. At some point, adding additional uninformative observations adversely affects the classification power of the random forest.

\begin{table}
	\centering
	\caption{Average misclassification error rates for optimal designs obtained by tree classification (cross-validated), random forest classification (using out-of-bag class predictions), and ABC approaches under the 0--1 loss (01L) or multinomial deviance loss (MDL) as well as for the equidistant designs for the infectious disease example. The average misclassification error rates were calculated by repeating the random forest classification procedure 100 times (see text) and taking the average. The standard deviations are given in parentheses \label{tab:epi_classvalidation}}
	\begin{tabular}{lccccc}
		\hline\noalign{\smallskip}
		Design & \multicolumn{1}{c}{$n = 1$} & \multicolumn{1}{c}{$n = 2$} & \multicolumn{1}{c}{$n = 3$} & \multicolumn{1}{c}{$n = 4$} & \multicolumn{1}{c}{$n = 5$} \\
		\noalign{\smallskip}\hline\noalign{\smallskip}
		Tree 01L & 0.5554 & 0.5158 & 0.5133 & 0.5116 & 0.5129 \\
		& (0.0023) & (0.0024) & (0.0026) & (0.0022) & (0.0025) \\
		RF 01L & 0.5548 & 0.5160 & 0.5132 & 0.5113 & 0.5138 \\
		& (0.0027) & (0.0025) & (0.0025) & (0.0025) & (0.0025) \\
		ABC 01L  & 0.5547 & 0.5161 & 0.5196 & 0.5046 & 0.5339 \\
		& (0.0023) & (0.0025) & (0.0030) & (0.0024) & (0.0027) \\
		\noalign{\smallskip}\hline\noalign{\smallskip}
		Tree MDL & 0.5547 & 0.5178 & 0.5152 & 0.5183 & 0.5159 \\
		& (0.0023) & (0.0030) & (0.0028) & (0.0028) & (0.0025) \\
		RF MDL & 0.5550 & 0.5179 & 0.5118 & 0.5104 & 0.5128 \\
		& (0.0022) & (0.0026) & (0.0026) & (0.0026) & (0.0026) \\
		ABC MDL  & 0.5553 & 0.5221 & 0.5216 & 0.5226 & 0.5416 \\
		& (0.0020) & (0.0028) & (0.0028) & (0.0029) & (0.0025) \\
		\noalign{\smallskip}\hline\noalign{\smallskip}
		Equidistant & 0.6592 & 0.6200 & 0.5760 & 0.5537 & 0.5519 \\
		& (0.0029) & (0.0025) & (0.0027) & (0.0029) & (0.0032) \\
		\noalign{\smallskip}\hline
	\end{tabular}
\end{table} 

\begin{figure}
	\centering
	\includegraphics[height=0.25\textheight,width=1\textwidth]{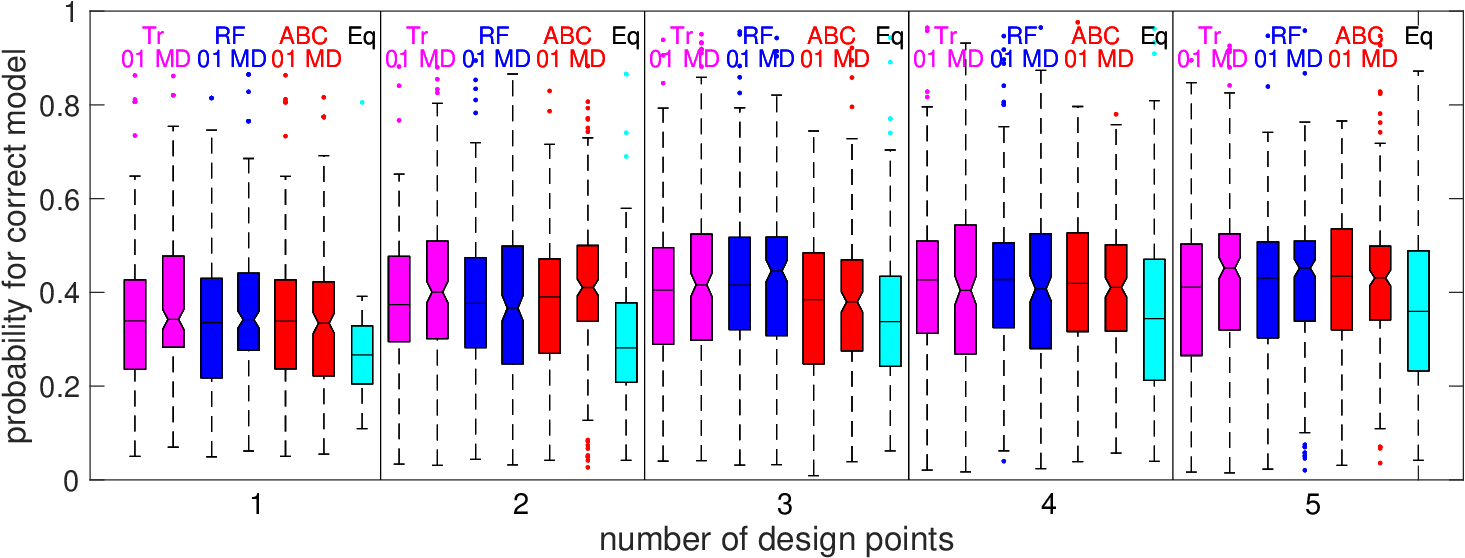}
	\caption{For each of the optimal designs obtained by the different approaches for 1 -- 5 observations in the infectious disease example, display the distribution of estimated ABC posterior model probabilities of the correct model over 200 process realisations (50 from each of the four models) simulated from the prior predictive distribution at the respective optimal design. For each number of design points, from left to right there are two magenta boxplots for the cross-validated tree classification designs, two blue boxplots for the random forest classification designs, two red boxplots for the ABC classification designs, and one cyan boxplot for the equispaced design. Boxplots for the 0--1 loss and for the equispaced designs do not have a notch, whereas boxplots for the multinomial deviance loss are notched
		\label{fig:epi_design_results}}
\end{figure}

Finally, we compare the optimal designs obtained by the different methods based on approximate posterior model probabilities estimated using ABC, as described in Section~\ref{subsec:epi_ABC}. To that end, for each design to evaluate we simulate 50 process realisations from the prior predictive distribution of each of the four models at that design and estimate the posterior model probability of the true model using ABC rejection. To get precise estimates of the posterior model probabilities for each of the 200 process realisations, we generate 10 million simulations from the prior predictive distribution to build the reference table. To estimate the posterior probabilities for each generated process realisation, we retain the 40K simulations from the reference table closest to that process realisation with respect to the Manhattan distance of the standardised observations. 
Boxplots showing the distributions of the estimated model probabilities over the 200 prior predictive process realisations for all the optimal designs as well as for the equispaced designs for 1 -- 5 observations are plotted in Figure~\ref{fig:epi_design_results}.  It can be seen that the results for all the different optimal designs are very similar, even though the approaches using the 0--1 loss criterion do not directly target the improvement in the posterior model probabilities. The equispaced designs perform appreciably worse up until $n = 4$ observations. It is also evident that, given the prior information in this example, not much gain can be achieved by collecting more than two observations, which is similar to the random forest classification results obtained in Table \ref{tab:epi_classvalidation}.  Assessing the optimal designs using random forests is much faster than performing this ABC simulation study.

A more detailed investigation of the classification performance at the optimal designs can be found in Appendix~\ref{app:epi_misclass}.

\subsection{Two-model Epidemiological Models with True Likelihood Validation} \label{subsec:epi2}

\subsubsection{Aims and Model Setup}

For the example in this section we use the same infectious disease model setup as in the previous Section~\ref{subsec:epi}. However, only the death and SI models (models 1 and 2) from Table~\ref{tab:epi_models} are considered. The reason is that for these two models the computation of the likelihood function is efficient enough to be able to compute likelihood-based posterior model probabilities for a sufficiently large amount of prior predictive samples. Therefore, we can compare the results for our likelihood-free approach using supervised classification methods to the results obtained by using the true likelihood functions to estimate the design criterion. Furthermore, we can assess the resulting optimal designs by computing the expected posterior model probabilities and misclassification error rates based on the true likelihood functions.

Another aim of this example is to demonstrate that the classification approach can easily cope with higher-dimensional designs, where other methods would fail to produce reasonable results in an acceptable amount of time. For the epidemiological example with four models in Section~\ref{subsec:epi}, one can see that there is hardly any gain in increasing the number of design points beyond three, so it makes no sense to consider any higher-dimensional designs. However, in Section~\ref{subsec:epi} we assume that we can only observe one realisation of the infectious disease process. In this section, in order to explore the performance of our methods for high-dimensional designs, we assume that several independent realisations of the stochastic process can be observed. For example, these independent realisations may pertain to independent populations of individuals. We allow each realisation to be observed at potentially different time points.

For simplicity, we assume that the same number of observations, $n_d$, is collected for each realisation. If there are $q$ realisations, then the total number of observations and therefore the design dimension is $n = q \cdot n_d$.

The prior distributions for the parameters are $b_1 \sim \mathcal{LN}(\mu = -0.48, \sigma = 0.3)$ for the death model and $b_1 \sim \mathcal{LN}(\mu = -1.1, \sigma = 0.4)$, $b_2 \sim \mathcal{LN}(\mu = -4.5, \sigma = \sqrt{0.4})$ for the SI model. 

In this example, we will only consider designs based on using the misclassification error rate as the design criterion. In order to compute the misclassification error rates based on the likelihoods, it is necessary to compute the marginal likelihoods for both models. When searching for the optimum design, we employ a relatively fast Laplace-type approximation to the marginal likelihood. For validating the resulting designs using the likelihood-based approach, we use a more expensive Gauss-Hermite quadrature scheme to obtain the marginal likelihoods. Details on both integral approximation methods can be found in Appendix~\ref{app:epi2_marglikelihood}.

\subsubsection{Example Settings and Results}

When searching for the optimal designs, we employ trees with cross-validation as well as random forests using out-of-bag class predictions for our supervised classification approach. For both classification approaches we use simulated samples of size 10K (5K per model).

For the likelihood-based approach, the expected 0--1 loss (= misclassification error rate) is estimated by averaging the computed 0--1 loss over a sample of size $400$ (200 per model) from the prior predictive distribution. The size of this prior predictive sample is considerably smaller than for the two supervised classification approaches due to computational limitations. Therefore, the volatility of our likelihood-based misclassification error rate estimates is much higher than for the supervised classification methods, so we expect our optimisation procedure to be less stable. The expected loss surface for the one-dimensional design is depicted in Figure~\ref{fig:1d_results_2models} in Appendix~\ref{app:epi2_results}.

However, setting the prior predictive sample size for the likelihood-based approach to 10K as well would have made it infeasible to find an optimal design in a reasonable amount of time. Running the classification methods is still much more time-efficient than evaluating the likelihood function many times, especially for the SI model in high dimensions, see also Appendix~\ref{app:performance}. Therefore, we only used the likelihood-based approach to find designs up to a total design dimension of $n = 8$. Furthermore, for the design search we used a relatively coarse grid with a spacing of $0.5$ days between the limits $0.5$ and $10$ days. We used the same design grid for all approaches.

We consider various combinations of the number of realisations, $q$, and the number of observations per realisation, $n_d$. All the design methods described in this section are applied to all integer combinations of $1 \leq n_d \leq 4$ and $1 \leq q \leq 4$ for which the total number of observations $n = q \cdot n_d$ does not exceed 8. We also investigate higher-dimensional designs, where we only employ the supervised classification approaches but not the likelihood-based approach. As higher-dimensional settings we consider all integer combinations of $q$ and $n_d$ which amount to a total number of observations of either $n = 12$, $24$, $36$, or $48$, and where $1 \leq n_d \leq 4$.

The optimal designs found with the different methods are validated in two ways. Firstly, for each observation from a sample of size 2K (1K per model) from the prior predictive distribution, the posterior model probabilities are computed using the generalised Gauss-Hermite quadrature approximation to the marginal likelihood with $Q=30$ quadrature points for the death model and up to $Q=30^2$ quadrature points (minus some pruned points) for the SI model. The resulting distributions of posterior model probabilities are displayed in Appendix~\ref{app:epi2_results}.

\begin{figure}[hbtp!]
	\centering
	\begin{tabular}{cc}
		\includegraphics[height=0.17\textheight]{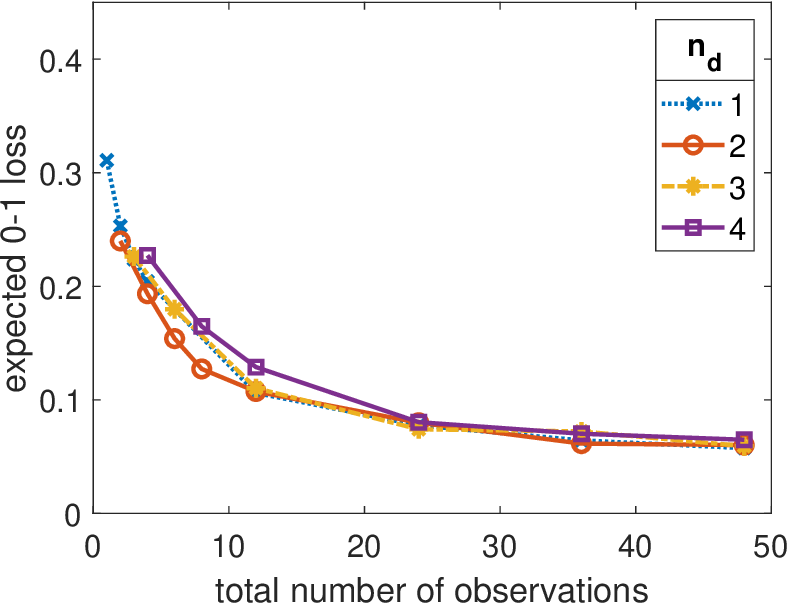} & \includegraphics[height=0.17\textheight]{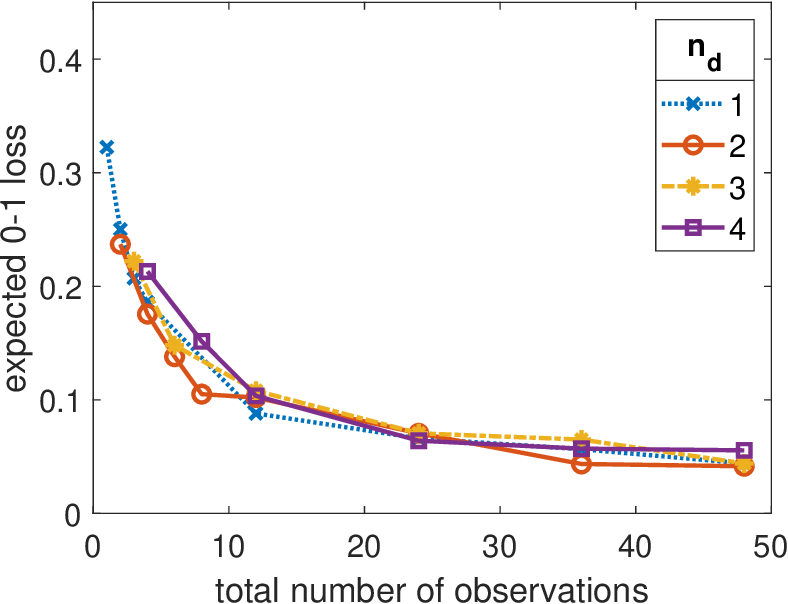}  \\
		\multicolumn{2}{c}{tree classification with cross-validation} \\[1ex]
		\includegraphics[height=0.17\textheight]{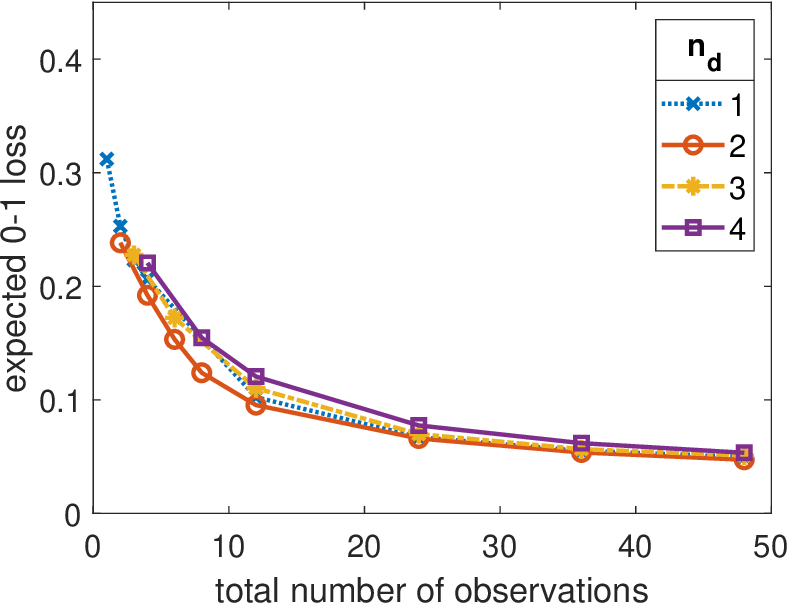} &  \includegraphics[height=0.17\textheight]{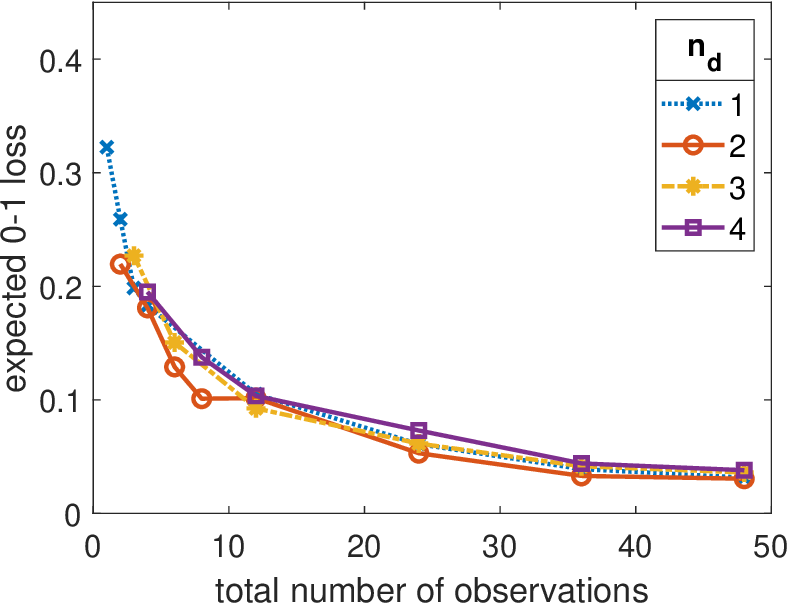} \\ 
		\multicolumn{2}{c}{random forest classification} \\[1ex]
		\includegraphics[height=0.17\textheight]{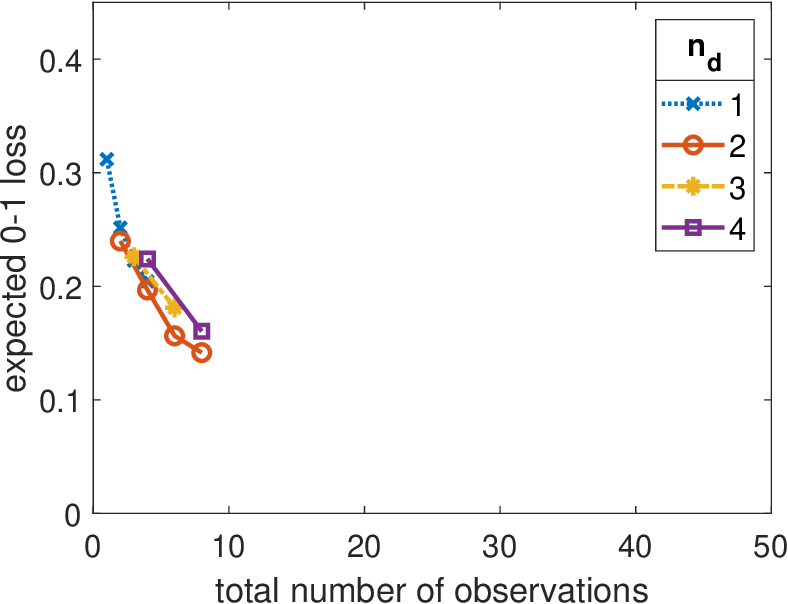} & \includegraphics[height=0.17\textheight]{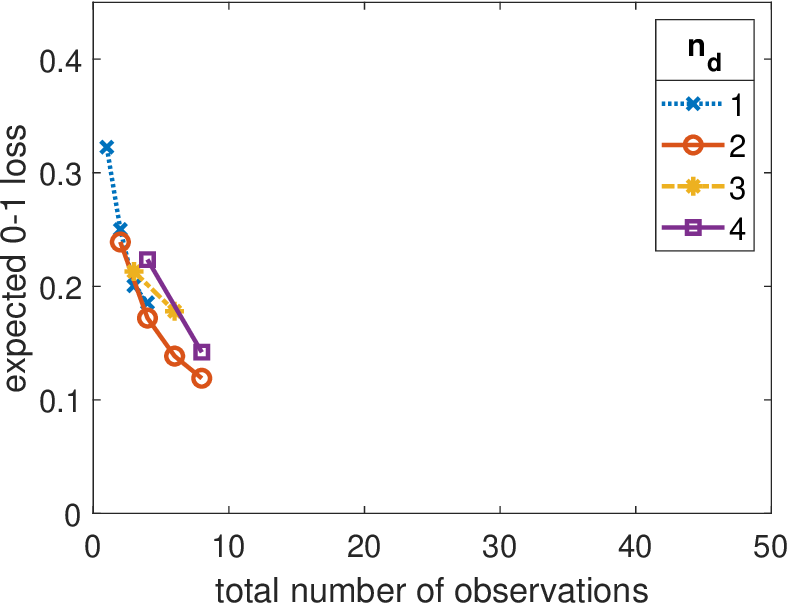} \\
		\multicolumn{2}{c}{using Laplace approximation to true likelihood} \\[1ex]
		\includegraphics[height=0.17\textheight]{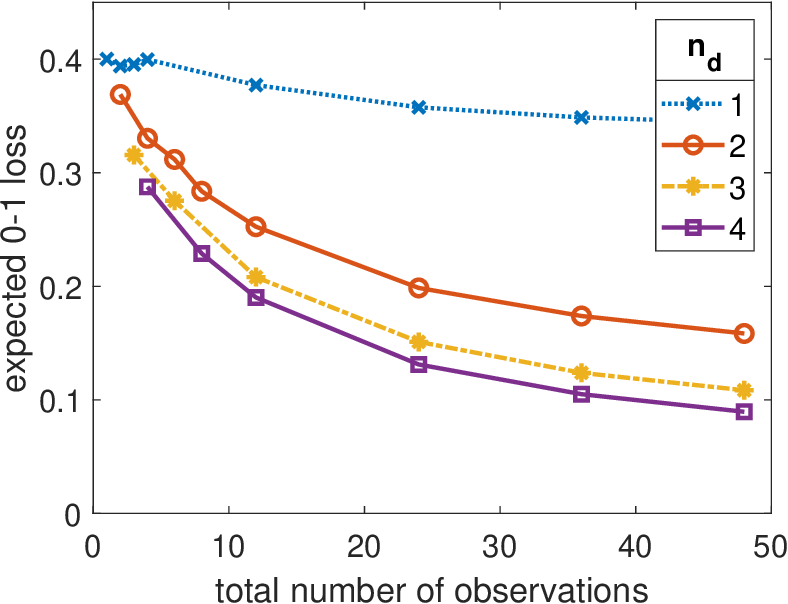} & \includegraphics[height=0.17\textheight]{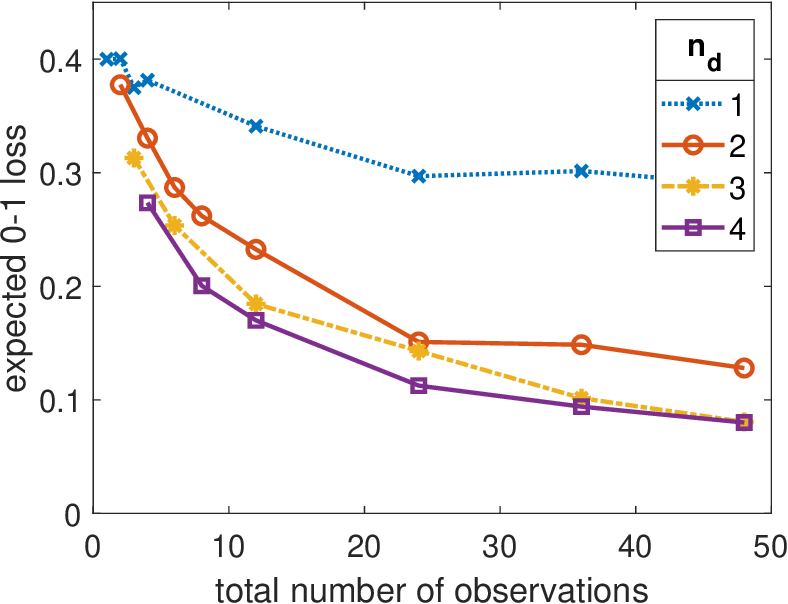} \\
		\multicolumn{2}{c}{equispaced designs}
	\end{tabular}
	\caption{Misclassification error rates computed using random forest classification with training and test samples of size 20K, averaged over 100 repetitions of the classification procedure (left column) and misclassification error rates computed using the Gauss-Hermite quadrature approximation to the marginal likelihood over 2K prior predictive simulations (right column) evaluated at various optimal designs for different methods (in the rows) for the infectious disease example with two models. The total number of observations ($n = q \cdot n_d$) is plotted on the x-axis of each graph. Each line connects the observed values of the  misclassification error rate as the number of realisations $q$ increases for a particular value of $n_d$. \label{fig:results_epi_2models_losses}}
\end{figure}

We can also use the estimates for the posterior model probabilities to compute estimates of the misclassification error rates for each of the methods and dimension settings. These estimates are provided in the plots on the right-hand side of Figure~\ref{fig:results_epi_2models_losses}, where each row contains the results for one design method, the x-axis of each plot shows the total number of observations, $n$, and each line within each plot displays the results for a particular setting of $n_d$. Alternatively, one can use a supervised classification method to estimate the misclassification error rates for the optimal designs. In our case, we use a random forest with training and test sets of size 20K (10K per model). The random forest classification procedure is repeated 100 times and the average misclassification error rate over the 100 repetitions is taken. The random forest-based validation results are shown in the plots on the left-hand side of Figure~\ref{fig:results_epi_2models_losses} analogous to the likelihood-based validation results.

From Figure~\ref{fig:results_epi_2models_losses}, it is evident that the misclassification error rates computed by the random forests are very close to the likelihood-based misclassification error rate computations. In most cases, the random forest-based estimates of the misclassification error rate are  a little higher than the likelihood-based estimates. This is no surprise since the likelihood-based estimates are directly targeting the Bayes error rate. However, the trajectories of the misclassification error rates as a function of $n$ are very similar for both validation methods. This suggests that for this example random forests are suitable to validate and compare the efficiency of the resulting designs. In addition, it is reasonable to expect that the designs which are optimal for the random forest classification approach are close to the true optimal designs.

One can also see from Figure~\ref{fig:results_epi_2models_losses} that for a fixed total number of observations there is not much difference in the performance of the different design configurations, at least for the small values of $n_d$ that we considered. It seems that having $n_d = 2$ observations per realisation is the most optimal choice, but only by a small margin.

\subsection{Macrophage Model} \label{subsec:macro}

\subsubsection{Aim of Experiment}

A common challenge in experimental biology is identifying the unobserved heterogeneity in a system. Consider for example the experimental system in \cite{Restif:2012}. In this system, the authors wished to identify the role of antibodies in modulating the interaction of  intracellular bacteria -- in particular, \emph{Salmonella enterica} serovar Typhimurium (\emph{S.} Typhimurium) -- with human phagocytes, inside which they can replicate. The experiments assessed the effect of a number of different human immunoglobulin subclasses on the intracellular dynamics of infection by combining observed numbers of bacteria per phagocyte with a mathematical model representing a range of different plausible scenarios. These models were fit to experimental data corresponding to each human immunoglobulin subclass in order to determine the underlying nature of the interactions between the antibodies and bacteria. In these experiments, the data demonstrated bimodal distributions in the number of intracellular bacteria per phagocytic cell. The aim was to identify the source of the unobserved heterogeneity in the system that caused the observed patterns. Specifically, is there underlying heterogeneity in the bacteria's ability to divide inside phagocytes, or is it the phagocyte population which is heterogeneous in its ability to control bacterial division? In this context, the classification approach allows us to find the experimental design which best enables us to discriminate between these competing hypotheses -- (1) unobserved heterogeneity in the bacteria, (2) in the cells, or (3) no heterogeneity.

\subsubsection{Experimental Procedure}

We give a brief account of the experimental procedure:
\begin{itemize}
	\item After bacterial opsonisation (i.e., the process by which bacteria are coated by antibodies), the bacteria are exposed to the phagocytic cells for a total of $t_{exp}$ hours, which can take the values $t_{exp} \in \{0.10, 0.20, \ldots, 1.50\}$ hours. During this time, phagocytosis occurs, i.e., the bacteria are internalised by the phagocytic cells.
	\item Next, the cells are treated with gentamicin, an antibiotic that kills extracellular bacteria, so that phagocytosis stops.
	\item At each of the $n$ observation times $\bm{t}_{obs}=(t_1,\dots,t_n)$ hours post-exposure, two random samples of $S$ cells each are taken from the overall population of cells: one sample to count the proportion of infected cells (under a low-magnification microscope), and one sample of infected cells to determine the distribution of bacterial counts per infected cell (at higher magnification).
\end{itemize}
That is, a design is composed of $\vectd=(t_{exp}; \bm{t}_{obs})$. The full experimental procedure is detailed in \cite{Restif:2012}.

For the purpose of our example, we consider a realistic scenario where we have the resources to count a fixed number of cells, $N_{cells}=200$. These cells are then equally split between all the observation times and the two independent observational goals at each observation time, so $S = \lfloor N_{cells} / (2 \, n) \rfloor$.

\subsubsection{Model}

We consider three mathematical models, based on \cite{Restif:2012}, to represent the three competing hypotheses about heterogeneity. These models are continuous-time Markovian processes that simulate the dynamics of intracellular bacteria within macrophages. Model (1) tracks the joint probability distribution of the number of replicating and non-replicating bacteria within a single macrophage, assuming all macrophages in a given experiment are from the same type. In model (2), each macrophage has a fixed probability $q$ of being refractory, in which case it only contains non-replicating bacteria, and a probability $1-q$ of being permissive, in which case it only contains replicating bacteria. In model (3), all macrophages are permissive and all bacteria are replicating.

Simulations from the models are based on simulations of bacterial counts for the individual macrophages. As for our infectious disease examples, we can use the efficient Gillespie algorithm \citep{gillespie1977}. The outcomes for the individual macrophages are then aggregated to obtain the same type of data as observed in the real experiment.

It is possible but cumbersome to compute the likelihood functions for all the models. Computing the likelihood involves solving a system of linear differential equations, which can be achieved by using matrix exponentials. However, these operations are quite expensive so that computing the posterior model probabilities becomes very costly. Computing the expected losses and searching for an optimal design can be considered intractable in these circumstances. In contrast, simulations from the models can be obtained very quickly.

Appendix~\ref{app:macro_models} contains a more detailed description of the Markov process models, the simulation procedure, the likelihood function, and the prior distributions. 

\subsubsection{Results}

We use the machine learning classification approach using classification trees with cross-validation or random forests to determine the optimal designs for discriminating between the three competing models (one model corresponding to each hypothesis) with respect to the misclassification error rate. 
It is assumed \emph{a priori} that the models are equally likely. We use 5K simulations from the prior predictive distribution of each model during the design process. The design grid for $t_{obs}$ goes from $0.25$ to $10$ with a spacing of $0.25$. 
The optimal designs are given in Appendix~\ref{app:macro_designs}. The tree and the random forest classification approaches lead to very similar designs.

Similar to the other examples, we assess each design by producing 10K new simulations under each model at that design and using these to train a random forest with 100 trees. A further 10K new simulations per model are then used to estimate the misclassification error rate. This is repeated 100 times for each design. The estimated misclassification error rates for the designs found under the tree and random forest classification approaches are shown in Table~\ref{tab:macro_misclass_error}. For comparison, we also include the estimated error rates for the equispaced designs.

\begin{table}
	\centering
	\caption{Average misclassification error rates for the optimal designs obtained under the classification approaches using trees or random forests and for the equispaced designs for the macrophage model. The average misclassification error rates were calculated by repeating the random forest classification procedure 100 times and taking the average. The standard deviations are given in parentheses\label{tab:macro_misclass_error}}
	\begin{tabular}{cccccc}
		\hline\noalign{\smallskip}
		Design & \multicolumn{1}{c}{$n = 1$} & \multicolumn{1}{c}{$n = 2$} & \multicolumn{1}{c}{$n = 3$} & \multicolumn{1}{c}{$n = 4$} & \multicolumn{1}{c|}{$n = 5$} \\
		\noalign{\smallskip}\hline\noalign{\smallskip}
		Tree  & 0.1928 & 0.1323 & 0.1433 & 0.1469 & 0.1483 \\
		& (0.0024) & (0.0021) & (0.0022) & (0.0019) & (0.0022) \\
		RF    & 0.1925 & 0.1325 & 0.1408 & 0.1410 & 0.1465 \\
		& (0.0022) & (0.0021) & (0.0022) & (0.0021) & (0.0021) \\
		Equi & 0.2442 & 0.1974 & 0.1928 & 0.1912 & 0.1935 \\
		& (0.0027) & (0.0023) & (0.0023) & (0.0025) & (0.0024) \\
		\noalign{\smallskip}\hline
	\end{tabular}
\end{table} 

\begin{figure}
	\centering
	\includegraphics[height=0.25\textheight,width=0.85\textwidth]{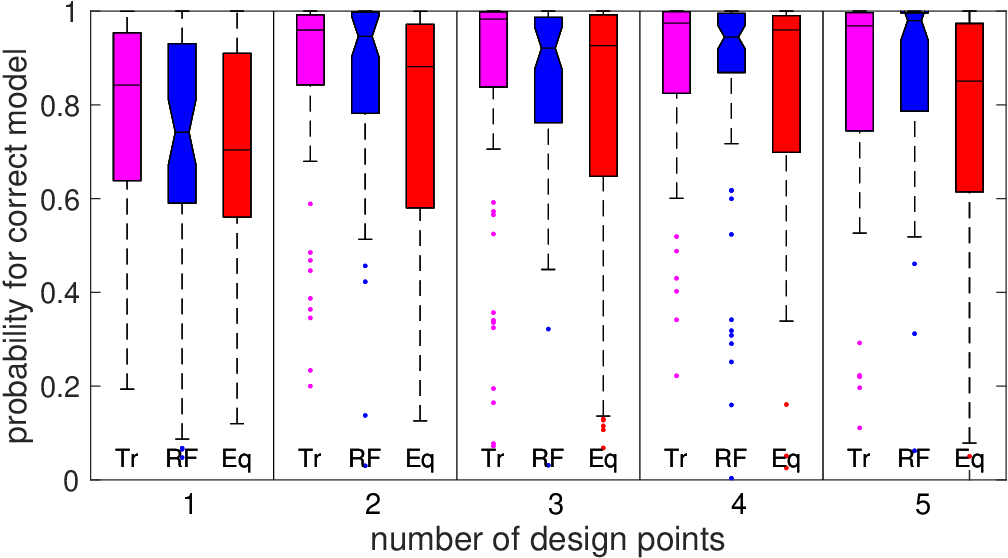}
	\caption{For each of the optimal designs obtained by the different approaches for 1 -- 5 observations in the macrophage example, display the distribution of estimated posterior model probabilities of the correct model over 80 process realisations (20 from each of the three models) simulated from the prior predictive distribution at the respective optimal design. For each number of design points, the magenta boxplot on the left-hand side is for the tree classification design, the notched blue boxplot in the middle is for the random forest classification design (rf), and the red boxplot on the right-hand side is for the equispaced design} \label{macro:model_probs}
\end{figure}

We are also interested in the posterior model probabilities at the different optimal designs. At each optimal design, we simulate 20 process realisations under the prior predictive distribution of each model. For each process realisation, we approximate the posterior model probability of the model that generated the data using importance sampling \citep[see, e.g.,][]{Liu2001} with 50K simulations from the importance distribution. In our case, the prior distribution serves as the importance distribution. Figure~\ref{macro:model_probs} shows boxplots for the distributions of the posterior model probabilities of the correct model over the prior predictive simulations for the different optimal designs. The computations required to generate one of these boxplots ranged from $5.9$ hours to $17.2$ hours using up to 24 parallel threads. In contrast, it took less than two minutes to obtain one estimate of the misclassification error rate using a random forest with training and test samples of size 30K each.

Table~\ref{tab:macro_misclass_error} indicates that $n = 2$ observation times yield the optimal  classification power when using trees and random forests, even though the posterior model probabilities of the correct model keep increasing until at least $n = 4$ (see Figure~\ref{macro:model_probs}). For more than two observations, the higher data dimension impedes the classification accuracy of those classification methods and more than offsets the gains from having marginally more information in the data due to the more optimal allocation of resources to the different observation times. However, there are no substantial improvements in the posterior probabilities after $n = 2$. Both machine learning classification approaches lead to very efficient designs for all design sizes. 

Overall, the ability to correctly classify output from the three models and thus to decide between the three competing hypotheses is very good at all the optimal designs. This suggests that we are able to identify with high certainty if heterogeneity is present, and if so, whether the bacteria or the human cells are the source of this heterogeneity.

\section{Discussion} \label{sec:discussion}

We introduce a new simulation-based Bayesian experimental design approach for model discrimination where the expected loss is estimated via a supervised classification procedure. This approach requires significantly fewer simulations than other simulation-based approaches based on ABC. Furthermore, efficient, flexible and fast classification methods such as classification trees or random forests can cope with medium to high data dimensions without imposing strict structural assumptions. Therefore, the classification approach significantly increases the scope of design problems which can be tackled compared to previous approaches. For example, optimal designs for the hierarchical logistic regression example could previously only be obtained by assuming normal-based approximations \citep{Overstall2018}. The high dimensions of the summary statistics for the macrophage and the spatial extremes example render the ABC approach unsuitable or even infeasible (see the limitations encountered by \citet{hainy2016} in a parameter estimation design problem for spatial extremes). For all the examples in this paper the classification approach is significantly more time-efficient than any of the other approaches we have considered. 
The most crucial requirement for the applicability of the classification approach is that efficient samplers are available for all the models.

The methodology we present is rather general. We find that classification trees and random forests work very well in conjunction with the 0--1 loss. They are less suitable for loss functions that directly depend on the posterior model probability such as the multinomial deviance loss. However, one may use any other classification method that is quick and leads to accurate predictions for the application at hand. For example, logistic regression provides natural and smooth estimates for the posterior model probabilities, but it is also less flexible due to the linear form of the predictor. Generalised additive models may improve the accuracy of logistic regression at the expense of a higher computing time. Other fast classification methods include linear discriminant analysis and its extensions like mixture and flexible discriminant analysis. If a higher computing time for the classifier is acceptable and a high predictive power is desired, more elaborate methods such as neural networks may be applied.
In general, for most applications it will be preferable to use a classification method where the optimal choice of the tuning parameters is insensitive to the selected design or where standard settings are available that work reasonably well in most circumstances. Otherwise the optimal tuning parameters have to be determined for each new design, for example via cross-validation. Apart from choosing different classification methods, one may also consider different loss functions. The choice of the loss function determines the functional form of the penalty for not correctly estimating the true class. Alternatives to the 0--1 loss and multinomial deviance loss include the exponential, logit, and hinge loss functions. For an overview of all the aforementioned methods and loss functions, see \citet{Hastie2009}.

One disadvantage of any simulation-based design approach is that the objective function to optimise over is stochastic. Even though the classification approach reduces the stochastic noise compared to ABC, the optimisation algorithm needs to take the noise into account. Our focus in this paper is not on optimisation, so we use a simple coordinate exchange algorithm on a discretised design space. However, our design algorithm may get stuck at suboptimal solutions if the noise is too large. We try to alleviate that problem by using parallel runs with randomly selected initial designs and by reconsidering the last few designs visited in each run, where the noise is reduced at these designs by evaluating the objective function several times. Furthermore, we employ a Gaussian process regression post-processing step where we use the data collected during the coordinate exchange procedure to train a Gaussian process in order to obtain a smooth estimate of the loss surface. This estimate of the loss surface is then minimised to find another candidate for the optimal design. Our algorithm leads to plausible optimal designs in our examples. For all our examples, the efficiencies of the optimal designs follow a reasonable trajectory as the design sizes are increased. Furthermore, the differences between the design approaches are consistent across the design sizes.  For high-dimensional designs with a continuous design space and noisy objective functions, the approximate coordinate exchange algorithm \citep{overstall_woods_2017} is a theoretically sound and efficient alternative. \citet{Price2018} present an `induced natural selection heuristic' algorithm that can cope with moderate to high dimensions and noisy objective functions. Other possible optimisation algorithms suited for noisy objective functions in small to moderate dimensions include `simultaneous perturbation, stochastic approximation' \citep{Spall1998-1} and the rather robust Nelder-Mead algorithm \citep{NelderMead}.

For future work, we will consider extending our approach to Bayesian parameter estimation designs. Another possibility is to attempt to utilize our classification approach within a sequential design setting in the spirit of \citet{kleinegesse2020}.

\section*{Acknowledgements}

MH was funded by the Austrian Science Fund (FWF): J3959-N32. DJP and OR acknowledge funding support from the Biotechnology and Biological Sciences Research Council grant BB/M020193/1 awarded to OR. CD was supported by an Australian Research Council Discovery Project (DP200102101). Computational resources and services used in this work were provided by the HPC and Research Support Group, Queensland University of Technology, Brisbane, Australia, and by the Scientific Computing Administration of the Johannes Kepler University, Linz, Austria.

\bibliographystyle{apalike} 
\bibliography{biblio}

\begin{appendices}

\clearpage

\section{Properties and Estimation of CARTs and Random Forests}
\label{app:CART_RF}

The CART algorithm generates a binary tree where each internal node consists of one binary rule that involves exactly one of the features, e.g., $y_3 < 10$. The feature space is split recursively at the internal nodes according to the binary rules, thereby creating a partition of the feature space consisting of hyperrectangles aligned along the feature axes. Each terminal node or leaf contains all the observations in the training sample which fall into the associated hyperrectangle. The hyperrectangle region of the feature space associated to a leaf is defined by the binary rules in the nodes leading to that leaf. For a classification tree, the class label which is assigned to a particular region of the feature space is determined by majority vote of the training samples in the corresponding leaf. The class proportions of the training samples in a leaf can be used to obtain crude estimates of the posterior class probabilities for observations falling into the associated feature space region.

Trees are constructed recursively beginning at the root. Each leaf contains those training samples that meet all the conditions leading down the path from the root to that leaf. If no stopping criterion is met and the leaf's sample contains more than one distinctive feature value, the leaf is split into two daughter nodes and becomes an internal node. To that end, the binary rule that splits the sample at the node into two subsets for the two new leaves has to be determined. The feature variable and the split point are selected such that a given criterion is minimised across all subsets. For classification, the default criterion of node impurity used for growing the tree is the Gini index $\sum_{m=1}^K \hat{p}_m (1 - \hat{p}_m)$, where $\hat{p}_m$ is the proportion of training samples from class $m$ in the node.

As noted for example by \citet{Hastie2009}, fully grown trees, where no further splits are possible, usually overfit the data. Therefore, one might stop earlier and define a minimum size of a node or a parent node. More preferably, one can grow a full tree and prune it afterwards according to a cost-complexity criterion that incorporates the node impurities and the number of terminal nodes. For an efficient algorithm to find the optimal pruned tree see \citet{Breiman1984}. The optimal choice of the minimum node size or the tuning parameters for cost-complexity pruning can be determined by cross-validation.

Exploiting the similarities between trees and nearest neighbour classifiers, \citet{Breiman1984} show that the misclassification error rate of a fully grown tree is bounded above by twice the Bayes error rate, which has been shown for 1-nearest neighbour classification by \citet{Cover1967}. It also follows from \citet{Breiman1984} that the misclassification error rate of a classification tree attains the Bayes error rate as the sample size tends to infinity.

The CART algorithm automatically assumes equal prior class probabilities, even if the training sample is not balanced. This is achieved by dividing the class counts in the leaves by the overall class counts in the training sample. Therefore, a given leaf is classified as 
\begin{equation}
\underset{m \in \{1,\ldots,K\}}{\arg \max} \frac{N_m(\mathrm{leaf})}{N_m(\mathrm{root})}, \label{eq:count_fractions}
\end{equation}
where $N_m(\mathrm{leaf})$ and $N_m(\mathrm{root})$ are the number of observations from class $m$ in the leaf and in the entire training sample, respectively. One may switch off this mechanism if the training sample reflects the true prior class probabilities. It is also possible to provide user-defined prior class probabilities. In that case the fractions in \eqref{eq:count_fractions} are multiplied by these user-defined prior probabilities.

Due to the recursive nature of their construction, trees exhibit a high variance. A suboptimal split at a top node affects the whole tree structure below that node, so slight changes in the data might lead to widely different trees. To reduce the variance, an ensemble method called \emph{bagging} was proposed by \citet{Breiman1996}.

Bagging means to draw $B$ bootstrap samples from the training sample and to apply the classification method to each bootstrap sample. As a result, one obtains $B$ different classifiers trained on the $B$ bootstrap samples. The class of a new observation $\vecty_*$ is predicted by casting a majority vote among the class predictions returned by the $B$ classifiers. Bagging has been shown to be particularly useful for classification methods that are unstable and exhibit a high variance such as trees and neural networks, where bagging can lead to a substantial reduction of the variance.

An ensemble of bagged trees might be highly correlated, which has a negative effect on the variance of the bagged predictor. To reduce the variance further, \emph{random forests} \citep{Breiman2001} seek to de-correlate the trees by considering only a random subset of the feature variables  for splitting the tree at each node when the trees are grown. For classification, the default setting is to consider $\lfloor \sqrt{p} \rfloor$ variables at each node, where $p$ is the total number of feature variables. The random selection of feature subsets reduces the correlation between the trees but it also increases the bias of the trees. On the other hand, the trees used in random forests are normally not pruned, and unpruned trees have less bias than pruned trees.

Random forests are able to account for overfitting when computing the misclassification error rate without the need to employ cross-validation or to generate a separate test set. Each tree is constructed from a bootstrap sample of the training set. The bootstrap samples are drawn from the training set with replacement. It follows that about one third of the training set is omitted in each bootstrap sample. It is therefore possible to make predictions for each training sample $\vecty_i$ based on those trees where $\vecty_i$ does not appear. These \emph{out-of-bag} class predictions can then be used to estimate the misclassification error rate. Out-of-bag estimation is qualitatively similar to leave-one-out cross-validation.

Random forests also provide estimates for the posterior model probabilities $p(m|\vecty,\vectd)$. The estimates are formed by simply averaging the posterior model probability estimates obtained from the trees in the forest. Due to the averaging, the posterior model probability estimates of the random forest are much more stable than those given by a single tree.

Unfortunately, there are some difficulties when trying to estimate the expected multinomial deviance loss by classification trees or random forests using cross-validation, independent test samples or out-of-bag class predictions. For a single tree, the lack of smoothness of its posterior model probability estimates means that in an independent test sample there are almost certainly some observations for which the estimated posterior model probability of the true model is 0. Therefore, minus the logarithm of the posterior model probability is $\infty$ and the expected multinomial deviance loss is also $\infty$. When evaluating a random forest on a test sample or when using out-of-bag class predictions, it is also very likely that some probability estimates are 0. In our examples, we therefore set the estimated posterior model probability to a value of $\varepsilon = 0.001$ whenever the posterior model probability is estimated to be 0. The lower the value of $\varepsilon$, the higher the variance of the expected loss estimate, because it becomes very sensitive to the number of posterior model probabilities estimated to be 0. In our experience, setting $\varepsilon$ to $0.001$ was striking a good balance between being reasonably close to 0 while not exhibiting excessive variability.

For all our examples except the spatial extremes example, we use the Matlab functions \texttt{fitctree} and \texttt{TreeBagger} to train classification trees and random forests, respectively. We mostly use the default settings of those functions. That is, for classification trees the maximum number of splits is set to the sample size -- 1, the minimum leaf size is 1 and the minimum internal node size is 10. This leads to rather deep trees. The trees are not pruned. The default settings of \texttt{TreeBagger} amount to following the standard methodology of random forests as outlined in this section. The random forests we employ are generally made up of 100 trees and utilise out-of-bag class predictions.


\clearpage

\section{Modification of Coordinate Exchange Algorithm}\label{app:mce_algo}

\newcommand{\availset}{\mathcal{A}}
\newcommand{\candidateset}{\mathcal{C}}
\newcommand{\visitset}{\mathcal{V}}
\newcommand{\designset}{\mathcal{D}}
\newcommand{\lossset}{\mathcal{L}}
\newcommand{\avglossset}{\mathcal{AL}}

\begin{footnotesize}
	\begin{algorithm}[H]
		\caption{Modification of coordinate exchange algorithm (one parallel instance)\label{algo:CE}}
		\SetKw{TRUE}{true}
		\SetKw{FALSE}{false}
		\KwIn{Set of available design points $\availset$; initial design $\vectd = \{d_1,\ldots,d_n\}$ consisting of $n = \text{card}(\vectd)$ design points; function $\mathtt{estimate\_loss}(\vectd)$ that estimates the expected loss for a given design $\vectd$; numbers $p$ and $q$: for the last (at most) $p$ designs visited, the expected loss is estimated $q$ times.}
		\KwOut{Set $\visitset_{\mathrm{GP}}$ containing the last designs visited; set $\lossset$ containing $q$ expected loss value estimates for each design in $\visitset_{\mathrm{GP}}$; preliminary optimal design $\vectd_{\mathrm{CE}}^*$ after running one instance of the modified coordinate exchange algorithm and the corresponding expected loss value $l_{\mathrm{CE}}^*$.}
		
		\texttt{swaps} = \TRUE \;
		\texttt{loss} = $\mathtt{estimate\_loss}(\vectd)$\;
		No designs visited so far: $\visitset = \{\}$\;
		\While{\texttt{swaps}}{ \label{algo_start_swaps}
			\texttt{swaps} = \FALSE \; 
			\For{$i = 1$ \KwTo $n$}{ \label{algo_start_sweep}
				Determine the set of candidate design points $\candidateset \subseteq \availset$\;	\label{algo_line}
				$m = \text{card}(\candidateset)$\;
				Clear \texttt{lossvec}\;
				\For{$j = 1$ \KwTo $m$}{
					$\vectd^{\text{try}} = \vectd$\;
					Replace element $i$ of $\vectd^{\text{try}}$ with element $j$ of $\candidateset$\;
					$\mathtt{lossvec}[j] = \mathtt{estimate\_loss}(\vectd^{\text{try}})$\;
				}
				Let $\mathtt{minloss} = \min(\mathtt{lossvec})$ and $k$ be the index for which $\mathtt{lossvec}[k]$ is equal to \texttt{minloss}\;
				\If{$\mathtt{minloss} < \mathtt{loss}$}{
					Replace element $i$ of $\vectd$ with element $k$ of $\candidateset$\;
					$\mathtt{loss} = \mathtt{minloss}$\;
					\texttt{swaps} = \TRUE \;
					Add $\vectd$ to $\visitset$, the history of designs visited so far;	
				}
			} \label{algo_end_sweep}
		} \label{algo_end_swaps}
		Let $h = \text{card}(\visitset)$ be the number of designs visited, where $\visitset = \{\vectd_1, \ldots, \vectd_h\}$\;
		Let $r = \min(h,p)$\;
		Let $\lossset = \{\}$ and $\avglossset = \{\}$\;
		\For{$i = 1$ \KwTo $r$}{
			Clear \texttt{lossvec}\;
			\For{$j = 1$ \KwTo $q$}{
				$\mathtt{lossvec}[j] = \mathtt{estimate\_loss}(\vectd_{h-i+1})$\;
			}
			Add $\mathtt{lossvec}$ as $i$th element to $\lossset$\;
			Add $\text{mean}(\mathtt{lossvec})$ as $i$th element to $\avglossset$\;
		}
		Let $s$ be the index of the smallest element in $\avglossset$ and $l_{\mathrm{CE}}^*$ be the corresponding value\;
		
		Return $\vectd_{\mathrm{CE}}^* = \vectd_{h-s+1}$, $l_{\mathrm{CE}}^*$, $\visitset_{\mathrm{GP}} = \{\vectd_h,\vectd_{h-1},\ldots,\vectd_{h-r+1}\}$, $\lossset$\; 
	\end{algorithm}
\end{footnotesize}

\newpage

\begin{footnotesize}
	\begin{algorithm}[H]
		\caption{Gaussian process regression post-processing step\label{algo:GPR}}
		\SetKw{TRUE}{true}
		\SetKw{FALSE}{false}
		\KwIn{Sets of visited designs $\visitset_{\mathrm{GP},i}$ and sets of corresponding expected loss estimates $\lossset_i$ ($q$ values for each design in $\visitset_{\mathrm{GP},i}$) for $i = 1,\ldots,I$ parallel runs of the modified coordinate exchange algorithm (Algorithm~\ref{algo:CE}); preliminary optimal designs $\vectd_{\mathrm{CE},i}^*$ and corresponding estimated expected loss values $l_{\mathrm{CE},i}^*$ for $i = 1,\ldots,I$ parallel runs of Algorithm~\ref{algo:CE}; function $\mathtt{estimate\_loss}(\vectd)$ that estimates the expected loss for a given design $\vectd$.}
		\KwOut{Overall optimal design $\vectd^*$.}
		
		Combine sets of visited designs $\visitset_{\mathrm{GP},i}$ for all $i = 1,\ldots,I$ parallel runs into one set $\visitset_{\mathrm{GP}}$. Do the same for the sets of expected loss estimates $\lossset_i$ and combine them into $\lossset$\;
		
		Train Gaussian process with the expected loss values in $\lossset$ as (univariate) response variable and the visited designs $\visitset_{\mathrm{GP}}$ as predictors (each design is repeated $q$ times)\;
		
		Find the minimum value of the predictive mean function of the Gaussian process over the design space using some generic optimisation function. Let the design at the minimum be denoted by $\vectd_{\mathrm{GP}}^*$\;
		
		Set $\vectd_{\mathrm{CE}}^*$ to the design $\vectd_{\mathrm{CE,i}}^*$ from parallel run $i$ with the lowest value for $l_{\mathrm{CE},i}^*$\;
		
		Clear \texttt{lossvec\_CE}, \texttt{lossvec\_GP}\;
		\For{$j = 1$ \KwTo $100$}{
			$\mathtt{lossvec\_CE}[j] = \mathtt{estimate\_loss}(\vectd_{\mathrm{CE}}^*)$\;
			$\mathtt{lossvec\_GP}[j] = \mathtt{estimate\_loss}(\vectd_{\mathrm{GP}}^*)$\;
		}	
		
		\uIf{$\mathrm{mean}(\mathtt{lossvec\_GP}) < \mathrm{mean}(\mathtt{lossvec\_CE})$}{
			Return $\vectd^* = \vectd^*_{\mathrm{GP}}$\;
		}
		\Else{
			Return $\vectd^* = \vectd^*_{\mathrm{CE}}$\;
		}
		
	\end{algorithm}
\end{footnotesize}

In all our examples we set $p = 6$ and $q = 10$.

Algorithm~\ref{algo:CE} can be run in parallel for different initial designs $\vectd$ to account for multimodality and local optima. We conduct 20 parallel runs in all our examples.

The selection of the candidate design points in Line~\ref{algo_line} of Algorithm~\ref{algo:CE} depends on the example. For the logistic regression and the macrophage example, there is no restriction and $\candidateset = \availset$. For the other examples, the current design points $d_1,\ldots,d_n$ in $\vectd$ have to be excluded since each design point can only be selected once. Furthermore, for the spatial extremes example we only consider design points with the same x- or y-coordinate as the current design point.

The sets of best designs found in each of the parallel runs of Algorithm~\ref{algo:CE} and their associated estimated expected loss values are combined and used as inputs for Algorithm~\ref{algo:GPR}. In Algorithm~\ref{algo:GPR}, a Gaussian process \citep[GP; see, e.g.,][]{Rasmussen2006} is trained on the combined data from all the parallel runs in order to obtain a smooth estimate of the expected loss surface by means of the predictive mean function of the GP. The predictive mean function is minimised and a new candidate for the optimal design is obtained. Since the predictive variance is relatively high in our examples, we compare this design to the best design found through Algorithm~\ref{algo:CE} without the GP post-processing step of Algorithm~\ref{algo:GPR}. To reduce the uncertainty for this comparison, we estimate the expected loss 100 times at each of the two designs and take the design with the lower average expected loss value as the overall optimal design. We do not perform the GP post-processing step for the spatial extremes example.

For Gaussian process regression, we use the default settings of the Matlab function \texttt{fitrgp} except that all the predictors are standardised. The default kernel function used is the squared exponential kernel and a constant GP prior mean is assumed. To find the optimal value for the initial value of the prior noise variance parameter, Bayesian optimisation is conducted with respect to the cross-validation loss.

For finding the minimum of the GP's predictive mean function, we use the Nelder-Mead simplex algorithm \citep{NelderMead}. Restrictions of the design space are considered by employing suitable transformations. For example, design points with the restriction $d_i \in (a,b]$ are transformed by the logit transformation to $\tilde{d}_i = \log\{z_i/(1-z_i)\}$, where $z_i = (d_i - a)/(b-a)$.


\section{Computational Performance Measures for Examples in Section~\ref{sec:examples}} \label{app:performance}

In this section, we provide some measures of computational performance for the design search algorithms used for the three examples in Section~\ref{sec:examples}. As explained in Appendix~\ref{app:mce_algo}, we ran Algorithm~\ref{algo:CE} twenty times in parallel, so there is a distribution of runtimes to consider. We focus on the exchange part of Algorithm~\ref{algo:CE} (lines \ref{algo_start_swaps} to \ref{algo_end_swaps}), because this is usually the most time-consuming part. However, it is not sensible to just compare the distributions of runtimes of the exchange part because the number of sweeps through the design grid until the algorithm converges (i.e., the number of passes through the \texttt{while}-loop) is random. Therefore, in Tables~\ref{tab:perform_epi} to \ref{tab:perform_macro} we provide the distributions for the runtimes per sweep for all the examples in Section~\ref{sec:examples}. More precisely, we state the mean and the standard deviation of the runtime per sweep over the parallel runs. As expected, one can see that these distributions exhibit little variation. The reason is that the number of calls of the \texttt{estimate\_loss} function in a sweep through the design grid is fixed for any given example and design configuration. Within any call to \texttt{estimate\_loss}, the simulated sample sizes are fixed (for details see the example settings for the respective models in Section~\ref{sec:examples}). The sample sizes for trees and random forests are always the same, so differences in runtimes can solely be attributed to the classification method. In general, for all our examples simulation is rather efficient, so the simulation effort is only a minor fraction of the total runtime. This is also true for ABC, where sorting the reference table for each draw from the outer sample is much more time-consuming than creating the reference table itself (see also Section~\ref{subsubsec:onedim_losscurve}).

It is interesting to note that there do not seem to be any systematic differences between the distributions of the number of sweeps between the different methods. Therefore, it is entirely sufficient to consider the runtimes per sweep or runtimes per call when analysing the differences between the methods.

In our examples, it was about four to five times faster to use cross-validated trees than to use random forests when the data dimension is small. However, as the data dimension increases, the tree method loses some of that advantage (see Tables~\ref{tab:perform_epi2_highdim} and \ref{tab:perform_macro}). Note that in the macrophage example the data consist of the various observed cell proportions at each design point and are therefore quite high-dimensional despite the low dimensionality of the designs. Furthermore, the higher the dimension, the bigger the advantage of the designs found through random forest classification in terms of discriminatory performance (see, e.g., Figure~\ref{fig:results_epi_2models_highdim} in this document). Therefore, for higher dimensions the recommendation is to use random forests.

Both classification approaches are many times faster than the other approaches investigated in Table~\ref{tab:perform_epi} (ABC) and Table~\ref{tab:perform_epi2_lowdim} (likelihood-based). Note that this is despite the relatively small simulation sizes for the outer Monte Carlo samples from the prior predictive distribution that we used for ABC (sample size 2000) as well as for the likelihood-based approach (sample size 800) to keep the runtimes within a tolerable range. These small outer sample sizes led to a considerably larger noise in the expected loss estimates for those two approaches compared to the classification approaches (see Figures~\ref{fig:1d_results} and \ref{fig:1d_results_2models}).

Runtimes are machine- and implementation-specific and should therefore be taken with caution. However, Tables~\ref{tab:perform_epi} to \ref{tab:perform_macro} can still give some clues on the relative efficiency of the different methods. All our examples were run on an SGI UV 3000 global shared memory system from Hewlett Packard Enterprises. It uses 12-core processors of type Intel Xeon E5-4650V3 that operate on 2.8 GHz and have an L3 cache of 30 MB.

We do not further analyse Algorithm~\ref{algo:GPR}.

\begin{table}[htbp!]
	\centering
	\caption{Several performance indicators for the infectious disease example of Section~\ref{subsec:epi} for design sizes of $n = 1$ to $n = 5$ time points: number of sweeps of coordinate exchange algorithm through design grid (minimum, median, maximum over 20 parallel runs), calls to loss estimation procedure per sweep (cs), mean and standard deviation of runtime per sweep over all parallel runs (in minutes).\label{tab:perform_epi}}
	\begin{tabular}{rlcccrD{.}{.}{1}D{.}{.}{2}}
		$n$ & Method & \multicolumn{3}{c}{\# sweeps} & cs & \multicolumn{2}{c}{runtime/sweep} \\
		& & min & med & max &  & \multicolumn{1}{r}{mean} & \multicolumn{1}{r}{std} \\ 
		\hline 
		\multirow{6}{*}{1} & Tree CV 01L & 1 & 2 & 4 & 39 & 0.8 & 0.03 \\ 
		& Tree CV MDL & 2 & 2 & 4 & 39 & 0.3 & 0.01 \\ 
		& RF 01L & 2 & 2 & 4 & 39 & 3.3 & 0.18 \\ 
		& RF MDL & 2 & 2 & 3 & 39 & 3.1 & 0.16 \\ 
		& ABC 01L & 1 & 2 & 3 & 39 & 33.2 & 1.75 \\ 
		& ABC MDL & 2 & 2 & 4 & 39 & 33.6 & 1.92 \\ 
		\hline
		\multirow{6}{*}{2} & Tree CV 01L & 2 & 2 & 5 & 76 & 1.6 & 0.05 \\ 
		& Tree CV MDL & 2 & 3 & 4 & 76 & 0.8 & 0.04 \\ 
		& RF 01L & 2 & 3 & 6 & 76 & 8.1 & 0.30 \\ 
		& RF MDL & 2 & 3 & 6 & 76 & 7.4 & 0.30 \\ 
		& ABC 01L & 2 & 3 & 5 & 76 & 95.2 & 4.88 \\ 
		& ABC MDL & 2 & 2.5 & 5 & 76 & 92.7 & 6.57 \\ 
		\hline
		\multirow{6}{*}{3} & Tree CV 01L & 2 & 2.5 & 6 & 111 & 2.7 & 0.10 \\ 
		& Tree CV MDL & 2 & 2 & 5 & 111 & 1.5 & 0.05 \\ 
		& RF 01L & 2 & 2.5 & 6 & 111 & 12.1 & 0.51 \\ 
		& RF MDL & 2 & 3 & 5 & 111 & 12.1 & 0.27 \\ 
		& ABC 01L & 2 & 3 & 5 & 111 & 179.0 & 5.81 \\ 
		& ABC MDL & 2 & 3 & 5 & 111 & 176.9 & 5.50 \\
		\hline 
		\multirow{6}{*}{4} & Tree CV 01L & 2 & 3 & 5 & 144 & 4.0 & 0.18 \\ 
		& Tree CV MDL & 2 & 3 & 5 & 144 & 2.3 & 0.08 \\ 
		& RF 01L & 2 & 2.5 & 4 & 144 & 17.1 & 1.18 \\ 
		& RF MDL & 2 & 3 & 4 & 144 & 16.4 & 1.39 \\ 
		& ABC 01L & 2 & 2 & 4 & 144 & 283.9 & 7.94 \\ 
		& ABC MDL & 2 & 3 & 5 & 144 & 281.4 & 8.88 \\ 
		\hline
		\multirow{6}{*}{5} & Tree CV 01L & 2 & 2 & 5 & 175 & 5.5 & 0.32 \\ 
		& Tree CV MDL & 2 & 3 & 7 & 175 & 3.0 & 0.16 \\ 
		& RF 01L & 2 & 3 & 5 & 175 & 25.9 & 2.39 \\ 
		& RF MDL & 2 & 3 & 5 & 175 & 23.8 & 1.09 \\ 
		& ABC 01L & 2 & 3 & 5 & 175 & 414.6 & 13.36 \\ 
		& ABC MDL & 2 & 3 & 5 & 175 & 408.4 & 16.98 \\ 
		\hline 
	\end{tabular}
\end{table}

\begin{table}[htbp!]
	\centering
	\caption{Several performance indicators for the two-model infectious disease example of Section~\ref{subsec:epi2}: number of sweeps of coordinate exchange algorithm through design grid (minimum, median, maximum over 20 parallel runs), calls to loss estimation procedure per sweep (cs), mean and standard deviation of runtime per sweep over all parallel runs (in minutes). The table contains the indicators for the lower-dimensional designs with $q$ independent realisations of the process, $n_d = 1, \ldots, 4$ observations per realisation and $q$ such that $q \cdot n_d \leq 8$. \label{tab:perform_epi2_lowdim}}
	\begin{tabular}{rrlcccrD{.}{.}{1}D{.}{.}{2}}
		$n_d$ & $q$ & Method & \multicolumn{3}{c}{\# sweeps} & cs & \multicolumn{2}{c}{runtime/sweep} \\
		& & & min & med & max &  & \multicolumn{1}{r}{mean} & \multicolumn{1}{r}{std} \\ 
		\hline 
		\multirow{12}{*}{1} & \multirow{3}{*}{1} & Tree CV & 2 & 2 & 2 & 19 & 0.2 & 0.01 \\ 
		&  & RF & 1 & 2 & 2 & 19 & 0.9 & 0.05 \\ 
		&  & ML & 1 & 2 & 3 & 19 & 4.9 & 0.07 \\ 
		\cdashline{2-9} 
		& \multirow{3}{*}{2} & Tree CV & 1 & 3 & 4 & 38 & 0.5 & 0.02 \\ 
		&  & RF & 2 & 3 & 5 & 38 & 2.2 & 0.12 \\ 
		&  & ML & 1 & 3 & 4 & 38 & 17.0 & 0.26 \\ 
		\cdashline{2-9}
		& \multirow{3}{*}{3} & Tree CV & 2 & 3 & 5 & 57 & 0.9 & 0.03 \\ 
		&  & RF & 2 & 3 & 5 & 57 & 3.8 & 0.16 \\ 
		&  & ML & 2 & 3 & 5 & 57 & 36.0 & 0.46 \\ 
		\cdashline{2-9}
		& \multirow{3}{*}{4} & Tree CV & 2 & 3 & 5 & 76 & 1.3 & 0.05 \\ 
		&  & RF & 2 & 2 & 4 & 76 & 5.6 & 0.22 \\ 
		&  & ML & 2 & 2.5 & 8 & 76 & 61.4 & 1.08 \\ 
		\hline
		\multirow{12}{*}{2} & \multirow{3}{*}{1} & Tree CV & 1 & 2 & 4 & 36 & 0.4 & 0.02 \\ 
		&  & RF & 2 & 2.5 & 5 & 36 & 2.0 & 0.12 \\ 
		&  & ML & 1 & 2 & 5 & 36 & 15.3 & 0.26 \\ 
		\cdashline{2-9}
		& \multirow{3}{*}{2} & Tree CV & 2 & 3 & 5 & 72 & 1.2 & 0.04 \\ 
		&  & RF & 2 & 3 & 5 & 72 & 5.2 & 0.14 \\ 
		&  & ML & 2 & 3 & 5 & 72 & 56.3 & 0.43 \\ 
		\cdashline{2-9}
		& \multirow{3}{*}{3} & Tree CV & 2 & 3.5 & 6 & 108 & 2.3 & 0.06 \\ 
		&  & RF & 2 & 3 & 6 & 108 & 10.1 & 0.47 \\ 
		&  & ML & 2 & 3 & 5 & 108 & 119.0 & 1.59 \\ 
		\cdashline{2-9}
		& \multirow{3}{*}{4} & Tree CV & 2 & 3 & 7 & 144 & 3.5 & 0.07 \\ 
		&  & RF & 2 & 3 & 6 & 144 & 13.6 & 0.54 \\ 
		&  & ML & 2 & 3 & 5 & 144 & 214.8 & 1.89 \\ 
		\hline
		\multirow{6}{*}{3} & \multirow{3}{*}{1} & Tree CV & 2 & 3 & 4 & 51 & 0.7 & 0.03 \\ 
		&  & RF & 2 & 2.5 & 6 & 51 & 3.1 & 0.15 \\ 
		&  & ML & 2 & 2.5 & 4 & 51 & 29.2 & 0.33 \\ 
		\cdashline{2-9}
		& \multirow{3}{*}{2} & Tree CV & 2 & 3 & 4 & 102 & 2.0 & 0.07 \\ 
		&  & RF & 2 & 3 & 9 & 102 & 9.5 & 0.34 \\ 
		&  & ML & 2 & 3 & 5 & 102 & 111.7 & 1.07 \\ 
		\hline
		\multirow{6}{*}{4} & \multirow{3}{*}{1} & Tree CV & 2 & 3 & 5 & 64 & 1.0 & 0.04 \\ 
		&  & RF & 2 & 2 & 5 & 64 & 4.2 & 0.26 \\ 
		&  & ML & 2 & 3 & 6 & 64 & 46.6 & 0.43 \\ 
		\cdashline{2-9}
		& \multirow{3}{*}{2} & Tree CV & 2 & 3 & 6 & 128 & 2.8 & 0.10 \\ 
		&  & RF & 2 & 4 & 6 & 128 & 11.8 & 0.35 \\ 
		&  & ML & 2 & 3 & 5 & 128 & 180.9 & 1.26 \\ 
		\hline 
	\end{tabular}
\end{table}

\begin{table}[htbp!]
	\centering
	\caption{Several performance indicators for the two-model infectious disease example of Section~\ref{subsec:epi2}: number of sweeps of coordinate exchange algorithm through design grid (minimum, median, maximum over 20 parallel runs), calls to loss estimation procedure per sweep (cs), mean and standard deviation of runtime per sweep over all parallel runs (in minutes). The table contains the indicators for the higher-dimensional designs with $q$ independent realisations of the process, $n_d = 1, \ldots, 4$ observations per realisation and $q$ such that $q \cdot n_d \in \{12, 24, 36, 48\}$.\label{tab:perform_epi2_highdim}}
	\begin{tabular}{rrlcccrD{.}{.}{1}D{.}{.}{2}}
		$n_d$ & $q$ & Method & \multicolumn{3}{c}{\# sweeps} & cs & \multicolumn{2}{c}{runtime/sweep} \\
		& & & min & med & max &  & \multicolumn{1}{r}{mean} & \multicolumn{1}{r}{std} \\  
		\hline 
		\multirow{8}{*}{1} & \multirow{2}{*}{12} & Tree CV & 2 & 3 & 6 & 228 & 7.5 & 0.19 \\ 
		&  & RF & 2 & 3 & 8 & 228 & 25.5 & 0.71 \\ 
		\cdashline{2-9}
		& \multirow{2}{*}{24} & Tree CV & 2 & 3 & 8 & 456 & 25.3 & 0.50 \\ 
		&  & RF & 2 & 3.5 & 6 & 456 & 63.1 & 1.47 \\ 
		\cdashline{2-9}
		& \multirow{2}{*}{36} & Tree CV & 2 & 2.5 & 9 & 684 & 61.4 & 1.26 \\ 
		&  & RF & 2 & 4 & 8 & 684 & 110.8 & 4.10 \\ 
		\cdashline{2-9}
		& \multirow{2}{*}{48} & Tree CV & 2 & 3 & 7 & 912 & 114.7 & 2.52 \\ 
		&  & RF & 2 & 3 & 6 & 912 & 169.8 & 5.30 \\ 
		\hline
		\multirow{8}{*}{2} & \multirow{2}{*}{6} & Tree CV & 2 & 3 & 7 & 216 & 7.1 & 0.27 \\ 
		&  & RF & 2 & 4 & 7 & 216 & 23.7 & 0.66 \\ 
		\cdashline{2-9}
		& \multirow{2}{*}{12} & Tree CV & 2 & 2.5 & 5 & 432 & 24.0 & 0.81 \\ 
		&  & RF & 2 & 3 & 6 & 432 & 58.1 & 1.91 \\ 
		\cdashline{2-9}
		& \multirow{2}{*}{18} & Tree CV & 2 & 3 & 7 & 648 & 51.6 & 1.04 \\ 
		&  & RF & 2 & 3 & 6 & 648 & 100.8 & 3.57 \\ 
		\cdashline{2-9}
		& \multirow{2}{*}{24} & Tree CV & 2 & 2.5 & 5 & 864 & 94.7 & 2.94 \\ 
		&  & RF & 2 & 3 & 7 & 864 & 153.4 & 6.85 \\ 
		\hline
		\multirow{8}{*}{3} & \multirow{2}{*}{4} & Tree CV & 2 & 3 & 5 & 204 & 6.5 & 0.30 \\ 
		&  & RF & 2 & 3 & 7 & 204 & 22.2 & 0.62 \\ 
		\cdashline{2-9}
		& \multirow{2}{*}{8} & Tree CV & 2 & 3 & 6 & 408 & 21.3 & 0.41 \\ 
		&  & RF & 2 & 4 & 7 & 408 & 53.8 & 1.15 \\ 
		\cdashline{2-9}
		& \multirow{2}{*}{12} & Tree CV & 2 & 3 & 7 & 612 & 44.0 & 1.27 \\ 
		&  & RF & 2 & 4 & 7 & 612 & 93.4 & 3.26 \\ 
		\cdashline{2-9}
		& \multirow{2}{*}{16} & Tree CV & 2 & 3 & 4 & 816 & 78.4 & 1.68 \\ 
		&  & RF & 2 & 3.5 & 5 & 816 & 138.9 & 5.88 \\ 
		\hline
		\multirow{8}{*}{4} & \multirow{2}{*}{3} & Tree CV & 2 & 3 & 6 & 192 & 5.9 & 0.10 \\ 
		&  & RF & 2 & 3 & 8 & 192 & 20.7 & 0.60 \\ 
		\cdashline{2-9}
		& \multirow{2}{*}{6} & Tree CV & 2 & 3 & 6 & 384 & 17.9 & 0.58 \\ 
		&  & RF & 2 & 4 & 8 & 384 & 49.0 & 1.59 \\ 
		\cdashline{2-9}
		& \multirow{2}{*}{9} & Tree CV & 2 & 3 & 5 & 576 & 35.0 & 1.14 \\ 
		&  & RF & 2 & 4 & 7 & 576 & 85.4 & 2.78 \\ 
		\cdashline{2-9}
		& \multirow{2}{*}{12} & Tree CV & 2 & 2.5 & 6 & 768 & 61.8 & 1.36 \\ 
		&  & RF & 2 & 3.5 & 6 & 768 & 127.4 & 5.16 \\ 
		\hline 
	\end{tabular}
\end{table}

\begin{table}[htbp!]
	\centering
	\caption{Several performance indicators for the macrophage example of Section~\ref{subsec:macro} for design sizes of $n = 1$ to $n = 5$ time points: number of sweeps of coordinate exchange algorithm through design grid (minimum, median, maximum over 20 parallel runs), calls to loss estimation procedure per sweep (cs), mean and standard deviation of runtime per sweep over all parallel runs (in minutes).\label{tab:perform_macro}}
	\begin{tabular}{rlcccrD{.}{.}{1}D{.}{.}{1}}
		$n$ & Method & \multicolumn{3}{c}{\# sweeps} & cs & \multicolumn{2}{c}{runtime/sweep} \\
		& & min & med & max &  & \multicolumn{1}{r}{mean} & \multicolumn{1}{r}{std} \\ 
		\hline 
		\multirow{2}{*}{1} & Tree CV & 2 & 3 & 6 & 55 & 10.4 & 0.6 \\ 
		& RF & 2 & 4 & 5 & 55 & 18.6 & 9.5 \\ 
		\hline
		\multirow{2}{*}{2} & Tree CV & 2 & 3 & 4 & 95 & 40.5 & 8.1 \\ 
		& RF & 2 & 3 & 6 & 95 & 53.4 & 19.4 \\ 
		\hline
		\multirow{2}{*}{3} & Tree CV & 2 & 3 & 5 & 135 & 72.6 & 18.5 \\ 
		& RF & 2 & 3 & 6 & 135 & 82.0 & 14.0 \\ 
		\hline
		\multirow{2}{*}{4} & Tree CV & 2 & 3 & 6 & 175 & 108.9 & 24.4 \\ 
		& RF & 2 & 3 & 8 & 175 & 159.0 & 33.2 \\ 
		\hline
		\multirow{2}{*}{5} & Tree CV & 2 & 3 & 5 & 215 & 156.1 & 57.1 \\ 
		& RF & 2 & 3 & 4 & 215 & 176.3 & 37.0 \\ 
		\hline 
	\end{tabular}
\end{table}

\clearpage

\section{Additional Details and Results for Epidemiological Example} \label{app:epi}

\subsection{Prior Distributions} \label{app:epi_prior}

The prior distributions for the four epidemiological Markov process models of Section~4.1 are given in Table~\ref{tab:epi_priors}.

\begin{table}[htbp!]
	\centering
	\caption{The prior distributions considered for the infectious disease example of Section~4.1.  Here $\mathcal{LN}(\mu,\sigma)$ denotes the lognormal distribution with location $\mu$ and scale $\sigma$.  $\mathcal{E}(\eta)$ denotes the exponential distribution with rate $\eta$.\label{tab:epi_priors}}
	\begin{tabular}{|ccc|}
		\hline
		Model Number & Parameter & Prior \\
		\hline
		Model 1  & $b_1^{(1)}$ & $\mathcal{LN}(-0.48,0.09)$ \\
		\hline
		Model 2  & $b_1^{(2)}$ & $\mathcal{LN}(-1.1,0.16)$ \\
		& $b_2^{(2)}$ & $\mathcal{LN}(-4.5,0.4)$ \\
		
		\hline
		Model 3  & $b_1^{(3)}$ & $\mathcal{LN}(-0.54,0.15)$ \\
		& $\gamma^{(3)}$ & $\mathcal{E}(0.01)$ \\
		\hline
		Model 4  & $b_1^{(4)}$ & $\mathcal{LN}(-1.34,0.41)$ \\
		& $b_2^{(4)}$ & $\mathcal{LN}(-4.26,0.25)$ \\
		& $\gamma^{(4)}$ & $\mathcal{E}(0.01)$ \\
		\hline
	\end{tabular}
\end{table}

\subsection{Optimal Designs for $n=4$ and $n=5$} \label{app:epi_designs}

\begin{table}[htbp!]
	\centering
	\caption{Optimal designs obtained by tree classification (cross-validated), random forest classification (using out-of-bag class predictions), and ABC approaches under the 0--1 loss (01L) or multinomial deviance loss (MDL) ($n = 4$ and $5$) for the infectious disease example. The equidistant designs are also shown. \label{tab:epi_designs2}}
	\begin{tabular}{|l|*{4}{D{.}{.}{3}}|*{5}{D{.}{.}{3}}|}
		\hline
		\multicolumn{1}{|l|}{Method/Loss} & \multicolumn{4}{c|}{$n = 4$} & \multicolumn{5}{c|}{$n = 5$} \\
		\hline
		Tree 01L & 0.750 & 4.250 & 9.750 & 10.000 & 0.910 & 4.304 & 8.671 & 10.000 & 10.000 \\
		RF 01L   & 0.750 & 4.250 & 8.500 & 9.750 & 0.750 & 4.250 & 8.250 & 9.000 & 9.250 \\
		ABC 01L  & 0.047 & 0.599 & 2.265 & 4.943 & 0.250 & 1.000 & 3.000 & 5.500 & 7.500 \\
		\hline
		Tree MDL & 0.750 & 5.000 & 9.791 & 10.000 & 0.750 & 4.750 & 9.566 & 9.750 & 10.000 \\
		RF MDL & 0.720 & 4.000 & 6.268 & 10.000 & 0.750 & 4.250 & 9.566 & 9.750 & 10.000 \\
		ABC MDL  & 0.222 & 0.703 & 3.120 & 5.753 & 0.500 & 1.500 & 3.000 & 4.750 & 7.250 \\
		\hline 
		Equidistant	 & 2.000 & 4.000 & 6.000 & 8.000 & 1.667 & 3.333 & 5.000 & 6.667 & 8.333 \\
		\hline
		
	\end{tabular}
\end{table}

\clearpage

\subsection{Misclassification Matrices} \label{app:epi_misclass}

The random forest classifiers and the corresponding random samples which we use to compute the misclassification error rates in Table~\ref{tab:epi_classvalidation} can also be used to compute misclassification matrices for the various optimal designs. A \emph{misclassification} or \emph{confusion matrix} contains for each combination of true model $m_i$ (in the rows) and predicted model $m_j$ (in the columns) the proportions of samples from true model $m_i$ that were classified as model $m_j$. In the case of random forests, the misclassification matrix is computed using out-of-bag class predictions. It provides a comprehensive picture of the classification accuracy at a given design.

\begin{figure}[htbp!]
	\centering
	\subfigure[1 design point]{\includegraphics[width=0.45\textwidth]{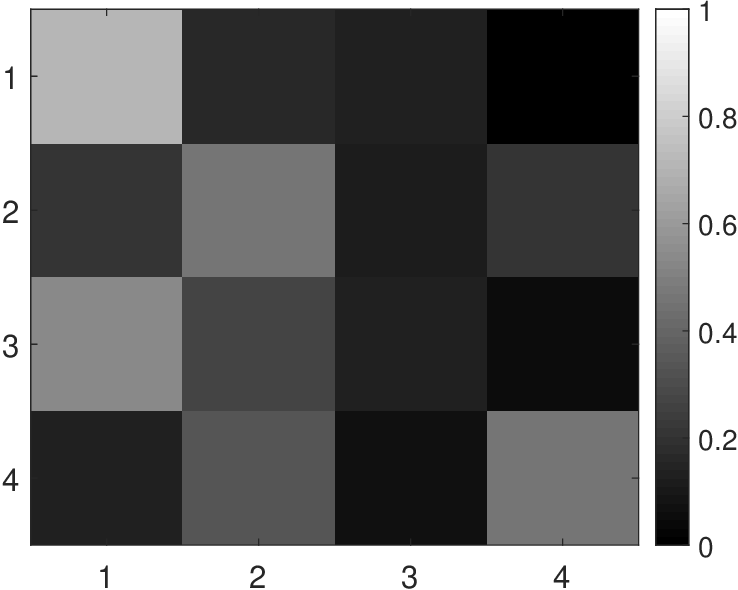}\label{figsub:epi_class1}} \quad
	\subfigure[2 design points]{\includegraphics[width=0.45\textwidth]{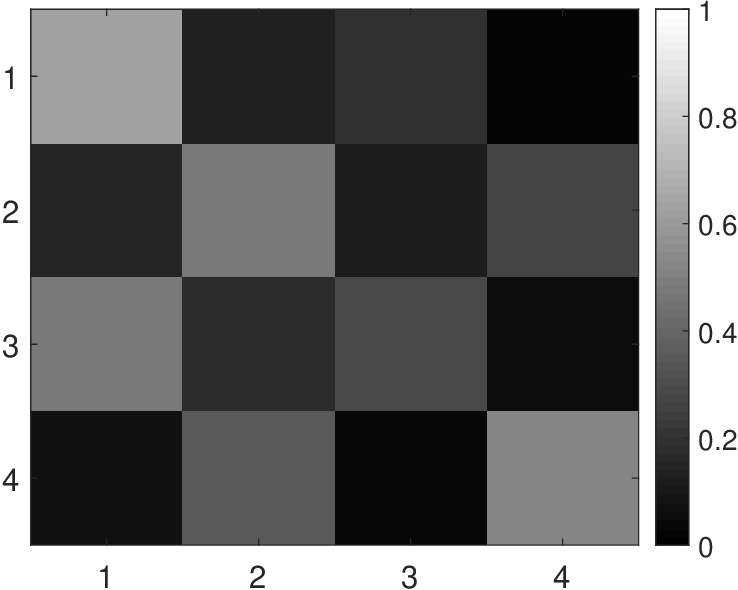}\label{figsub:epi_class2}}
	\vspace*{1ex}
	\subfigure[3 design points]{\includegraphics[width=0.45\textwidth]{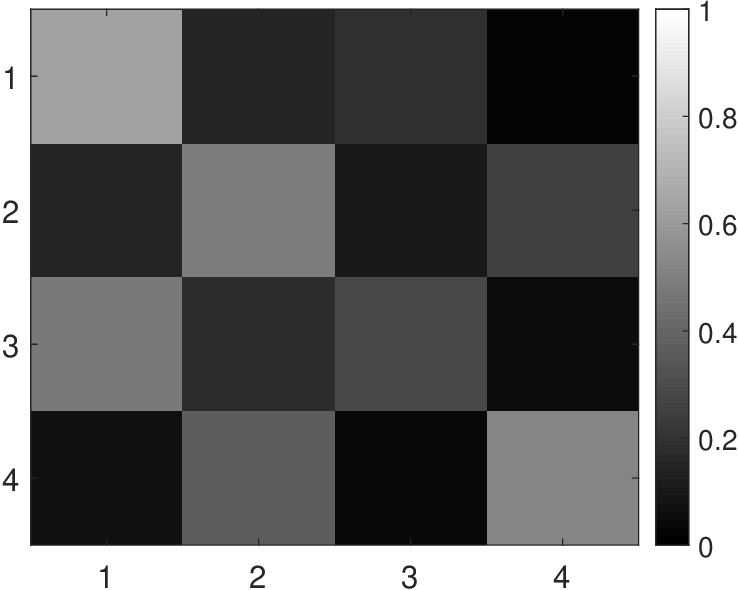}\label{figsub:epi_class3}} \quad
	\subfigure[4 design points]{\includegraphics[width=0.45\textwidth]{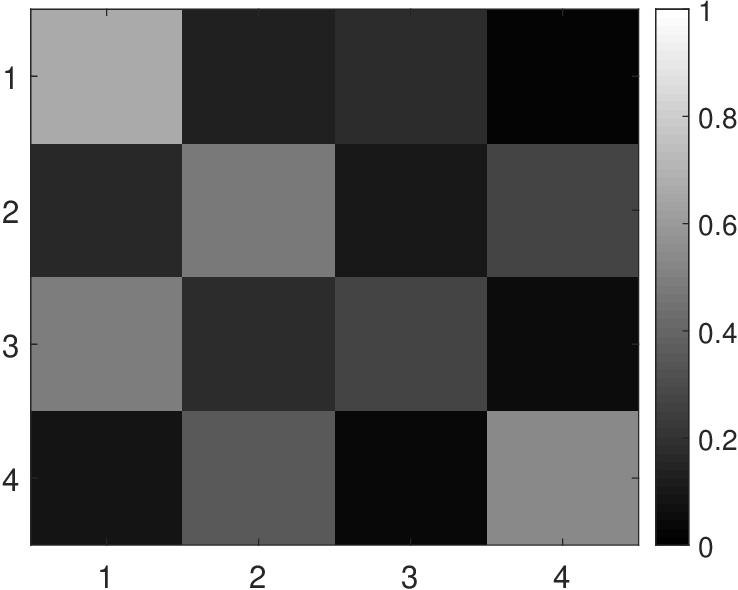}\label{figsub:epi_class4}}
	\caption{Misclassification matrices obtained for the \emph{tree classification designs (using cross-validation)} under the \emph{0--1 loss} for the infectious disease example. Designs for 1 -- 4 observations are considered.\label{fig:epi_misclass_tree}}
\end{figure}

For the optimal design obtained by the tree classification approach with cross-validation under the 0--1 loss, the misclassification matrices for 1 -- 4 time points are shown in Figure~\ref{fig:epi_misclass_tree}.  The figure suggests that it is difficult to discriminate between models 1 and 3 and also models 2 and 4.  This is not surprising given that we do not observe the exposed population. Especially model 3 is most often misclassified as model 1. The misclassification matrices for the other machine learning classification approaches and loss functions are qualitatively all very similar to Figure~\ref{fig:epi_misclass_tree}.

\begin{figure}[htbp!]
	\centering
	\subfigure[1 design point]{\includegraphics[width=0.45\textwidth]{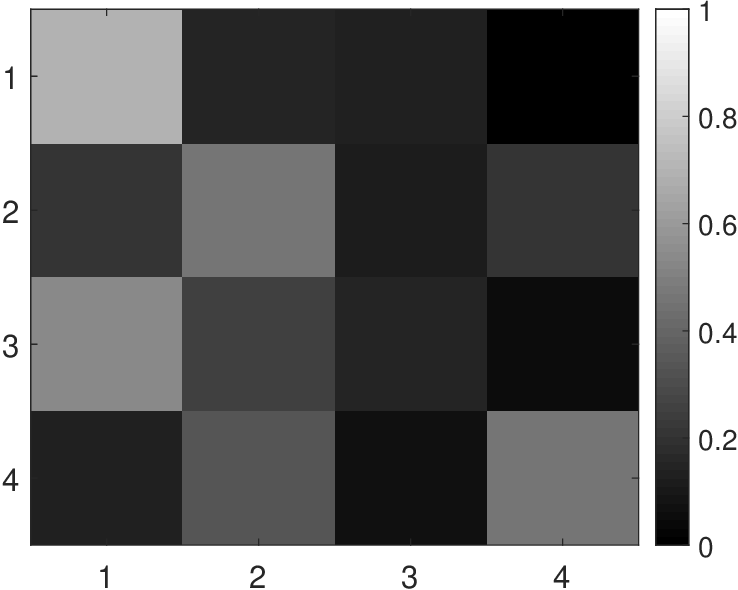}\label{figsub:epi_class1}} \quad
	\subfigure[2 design points]{\includegraphics[width=0.45\textwidth]{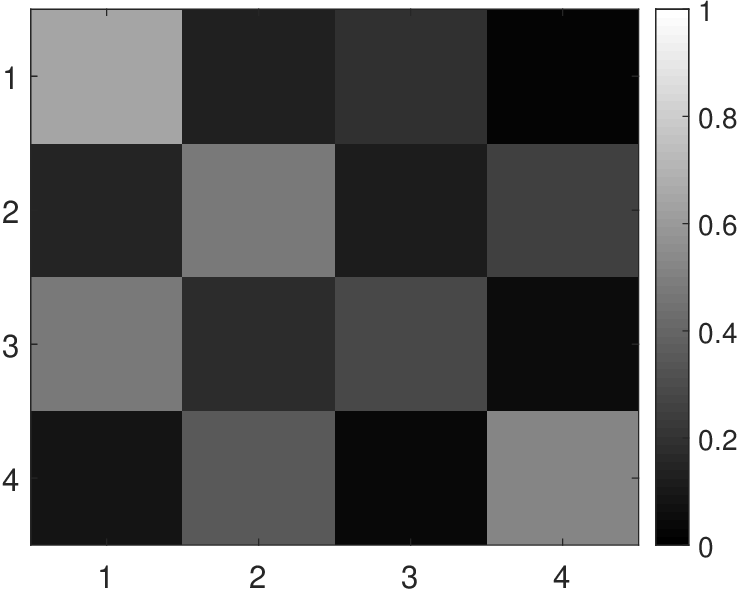}\label{figsub:epi_class2}}
	\vspace*{1ex}
	\subfigure[3 design points]{\includegraphics[width=0.45\textwidth]{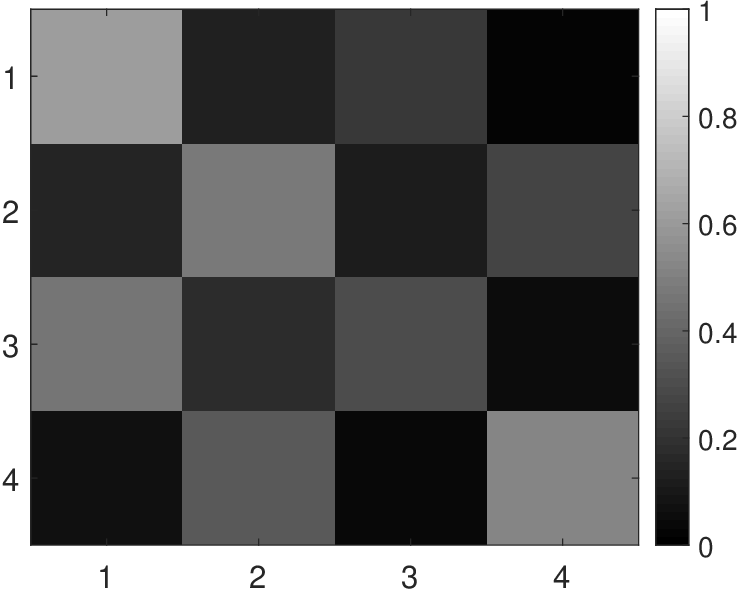}\label{figsub:epi_class3}} \quad
	\subfigure[4 design points]{\includegraphics[width=0.45\textwidth]{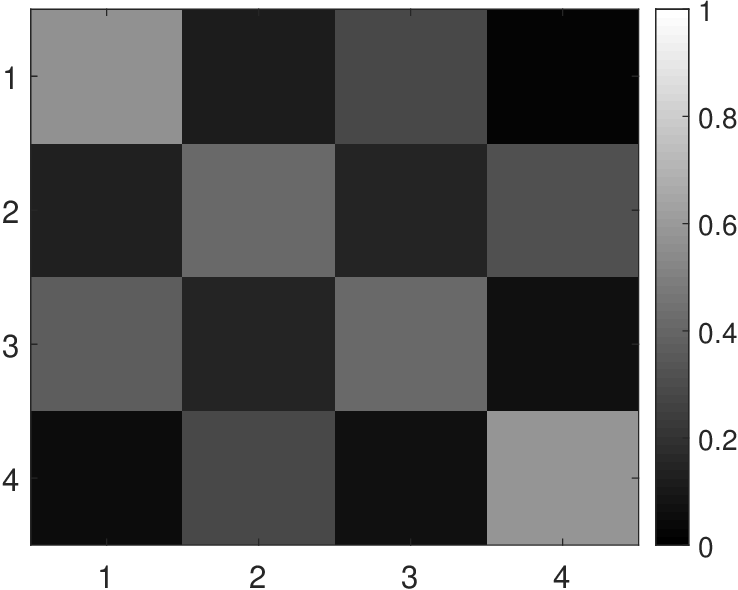}\label{figsub:epi_class4}}
	\caption{Misclassification matrices obtained for the \emph{ABC designs} under the \emph{0--1 loss} for the infectious disease example. Designs for 1 -- 4 observations are considered.\label{fig:epi_misclass_ABC}}
\end{figure}

In Figure~\ref{fig:epi_misclass_ABC}, the misclassification matrices for the ABC approach under the 0--1 loss are depicted. The ABC approach leads to designs with generally lower values for the design points than the machine learning approaches (see Tables~\ref{tab:epi_designs1} and \ref{tab:epi_designs2}). The overall misclassification error rates are similar, but one can see that the pattern is a bit different from Figure~\ref{fig:epi_misclass_tree}. At 4 design points, model~3 is more likely to be correctly classified, but the misclassification error of models~1 and 2 increases.

\clearpage

\section{Additional Details and Results for the Two-model Epidemiological Example}

\subsection{Model Description and Likelihood Functions} \label{app:epi2_likelihood}

Let the design for realisation $i$ ($i = 1,\ldots,q$) be given by $\vectd_i = (d_{i,1}, \ldots, d_{i,n_d})$ and the overall design be given by $\vectd = (\vectd_1, \ldots, \vectd_q)$, where $d_{i,j}$ is the $j^{\text{th}}$ observation time for realisation $i$. Denote the observed number of infected and susceptible subjects for realisation $i$ at time $d_{i,j}$ by $I(d_{i,j}) = I_{i,j}$ and $S(d_{i,j}) = S_{i,j}$, respectively, where $S_{i,j} = N - I_{i,j}$. Collect all the $S_{i,j}$ in the vector $\boldsymbol{S}$ in the same way as the design times $d_{i,j}$ have been collected in the vector $\vectd$. Each $S_{i,j}$ is a discrete random variable that can assume the $N+1$ values $0$ to $N$. The parameters are denoted by $\vectt = (\log(b_1), \log(b_2))$.

Since the death and SI models are continuous-time Markov processes, their likelihood functions have the form
\begin{equation}
p(\vectS| \vectt, \vectd) = \prod_{i=1}^q \prod_{j=1}^{n_d} \: \Pr \left( S_{i,j} | \, S_{i,j-1}, \vectt, d_{i,j-1}, d_{i,j} \right), \label{eq:likelihood_function}
\end{equation}
where $S_{i,0} = N$ is the number of susceptible individuals at time $d_{i,0} = 0$ $\forall \, i$ \citep*[see, e.g.,][]{cook2008}.

Let the $(N+1)$-dimensional vector $\boldsymbol{v}_{i,j|S_{i,j-1} = k}$ contain the probabilities of all the possible states of the random variable $S_{i,j}$ when the value of $S_{i,j-1}$ is known to be $k$. The $m^{\text{th}}$ element of $\boldsymbol{v}_{i,j|S_{i,j-1} = k}$ gives the probability that $S_{i,j} = m - 1$ when $S_{i,j-1} = k$. Since the value of $S_{i,j-1}$ is known and therefore certain, the state probability vector at observation time $d_{i,j-1}$ reduces to $\boldsymbol{e}_{k+ 1}$, where $\boldsymbol{e}_{m}$ denotes a vector for which the $m^{\text{th}}$ element is $1$ and the remaining elements are $0$.

Given the vectors and notation introduced above, the transition probabilities can be written as
\begin{equation*}
\Pr \left( S_{i,j} | \, S_{i,j-1}, \vectt, d_{i,j-1}, d_{i,j} \right) = \boldsymbol{v}_{i,j|S_{i,j-1}}^T \cdot \boldsymbol{e}_{S_{i,j}+1} = \boldsymbol{e}_{S_{i,j-1}+1}^T  \boldsymbol{A}_{\vectt,i,j} \, \boldsymbol{e}_{S_{i,j}+1},
\end{equation*}

where the matrix $\boldsymbol{A}_{\vectt,i,j}$ has dimension $(N+1) \times (N+1)$ and contains the transition probabilities for all pairs of states between observation times $d_{i,j-1}$ and $d_{i,j}$. This matrix follows from the solution of the Kolmogorov forward equations and can be calculated using the matrix exponential \citep[see][]{Higham2008},
\begin{equation}
\boldsymbol{A}_{\vectt,i,j} = \exp [(d_{i,j} - d_{i,j-1}) \, \boldsymbol{G}_{\vectt}], \label{eq:matrix_exponential}
\end{equation}
where $\boldsymbol{G}_{\vectt}$ is the infinitesimal generator matrix that is constructed from the transition rates given in Table~\ref{tab:epi_models}, see, e.g., \citet{grimmet2001}, pp.\ 258.

Let the $N+1$ rows of the generator matrix be numbered from $0$ to $N$. For the SI model, row $i$ ($i = 0,\ldots,N$) of the generator matrix is given by
\begin{equation*}
[\boldsymbol{G}_{\vectt}]_{i \cdot} =  \bigl( \underbrace{0}_{\times \max\{0,\,i-1\}}, \: \underbrace{[b_1 + b_2 (N - i)] \, i}_{\times \min\{1,\,i\}}, \: \underbrace{-[b_1 + b_2 (N - i)] \, i}_{\times 1}, \: \underbrace{0}_{\times (N-i)} \bigr).
\end{equation*}

Setting $b_2 = 0$ for the death model, the transition probabilities can be simplified to a binomial probability mass function \citep[see][]{cook2008}:
\begin{equation*}
\Pr \left( S_{i,j} | \, S_{i,j-1}, b_1, d_{i,j-1}, d_{i,j} \right)  = \mathcal{B}\left\{ S_{i,j} | \, S_{i,j-1}, \, \exp[-b_1 (d_{i,j} - d_{i,j-1})] \right\}.
\end{equation*}
Therefore, there is no need to numerically compute the matrix exponential for the death model, and so the likelihood function can be evaluated very quickly. However, for the SI model each of the $q \cdot n_d$ matrices $\boldsymbol{A}_{\vectt,i,j}$ in the likelihood function \eqref{eq:likelihood_function} is obtained by numerical computation of the matrix exponential \eqref{eq:matrix_exponential}.

\subsection{Approximating the Marginal Likelihood} \label{app:epi2_marglikelihood}

To obtain
\begin{equation*}
p(m|\vectS,\vectd) \propto p(\vectS|m,\vectd) \, p(m),
\end{equation*}
we need to compute the marginal likelihood 
\begin{equation}
p(\vectS|m,\vectd) = \int_{\vectt_{m}} p(\vectS | \vectt_m, m, \vectd) \, p(\vectt_m|m) \, \mathrm{d} \vectt_m. \label{eq:marginal_likelihood}
\end{equation}

We pursue two different approaches to approximating this integral. During the optimisation procedure, we use a comparatively quick Laplace-type approximation to the marginal likelihood, see \citet{gelman2013}, p.\ 318. Let 
\begin{equation}
\tilde{\vectt}_m = \arg \max_{\vectt_m} \, p(\vectS | \vectt_m, m, \vectd) \, p(\vectt_m|m) \label{eq:posterior_mode}
\end{equation}
be the posterior mode of model $m$. Performing a second-order Taylor expansion of \linebreak $p(\vectS | \vectt_m, m, \vectd) \, p(\vectt_m | m)$ around $\tilde{\vectt}_m$ and integrating out $\vectt_m$ yields
\begin{equation}
p(\vectS|m,\vectd) \approx (2 \pi)^{p_m/2} \, | \widetilde{\boldsymbol{\Sigma}}_{\vectS,\tilde{\vectt}_m,\vectd} |^{1/2} \, p(\vectS| \tilde{\vectt}_m, m, \vectd) \, p(\tilde{\vectt}_m|m), \label{eq:marginal_likelihood_Laplace}
\end{equation}
where $p_m$ is the number of parameters of model $m$ and
\begin{equation}
\widetilde{\boldsymbol{\Sigma}}^{-1}_{\vectS,\tilde{\vectt}_m,\vectd} = - \nabla_{\vectt_m} \nabla_{\vectt_m}^{T} \left[ \log p(\vectS|\vectt_m, m, \vectd) + \log p(\vectt_m | m) \right] \, \Bigr|_{\tilde{\vectt}_m} \label{eq:Hessian}
\end{equation}
is the Hessian of the negative log-posterior evaluated at the posterior mode.

When validating the optimal designs found by the different methods, we employ generalised Gauss-Hermite quadrature \citep{kautsky1982,elhay1987} with $Q$ sample points to compute the integral~\eqref{eq:marginal_likelihood}. As weighting kernel we use a multivariate normal density with mean and variance-covariance matrix given by the mean and twice the variance-covariance matrix, respectively, of the normal Laplace approximation to the posterior,
\begin{equation*}
\omega(\vectt_m) = \mathcal{N}(\vectt_m|\tilde{\vectt}_m, 2 \, \widetilde{\boldsymbol{\Sigma}}_{\vectS,\tilde{\vectt}_m,
	\vectd}),
\end{equation*}
where $\tilde{\vectt}_m$ is given by \eqref{eq:posterior_mode} and $\widetilde{\boldsymbol{\Sigma}}_{\vectS,\tilde{\vectt}_m,
	\vectd}$ is given by \eqref{eq:Hessian}. Using this weighting kernel, we expect that many sample points are in relevant regions where the integrand has high mass.
In the bivariate case, determining the sample points involves two steps, see \citet{jackel2005}. First, all combinations of sample points resulting from applying the standard univariate Gauss-Hermite quadrature rule to each dimension are considered. The sample weights are simply computed by multiplying the univariate weights. To account for the correlation and different scaling and location implied by the multivariate normal weighting kernel, the sample points are then transformed accordingly based on a spectral decomposition of the variance-covariance matrix, seeking to align the diagonals of the rectangle of sample points to the principal axes of the confidence ellipsoid. Furthermore, for the two-parameter SI model we drop sample points below a weight of $w_{1} \cdot w_{\lfloor(\sqrt{Q} + 1)/2\rfloor} / \sqrt{Q}$, where $\sqrt{Q}$ is the number of univariate sample points of the Gauss-Hermite quadrature rule and $w_{i}$ denotes the weight for the $i$th ordered univariate sample point.

After obtaining the $Q$ sample points $\boldsymbol{\theta}_{m,i}$ and quadrature weights $w_i$ ($i = 1,\ldots,Q$) according to the quadrature rule, the marginal likelihood can be approximated by
\begin{equation}
p(\vectS| m,\vectd) \approx \sum_{i=1}^Q w_i \, \frac{p(\vectS| \boldsymbol{\theta}_{m,i}, m,\vectd) \, p(\boldsymbol{\theta}_{m,i}|m)}{\mathcal{N}(\boldsymbol{\theta}_{m,i}|\tilde{\vectt}_m, 2 \, \widetilde{\boldsymbol{\Sigma}}_{\vectS,\tilde{\vectt}_m,
		\vectd})}. \label{eq:marginal_likelihood_quadr}
\end{equation}

\subsection{Further Results} \label{app:epi2_results}

\begin{figure}
	\centering
	\includegraphics[width=0.8\textwidth]{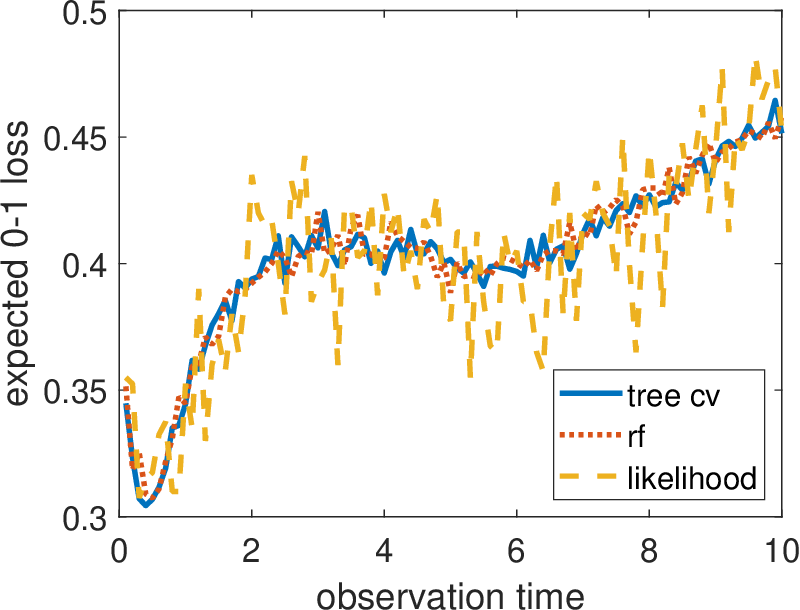}
	\caption{Plots of the approximated expected 0--1 loss functions produced by the tree classification approach with cross-classification (solid), the random forest classification approach (dotted), and the likelihood-based approach using a Laplace-type approximation to the marginal likelihood (dashed) for the infectious disease example with two models.}
	\label{fig:1d_results_2models}
\end{figure}

Figure~\ref{fig:1d_results_2models} shows the estimated expected 0--1 loss surface for the one-dimensional design obtained by the different approaches using the simulation sizes we used for the design search. The comparatively high volatility of the expected 0--1 loss under the likelihood-based approach is evident from Figure~\ref{fig:1d_results_2models}. To create Figure~\ref{fig:1d_results_2models} on our computer, it took about 17 seconds for the tree classification approach, about $2.7$ minutes for the random forest classification approach, but more than 18 minutes for the likelihood-based approach despite the low data dimension and the much smaller prior predictive sample size. 

\begin{figure}[hbtp!]
	\centering
	\subfigure[$n_d = 1$]{\includegraphics[height=0.15\textheight]{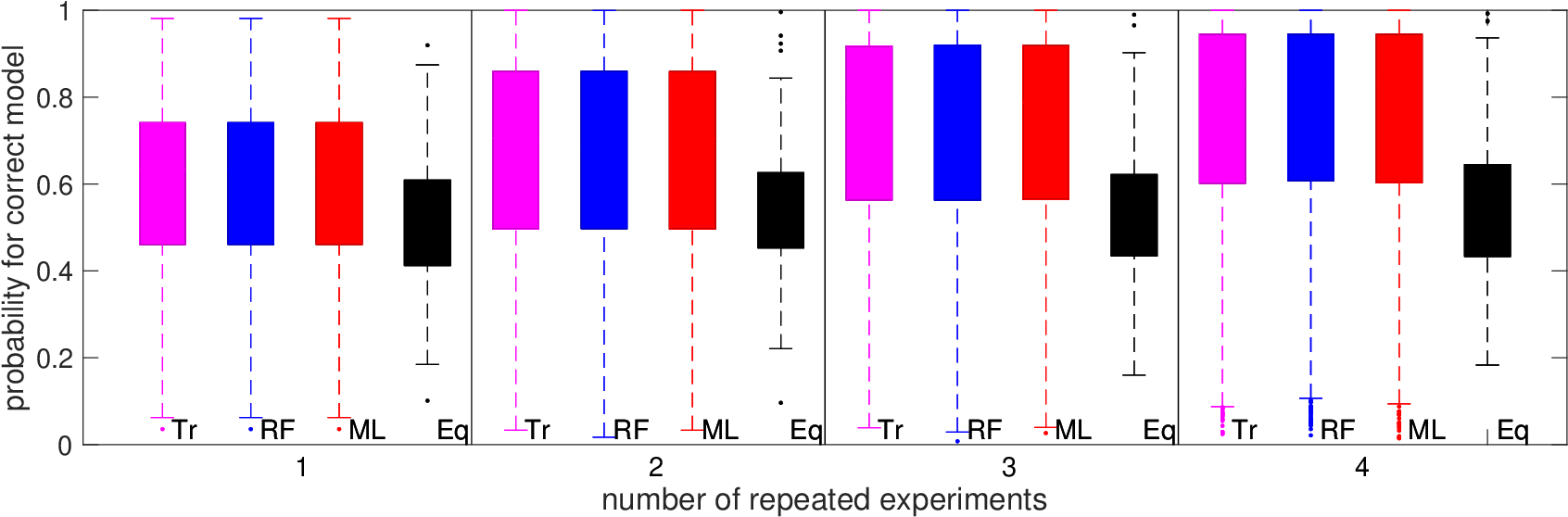}}
	\subfigure[$n_d = 2$]{\includegraphics[height=0.15\textheight]{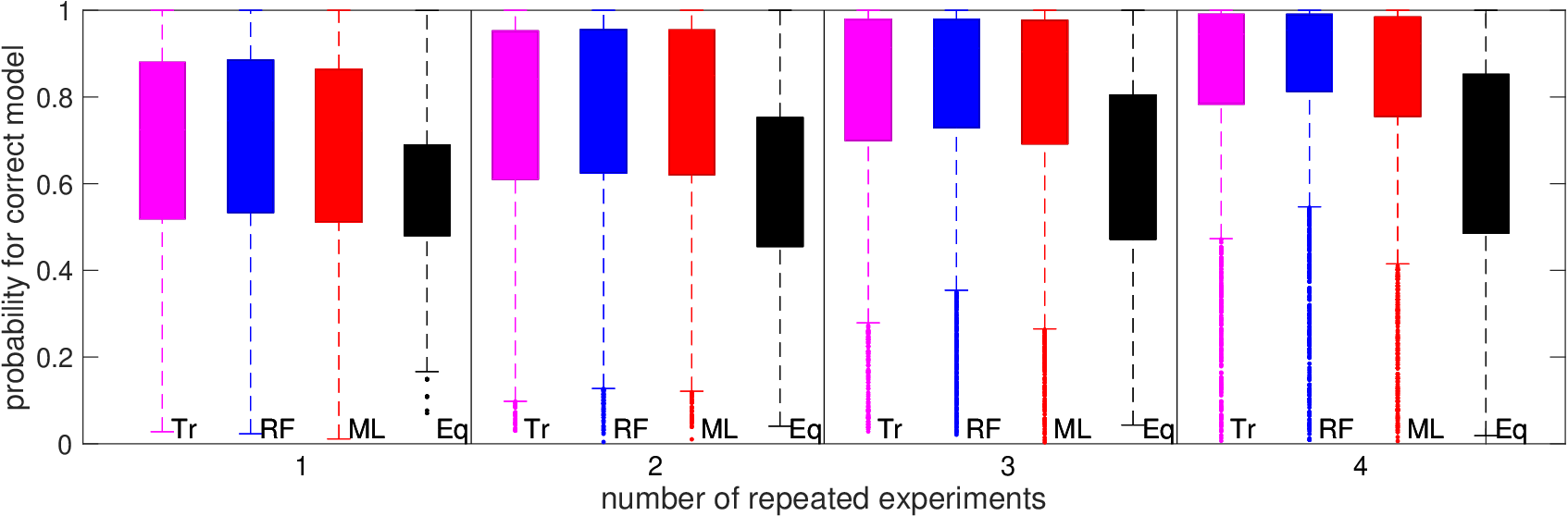}}
	\subfigure[$n_d = 3$]{\includegraphics[height=0.15\textheight]{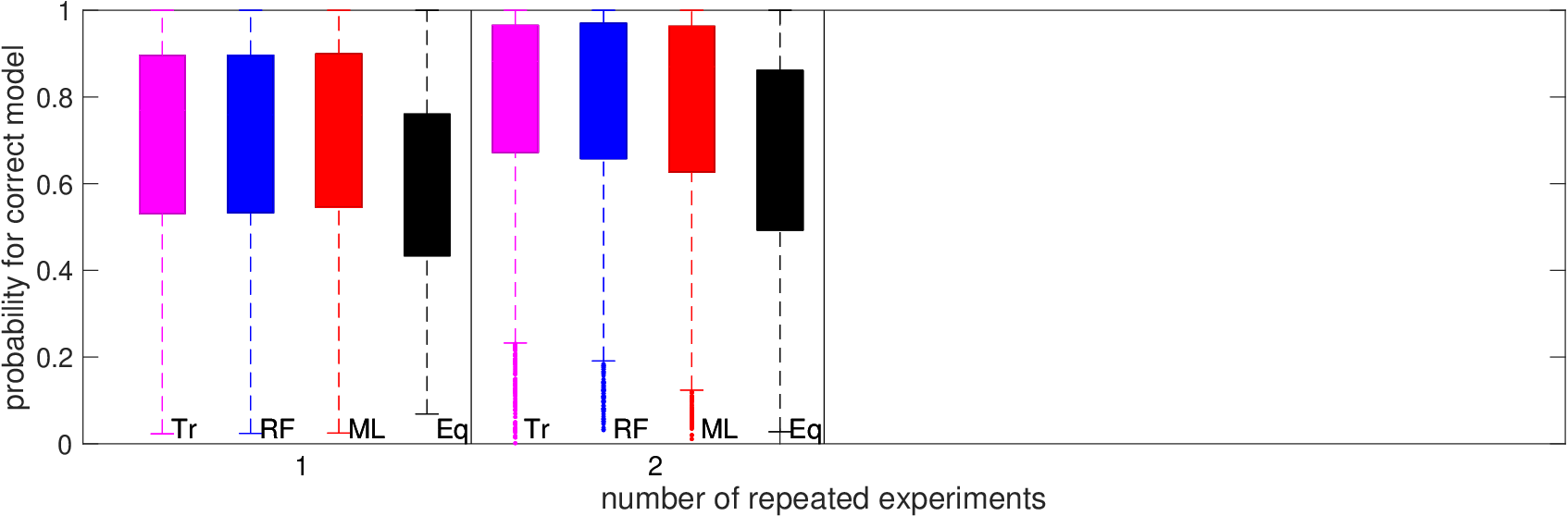}}
	\subfigure[$n_d = 4$]{\includegraphics[height=0.15\textheight]{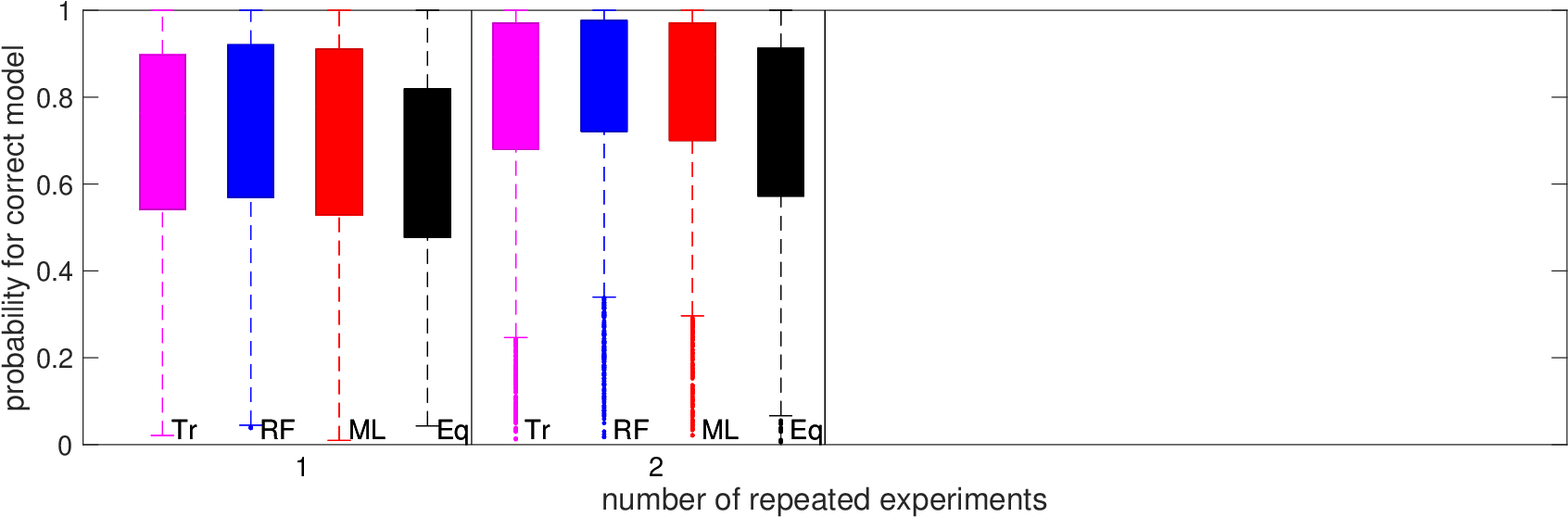}}
	\caption{Distributions of posterior model probabilities of the correct model for 2K prior predictive simulations (1K from each of the two models) for the infectious disease example with two models. The data are all simulated at the respective optimal designs for the different approaches. The 0--1 loss is used as criterion. Settings with $q = 1$ to $q = 4$ realisations and $n_d = 1$ to $n_d = 4$ observations per realisation are considered ($q \leq 2$ for $n_d = 3$ and $n_d = 4$). For each setting, from left to right the boxplots are for the cross-validated tree classification design (Tr; magenta), the random forest classification design (RF; blue), the design found using the Laplace approximations to the  marginal likelihoods (ML; red), and the equispaced design (Eq; black).  \label{fig:results_epi_2models}}
\end{figure}

\begin{figure}[hbtp!]
	\centering
	\subfigure[$n_d = 1$]{\includegraphics[height=0.15\textheight]{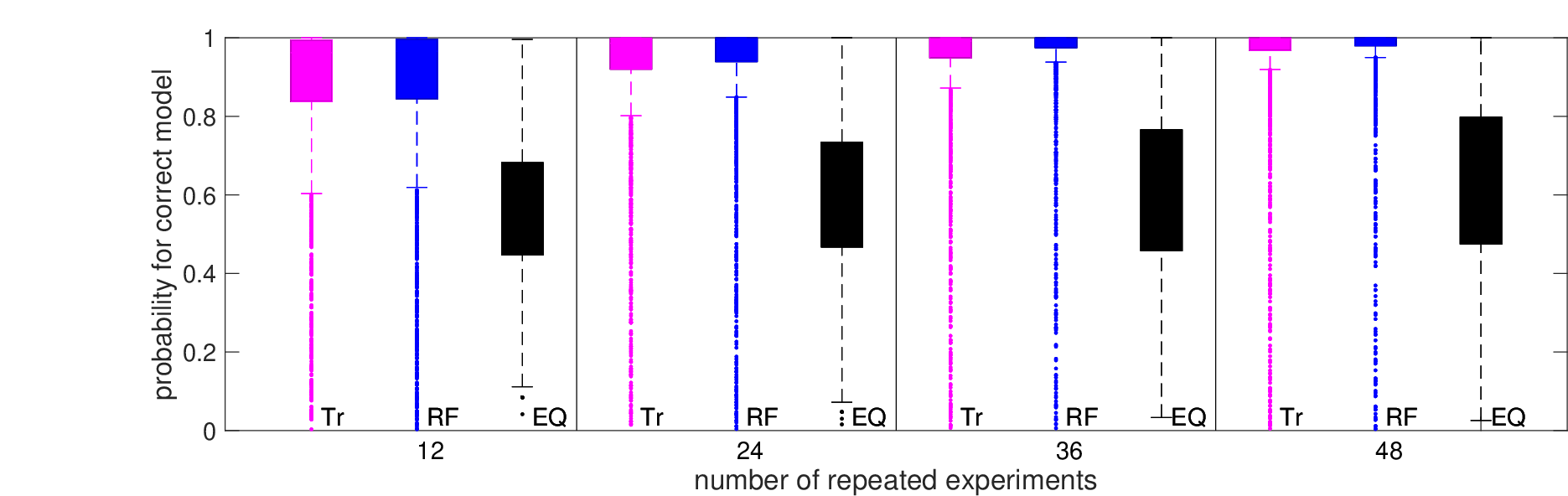}}
	\subfigure[$n_d = 2$]{\includegraphics[height=0.15\textheight]{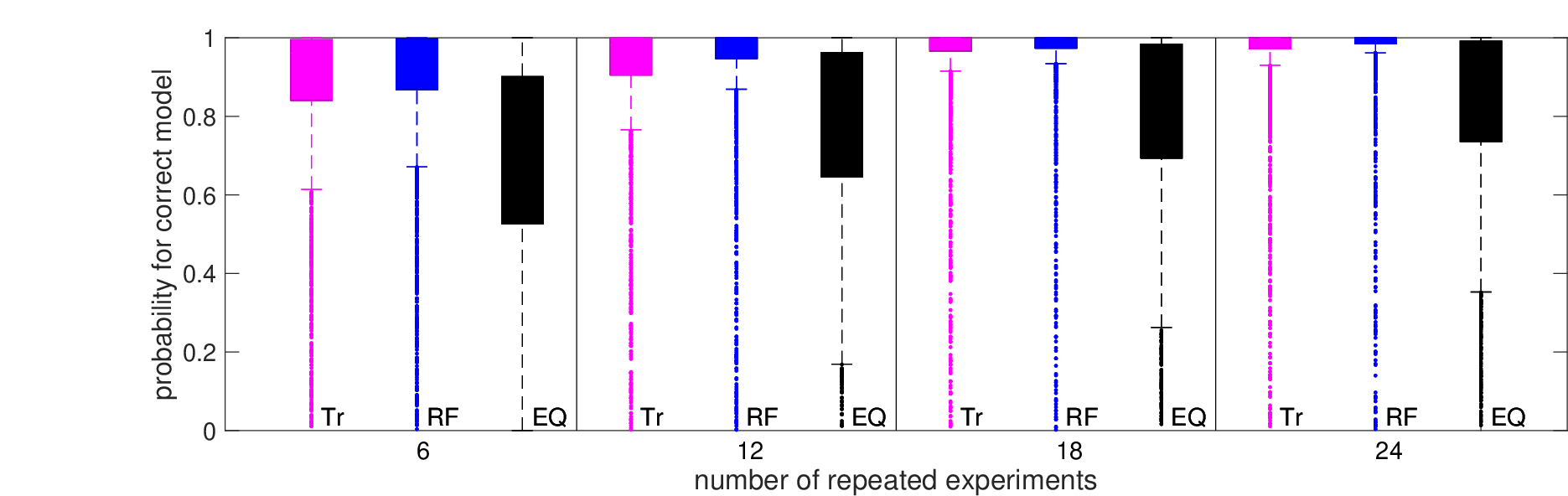}}
	\subfigure[$n_d = 3$]{\includegraphics[height=0.15\textheight]{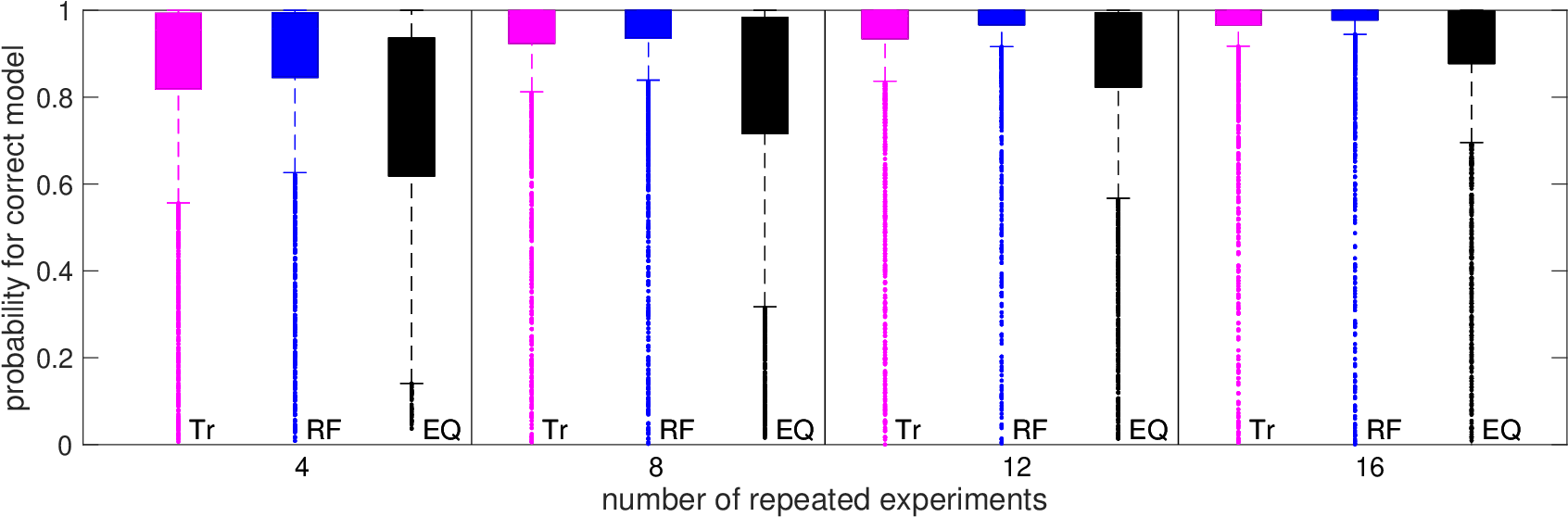}}
	\subfigure[$n_d = 4$]{\includegraphics[height=0.15\textheight]{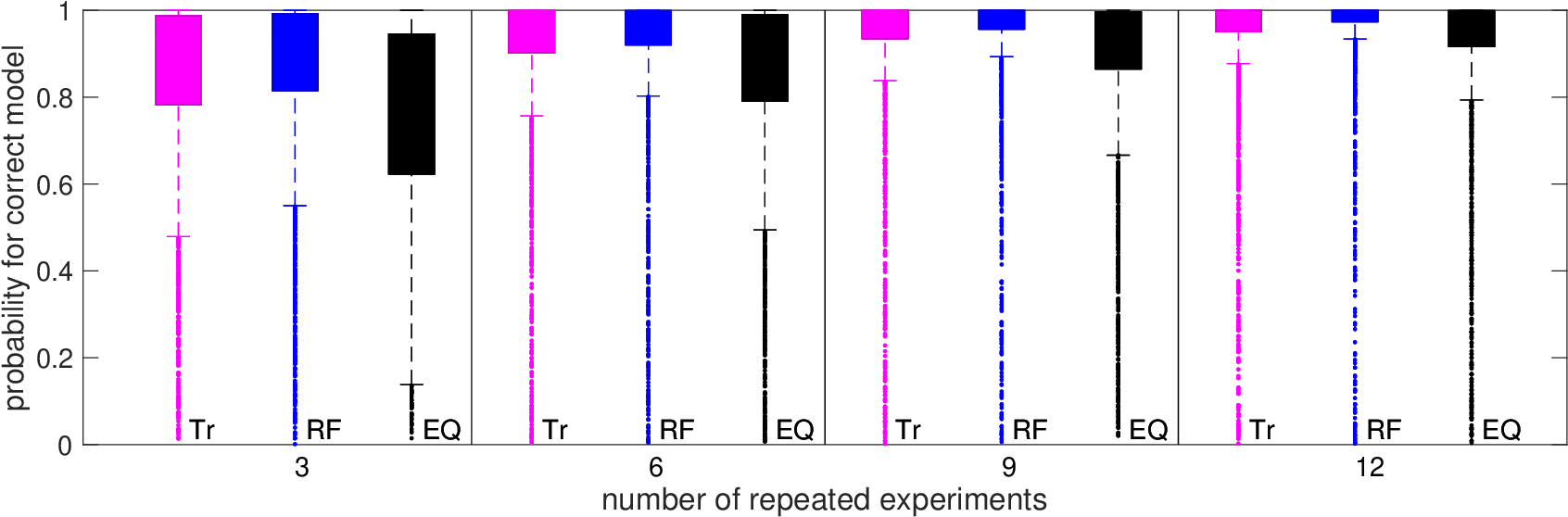}}
	\caption{Distributions of posterior model probabilities of the correct model for 2K prior predictive simulations (1K from each of the two models) for the infectious disease example with two models.  The data are all simulated at the respective optimal designs for the different approaches. The 0--1 loss is used as criterion. Settings with various numbers of realisations and $1 \leq n_d \leq 4$ observations per realisation are considered. The number of realisations were chosen such that the total number of observations $n = q \cdot n_d$ is equal to $n = 12$, $24$, $36$, or $48$. For each setting, from left to right the boxplots are for the cross-validated tree classification design (Tr; magenta), the random forest classification design (RF; blue), and the equispaced design (Eq; black).  \label{fig:results_epi_2models_highdim}}
\end{figure}

Figures~\ref{fig:results_epi_2models} (lower-dimensional designs) and \ref{fig:results_epi_2models_highdim} (higher-dimensional designs) display the distributions of posterior model probabilities for samples of size 2K (1K per model) from the prior predictive distribution at the various optimal designs found for all the dimension settings and the different methods. We also include equispaced designs for comparison. The marginal likelihoods are computed using the generalised Gauss-Hermite quadrature approximation \eqref{eq:marginal_likelihood_quadr} with $Q=30$ quadrature points for the death model and up to $Q=30^2$ quadrature points for the SI model.

Figure~\ref{fig:results_epi_2models} shows that for lower-dimensional designs all methods lead to designs with a very similar classification accuracy as measured by the distribution of the posterior model probabilities of the true model. For the higher-dimensional designs, Figure~\ref{fig:results_epi_2models_highdim} indicates that the designs found using random forests are performing slightly better than the designs found using cross-validated trees. This comes at the cost of a higher computing time.

\clearpage

\section{Additional Details and Results for Macrophage Example} \label{app:macro}

\subsection{Models and Prior Distributions} \label{app:macro_models}

In all three models, a macrophage can acquire a new bacterium with a constant rate $\phi$ while there is no antibiotic in the medium ($t<t_{exp}$); this rate then drops to 0 for the remainder of the simulations. In model~1, we assume that a proportion $p>0$ of available bacteria are non-replicating, so these are acquired by macrophages at rate $\phi\,p$, while replicating bacteria are acquired at rate $\phi(1-p)$. Intracellular bacteria are degraded at rate $d$ for replicating bacteria and rate $\epsilon$ for non-replicating bacteria. Within permissive macrophages containing $R>1$ replicating bacteria, the number of replicating bacteria increases by one every time one of these bacteria divides, but this division rate is assumed to be a decreasing function of $R$ (due to limited resources for bacterial growth within a macrophage), expressed as $a\,e^{-bR}$, where $a$ is the maximum division rate of bacteria and $b$ is a dimensionless scaling parameter. Finally, in model (1), replicating bacteria within permissive macrophages become non-replicating at rate $\delta$. All these transitions are listed in Table~\ref{tab:macro_models}.

\begin{table}[htbp!]
	\centering
	\caption{Three competing models considered in the macrophage example. $R(t)$ represents the number of replicating bacteria and $D(t)$ the number of non-replicating bacteria within a macrophage. In model~2, a proportion $q$ of macrophages are refractory and $1-q$ permissive.\label{tab:macro_models}}
	\begin{tabular}{|c|l|l|l|}
		\hline
		Model Number 	& Event Type 			& Update 		& Rate \\
		\hline
		(1)	 			& Acquisition of R 		& $R(t)+1$		& $\phi (1-p)$ \\
		& Acquisition of D 		& $D(t)+1$		& $\phi p$ \\
		& Division			& $R(t)+1$		& $a\,e^{-b\,R(t)}\, R(t)$ \\
		& Loss of R			& $R(t)-1$		& $d\,R(t)$ \\
		& Loss of D			& $D(t)-1$		& $\epsilon\,D(t)$ \\
		& Switch of R to D		& $R(t)-1,\,D(t)+1$	& $\delta\,R(t)$ \\
		\hline
		(2) Refractory		& Acquisition of D 		& $D(t)+1$		& $\phi$ \\
		& Loss of D			& $D(t)-1$		& $\epsilon\,D(t)$ \\
		(2) Permissive		& Acquisition of R 		& $R(t)+1$		& $\phi$ \\
		& Loss of R			& $R(t)-1$		& $d\,R(t)$ \\
		& Division			& $R(t)+1$		& $a\,e^{-b\,R(t)}\, R(t)$ \\
		\hline
		(3) 				& Acquisition of R 		& $R(t)+1$		& $\phi$ \\
		& Loss of R			& $R(t)-1$		& $d\,R(t)$ \\
		& Division			& $R(t)+1$		& $a\,e^{-b\,R(t)}\, R(t)$ \\
		\hline
	\end{tabular}
\end{table}

For each macrophage, numerical simulations of the three models are produced using the Gillespie algorithm \citep{gillespie1977}. 
In line with the general experimental setup, each macrophage is initially uninfected, but in model~2 it has a probability $q$ of being refractory. This state is set at the start of each simulation and does not change thereafter. To reproduce the data collection process described above, we produce two independent sets of simulations for each observation time $t_{obs}$ in a given experimental design. First, we run $S$ simulations of individual macrophages and record the proportion $\pi(t_{obs})$ of infected macrophages. Second, we run another set of simulations for the same duration until $S$ infected macrophages are obtained, from which we record the proportions $\{\mu_k(t_{obs}), k>0\}$ of infected macrophages containing $k$ bacteria. This can be repeated multiple times to generate multiple sets of observations from each model $m$, parameter vector $\vectt_m$ and experimental design $\vectd$.
Importantly, the simulations' results do not distinguish between replicating and non-replicating bacteria (model 1) or between refractory and permissive macrophages (model 2), as these cannot be told apart by microscopy alone.

The number of infected macrophages at time $t_{obs}$ has the binomial distribution \linebreak $\mathrm{Bin}(S; \: \mathrm{E}[\pi(t_{obs})])$. Likewise, the vector of numbers of infected macrophages containing $k = 1,\ldots,K_+$ bacteria has the multinomial distribution \newline $\mathrm{Mult}(S; \: \{\mathrm{E}[\mu_1(t_{obs})],\ldots,\mathrm{E}[\mu_{K_+}(t_{obs})]\})$. The last category $K_+$ contains all macrophages with at least $K_+$ bacteria. 

The most involved part is to obtain the expected proportions $\mathrm{E}[\pi(t_{obs})]$ and \newline $\mathrm{E}[\mu_1(t_{obs})],\ldots,\mathrm{E}[\mu_{K_+}(t_{obs})]$ for any particular set of parameters. A system of linear differential equations consisting of the Kolmogorov forward equations for the models in Table~\ref{tab:macro_models} has to be solved to determine the expected proportions of macrophages that contain a certain number of replicating and non-replicating bacteria (see \citet{Restif:2012}). The solution of this system can be computed using matrix exponentials. Considering only the total number of bacteria in a macrophage, the expected proportions $\mathrm{E}[\pi(t_{obs})], \mathrm{E}[\mu_1(t_{obs})],\ldots,\mathrm{E}[\mu_{K_+}(t_{obs})]$ can then be derived.

The prior distributions for each model were driven by the analysis of the experimental system in \citet{Restif:2012}. We assume truncated multivariate normal distributions, where the mean vector and variance-covariance matrix are based on the maximum likelihood estimates (MLEs) and the inverse of the Hessian obtained from the optimisation routine, respectively. All parameters are truncated below at $0$. The proportion parameters $p$ and $q$ are additionally truncated above at $1$.

In model~1, all macrophages are permissive, so $q = 0$. The mean vector and the variance-covariance matrix of the truncated normal prior for the remaining parameters of model~1 are given by

\begin{equation*}
\boldsymbol{\mu}^{\top}_1 = 
\begin{blockarray}{ccccccc}
a & b & d & \delta & \epsilon & p & \phi	\\
\begin{block}{(ccccccc)}
6.46 & 1.54 & 0.073 & 2.529 \cdot 10^{-10} & 0.035 & 0.097 & 0.25 \\
\end{block}
\end{blockarray}	
\end{equation*}

and

\begin{equation*}
\boldsymbol{\Sigma}_1 = 
\begin{blockarray}{lccccccc}
&  a & b & d & \delta & \epsilon & p & \phi	\\
\begin{block}{l(rrrrrrr)}
a 		& 32.8310 &  &  &  &  &  & \\
b 		& 0.6224 & 0.0696 &  &  &  &  & \\
d 		& 0.1991 & -0.0017 & 0.0487 &  &  &  & \\
\delta 	& 0.1258 & 0.0218 & -0.0164 & 0.0153 &  &  & \\
\epsilon & 0.0166 & 0.0048 & -0.0069 & 0.0052 & 0.0024 &  &  \\
p 		& 0.2142 & 0.0252 & -0.0061 & 0.0102 & 0.0039 & 0.0192 &  \\
\phi 	& -0.0101 & 0.0001 & -0.0029 & 0.0018 & 0.0011 & 0.0018 & 0.0030 \\
\end{block}
\end{blockarray}.
\end{equation*}
(The upper triangular part of the variance-covariance matrices is omitted.)

For model~2, where all bacteria are replicating and hence $\delta = p = 0$, the mean vector and the variance-covariance matrix are selected to be

\begin{equation*}
\boldsymbol{\mu}^{\top}_2 = 
\begin{blockarray}{cccccc}
a & b & d & \epsilon & \phi & q	\\
\begin{block}{(cccccc)}
8.54221 & 1.450254 & 0.09111 & 0.03 & 0.25948 & 0.266837 \\
\end{block}
\end{blockarray}	
\end{equation*}

and

\begin{equation*}
\boldsymbol{\Sigma}_2 = 
\begin{blockarray}{lcccccc}
&  a & b & d & \epsilon & \phi & q \\
\begin{block}{l(rrrrrr)}
a 		& 33.5250 &  &  &  &  &  \\
b 		& 1.1380 & 0.3586 &  &  &  &  \\
d 		& 0.8252 & -0.1213 & 0.0952 &  &  &  \\
\epsilon & 0.0253 & 0.0077 & -0.0023 & 0.1067 &  &  \\
\phi 	& -0.1471 & -0.0511 & 0.0197 & -0.0001 & 0.0355 & \\
q		& 0.9048 & 0.1962 & -0.0658 & 0.0097 & -0.0284 & 0.2765 \\
\end{block}
\end{blockarray}.
\end{equation*}

Finally, model~3 assumes that all macrophages are permissive and all bacteria are replicating, so $\delta = \epsilon = p = q = 0$. For this model the truncated normal prior's mean vector and variance-covariance matrix are

\begin{equation*}
\boldsymbol{\mu}^{\top}_3 = 
\begin{blockarray}{cccc}
a & b & d & \phi	\\
\begin{block}{(cccc)}
0.8161965 & 0.52672325 & 0.20740975 & 0.3203258 \\
\end{block}
\end{blockarray}	
\end{equation*}

and

\begin{equation*}
\boldsymbol{\Sigma}_3 = 
\begin{blockarray}{lcccc}
&  a & b & d & \phi\\
\begin{block}{l(rrrr)}
a 		& 0.7518 &  &  & \\
b 		& 0.1172 & 0.0506 &  & \\
d 		& 0.0720 & -0.0090 & 0.0228 &  \\
\phi 	& 0.0008 & -0.0106 & 0.0100 & 0.0287 \\
\end{block}
\end{blockarray}.
\end{equation*}

\subsection{Optimal Designs} \label{app:macro_designs}

Tables~\ref{macro:design_acc_loss_table1} and \ref{macro:design_acc_loss_table2} show the optimal designs for each classification method and for the different numbers of observation times. The tree and the random forest classification approaches lead to very similar designs.

\begin{table}[htbp]
	\centering
	\caption{\label{macro:design_acc_loss_table1}Optimal classification designs $(t_{exp}; \boldsymbol{t}_{obs})$ using trees or random forests under the 0--1 loss and equispaced designs for the macrophage model ($n = 1$, $2$, and $3$).}
	\begin{tabular}{|l|D{.}{.}{2}|D{.}{.}{2}|D{.}{.}{2}|*{2}{D{.}{.}{2}}|D{.}{.}{2}|*{3}{D{.}{.}{2}}|}
		\hline
		\multicolumn{1}{|l|}{Method} & \multicolumn{2}{c|}{$n = 1$} & \multicolumn{3}{c|}{$n = 2$} & \multicolumn{4}{c|}{$n = 3$}\\
		\cline{2-10}
		\multicolumn{1}{|l|}{} & \multicolumn{1}{c|}{$t_{exp}$} & \multicolumn{1}{c|}{$\boldsymbol{t}_{obs}$} & \multicolumn{1}{c|}{$t_{exp}$} & \multicolumn{2}{c|}{$\boldsymbol{t}_{obs}$} & \multicolumn{1}{c|}{$t_{exp}$} & \multicolumn{3}{c|}{$\boldsymbol{t}_{obs}$} \\
		\hline
		Tree  & 1.20 & 10.00 & 0.09 & 1.75 & 10.00 & 0.09 & 1.25 & 2.75 & 10.00 \\
		RF    & 1.11 & 10.00 & 0.10 & 1.75 & 10.00 & 0.10 & 1.50 & 10.00 & 10.00 \\
		Equi & 0.80 & 5.00  & 0.80 & 3.33 & 6.67  & 0.80 & 2.50 & 5.00 & 7.50 \\
		\hline 
	\end{tabular}
\end{table}

\begin{table}[htbp]
	\centering
	\caption{\label{macro:design_acc_loss_table2}Optimal classification designs $(t_{exp}; \boldsymbol{t}_{obs})$ using trees or random forests under the 0--1 loss and equispaced designs for the macrophage model ($n = 4$ and $5$).}
	\begin{tabular}{|l|D{.}{.}{2}|*{4}{D{.}{.}{2}}|D{.}{.}{2}|*{5}{D{.}{.}{2}}|}
		\hline
		\multicolumn{1}{|l|}{Method} & \multicolumn{5}{c|}{$n = 4$} & \multicolumn{6}{c|}{$n = 5$} \\
		\cline{2-12}
		\multicolumn{1}{|l|}{} & \multicolumn{1}{c|}{$t_{exp}$} & \multicolumn{4}{c|}{$\boldsymbol{t}_{obs}$} & \multicolumn{1}{c|}{$t_{exp}$} & \multicolumn{5}{c|}{$\boldsymbol{t}_{obs}$} \\
		\hline
		Tree  & 0.09 & 0.75 & 2.25 & 9.00 & 10.00 & 0.10 & 0.75 & 2.50 & 2.75 & 10.00 & 10.00 \\
		RF    & 0.10 & 1.25 & 2.75 & 10.00 & 10.00 & 0.09 & 1.50 & 2.50 & 9.75 & 10.00 & 10.00 \\
		Equi & 0.80 & 2.00 & 4.00 & 6.00 & 8.00 & 0.80 & 1.67 & 3.33 & 5.00 & 6.67 & 8.33 \\
		\hline 
	\end{tabular}
\end{table}

\subsection{Misclassification Matrix} \label{app:macro_misclass}

\begin{figure}[hbtp!]
	\centering
	\subfigure[1 design point]{\includegraphics[width=0.45\textwidth]{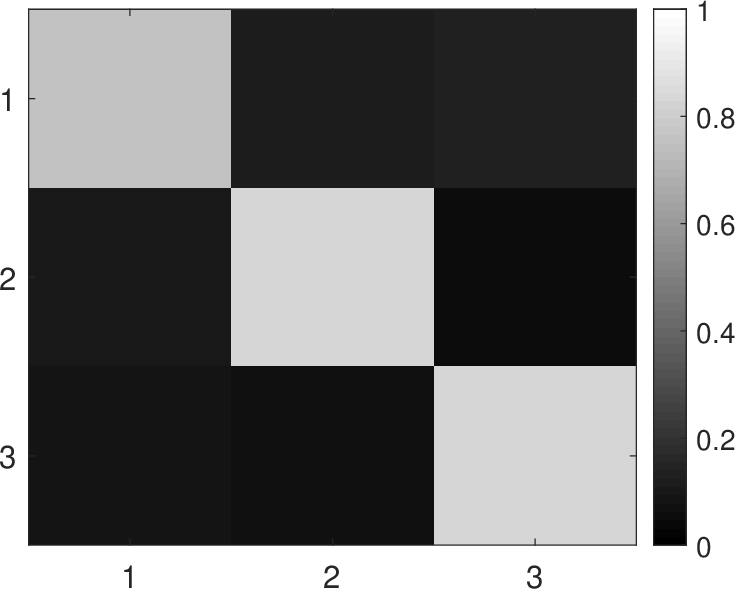}} \quad
	\subfigure[2 design points]{\includegraphics[width=0.45\textwidth]{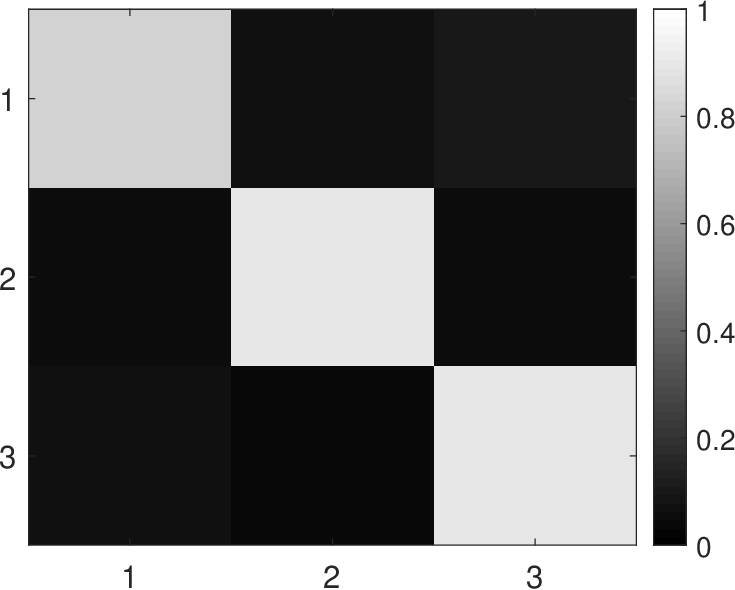}} 
	\vspace*{1ex}
	\subfigure[3 design points]{\includegraphics[width=0.45\textwidth]{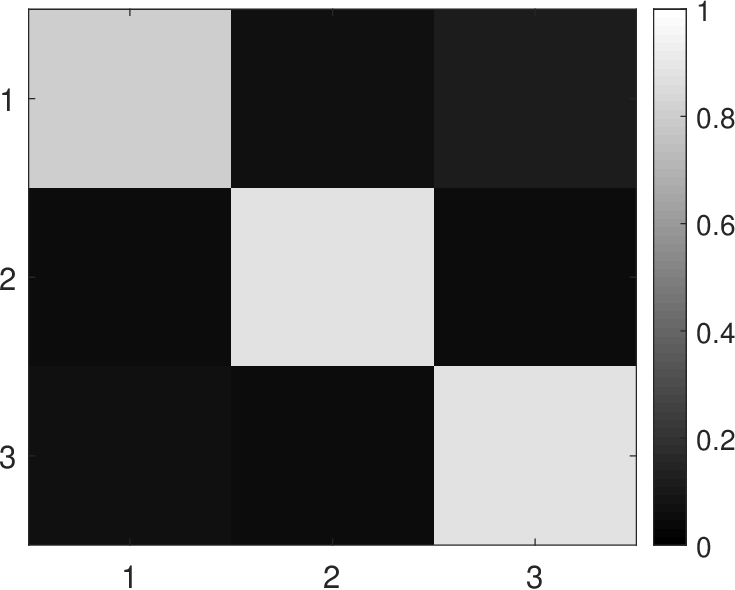}} \quad
	\subfigure[4 design points]{\includegraphics[width=0.45\textwidth]{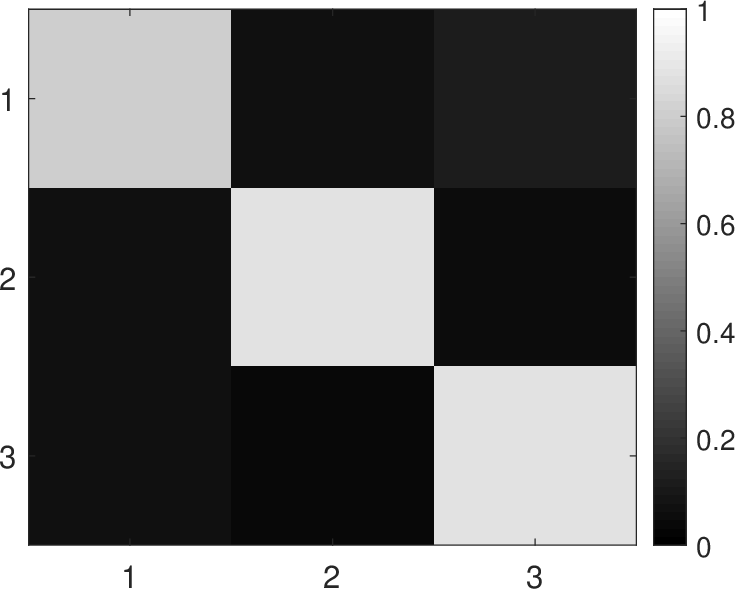}}
	\caption{Misclassification matrices obtained for the \emph{random forest classification designs} under the \emph{0--1 loss} for the macrophage example.  Designs for 1 -- 4 observation times plus the exposure duration are considered.\label{fig:macro_misclassmatrix}}
\end{figure}

We can use the same random forest classifiers and their associated samples that were created to estimate the misclassification error rates in Table~\ref{tab:macro_misclass_error} to compute the misclassification matrices. The misclassification matrices for the optimal designs obtained under the random forest classification approach are displayed in Figure~\ref{fig:macro_misclassmatrix}. The classification power is very high for all the models. One can see that it is slightly more difficult to detect heterogeneity between bacteria (model~1) than heterogeneity between macrophages (model~2). The misclassification matrices for the designs obtained under the tree classification approach are almost identical.

\section{Logistic Regression Example} \label{app:logreg}

We consider the logistic regression example of \citet{overstall_woods_2017} and \citet{Overstall2018}.  The response is binary, $y_{ij} \sim \mathcal{B}(p_{ij})$, and
\begin{align*}
\mbox{logit}(p_{ij}) &= \beta_0 + \gamma_{0i} + \sum_{a=1}^4 v_a (\beta_a + \gamma_{ai}) x_{aij},
\end{align*}
where $j = 1,\ldots,n_G$ and $i = 1,\ldots,G$.  Here $G$ is the total number of groups and $n_G$ is the number of observations per group.  The total number of observations is $n = G \times n_G$.   The model parameter of interest is $\vectt = (\beta_0,\beta_1,\beta_2,\beta_3,\beta_4)^\top$.  The random effect for the $i$th group is $\vect{\gamma}_i = (\gamma_{0i},\gamma_{1i},\gamma_{2i},\gamma_{3i},\gamma_{4i})^\top$.  The observed vector of responses for the $i$th group is $\vecty_i = (y_{i1},\ldots,y_{in_G})$ and the total dataset is denoted $\vecty = (\vecty_1,\ldots,\vecty_G)^\top$. The design vector is the concatenation of the controllable elements of the design matrix,  $\vectd = \{x_{aij}; a=1,\ldots,4, i = 1,\ldots,G, j=1,\ldots,n_G\}$ and is of length $n\times 4$.  Each design element is restricted, $x_{aij} \in [-1,1]$.  The variable $v_a$ is an indicator variable that is equal to 1 if the $a$th predictor is present in the model.  It may not be clear which of the four predictors should be included in the model, so there are $2^4 = 16$ possible models to choose from.  We aim to select the design $\vectd$ that maximises our ability to discriminate between all possible models under various prior assumptions as described below.  

As in \citet{Overstall2018}, two different model structures are considered.  The first structure is that all random effects (RE) are set to 0, resulting in the fixed effects (FE) structure.  The second structure is that the random effects are allocated a distribution (RE structure).  Within each chosen structure, there are 16 models to discriminate between.  In both the FE and RE structures, we use the priors $\beta_0 \sim \mathcal{U}(-3,3)$, $\beta_1 \sim \mathcal{U}(4,10)$, $\beta_2 \sim \mathcal{U}(5,11)$, $\beta_3 \sim \mathcal{U}(-6,0)$, $\beta_4 \sim \mathcal{U}(-2.5,3.5)$. We assume that all parameters are independent \emph{a priori}.  For the RE model we set $\gamma_{ai} \sim \mathcal{U}(-\zeta_a,\zeta_a)$ and allocate a triangular prior to $\zeta_a$, $p(\zeta_a) = 2(U_a - \zeta_a)/U_a^2$, $0 < \zeta_a < U_a$, where $(U_0,U_1,U_2,U_3,U_4) = (3,3,3,1,1)$. One possibility for the prior distribution placed on each model is a prior which depends on the number of predictors present in the model.  Let $(v_{m1},\ldots,v_{m4})$ denote the values of $(v_{1},\ldots,v_{4})$ for model $m$. A model prior accounting for Bayesian multiplicity \citep{Scott2010} is 
\begin{align}
p(m) &= \frac{1}{5 \binom{4}{\sum_{a=1}^4 v_{ma}}}. \label{eq:model_prior}
\end{align}
In order to estimate the misclassification error rate under the Bayes classifier (the Bayes error rate) for some design $\vectd$, we need to estimate posterior model probabilities for $J$ datasets simulated from the prior predictive distributions of all the models.  A common approach for rapid approximation of the evidence for model $m$, $p(\vecty|m,\vectd)$, in the context of Bayesian optimal design is importance sampling (IS), where the importance distribution is the prior (e.g.\ \citet{ryan_drovandi_thompson_pettitt_2014}).  However, if the data is informative (as might be the case in this example if $n$ is large), the number of IS samples to estimate the evidence with reasonable precision may be prohibitively large.  The situation is significantly worse for the RE structure, as an importance distribution is required over the space of both the parameter of interest and the random effects (see, e.g., \citet{ryan_drovandi_pettitt_2015}).  For the FE structure and $n=48$, using 100K importance samples from the prior and $J=800$ ($50$ per model), the time taken to approximate the misclassification error rate for a random design on a cluster using 24 parallel threads was almost $2.75$ minutes. This is very computationally intensive considering that we need to optimise over $48 \times 4$ design variables. Performing IS for the RE structure might be considered as completely intractable.  \citet{Overstall2018} propose the use of normal-based approximations to the posterior in the Bayesian design context to provide a convenient estimate of the evidence. They consider the same logistic regression example but use normal priors to facilitate the approximation of the evidence. In some applications, a normal-based approximation may not be adequate.

In contrast, our classification approach avoids computing posterior quantities and requires only simulation from all the models.  Interestingly, moving to the RE structure poses little additional difficulty for the classification approach as it remains trivial to simulate from the models.  This is a significant advantage of the classification approach.     

For the FE structure we consider $n \in \{6,12,24,48\}$ and for the RE structure we consider $n_G=6$ and $G \in {2,4,8}$ (to give $n \in \{12,24,48\}$).  Two prior distributions on the model indicator are trialled: (1) the prior where models are equally likely \emph{a priori} and (2) the prior in \eqref{eq:model_prior} that corrects for Bayesian multiplicity.  We refer to the first as the equal prior and the second as the unequal prior. For this example, the only design criterion that we consider is the misclassification error rate (the excepted 0--1 loss). During the design optimisation phase, we estimate the expected loss by employing cross-validated classification trees using a sample of size 80K (5K simulations per model). The observations are weighted within the trees according to their prior model probabilities. We consider a discretised design space for each $x_{aij}$ consisting of the five values $\{-1,-0.5,0,0.5,1\}$.

After having obtained the optimal designs for the different scenarios regarding model structure (FE or RE) and prior distributions (equal or unequal), we attempt to assess the classification performance of these optimal designs using random forests. For each optimal design, 10K simulations under each model are used to train a random forest with 100 trees.  A fresh set of $16 \times 10$K = 160K simulations is used to estimate the misclassification error rate and the misclassification matrix. The model proportions of this test sample reflect the prior model probabilities. The results for the optimal designs of the different scenarios are shown in the rows with bold row labels in Table~\ref{tab:classification_validation_logreg}.  For each scenario, results for optimal designs under different scenarios as well as a randomly generated design are also provided. For the randomly generated designs, each design point $x_{aij}$ equals $1$ or $-1$ with equal probability.

The results suggest that the optimal designs found for this example are remarkably robust with respect to the assumed model structure (FE or RE) and the assumed prior model probabilities (equal or unequal).  The random design has the worst performance under all scenarios.  We can also see a decrease in the misclassification error rate as the sample size is increased, as expected.

\begin{table}
	\centering
	\caption{Shown are the misclassification error rates obtained at various optimal tree classification designs for the different logistic regression models.  Four scenarios for the true model are considered: (1) FE structure under the equal prior, (2) FE structure under the unequal prior, (3) RE model under the equal prior and (4) RE model under the unequal prior. Rows with bold labels contain the results for the optimal designs under each scenario.  Also shown, for each scenario, are the results for various designs obtained under different wrong scenarios and the results for a random design.  The results suggest that the optimal designs found are robust to the model structure (FE or RE) and to the prior model probabilities (equal or unequal).  The random design has the worst performance under all scenarios.\label{tab:classification_validation_logreg}}
	\begin{tabular}{|c|cccc|}
		\hline
		\multicolumn{5}{|c|}{FE structure under the equal prior} \\
		\hline
		Design & \multicolumn{4}{c|}{Sample Size ($n$)} \\
		& 6 & 12 & 24 & 48 \\
		\hline
		\textbf{FE equal} & 0.616 &	0.494 &	0.407 &	0.354 \\
		FE unequal        & 0.665 & 0.535 & 0.431 & 0.386 \\
		RE equal          & NA    &	0.497 & 0.413 & 0.359 \\
		random            & 0.730 & 0.638 & 0.534 & 0.463 \\
		\hline
		\hline
		\multicolumn{5}{|c|}{FE structure under the unequal prior} \\
		\hline
		Design & \multicolumn{4}{c|}{Sample Size ($n$)} \\
		& 6 & 12 & 24 & 48 \\
		\hline
		FE equal            & 0.511 & 0.416 & 0.337 & 0.290 \\
		\textbf{FE unequal} & 0.480 & 0.409 & 0.340 & 0.307 \\
		RE unequal          & NA    & 0.410 & 0.341 & 0.313 \\
		random              & 0.553 & 0.456 & 0.401 & 0.352 \\
		\hline
		\hline
		\multicolumn{5}{|c|}{RE model under the equal prior} \\
		\hline
		Design & \multicolumn{4}{c|}{Sample Size ($n$)} \\
		& 6 & 12 & 24 & 48 \\
		\hline
		FE equal          & NA & 0.504 & 0.423 & 0.366 \\
		\textbf{RE equal} & NA & 0.506 & 0.424 & 0.369 \\
		RE unequal        & NA & 0.545 & 0.442 & 0.399 \\
		random            & NA & 0.629 & 0.538 & 0.462 \\
		\hline
		\hline
		\multicolumn{5}{|c|}{RE model under the unequal prior} \\
		\hline
		Design & \multicolumn{4}{c|}{Sample Size ($n$)} \\
		& 6 & 12 & 24 & 48 \\
		\hline
		FE unequal          & NA & 0.416 & 0.351 & 0.317 \\
		RE equal            & NA & 0.426 & 0.349 & 0.302 \\
		\textbf{RE unequal} & NA & 0.416 & 0.349 & 0.316 \\
		random              & NA & 0.483 & 0.406 & 0.362 \\
		\hline
	\end{tabular}
\end{table}

\begin{figure}[htbp!]
	\centering
	\subfigure[$n=6$]{\includegraphics[width=0.45\textwidth]{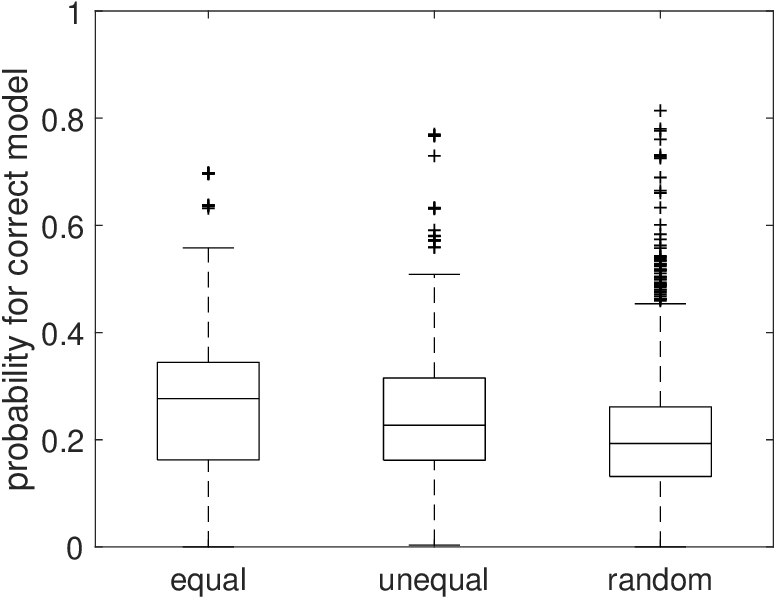}\label{figsub:validation_logreg_equal_n6}} \quad
	\subfigure[$n=12$]{\includegraphics[width=0.45\textwidth]{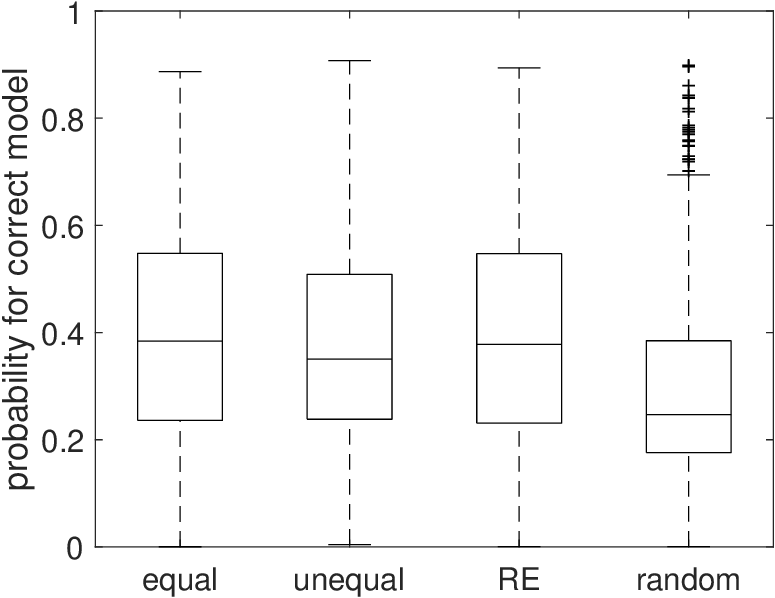}\label{figsub:validation_logreg_equal_n12}}
	\vspace*{1ex}
	\subfigure[$n=24$]{\includegraphics[width=0.45\textwidth]{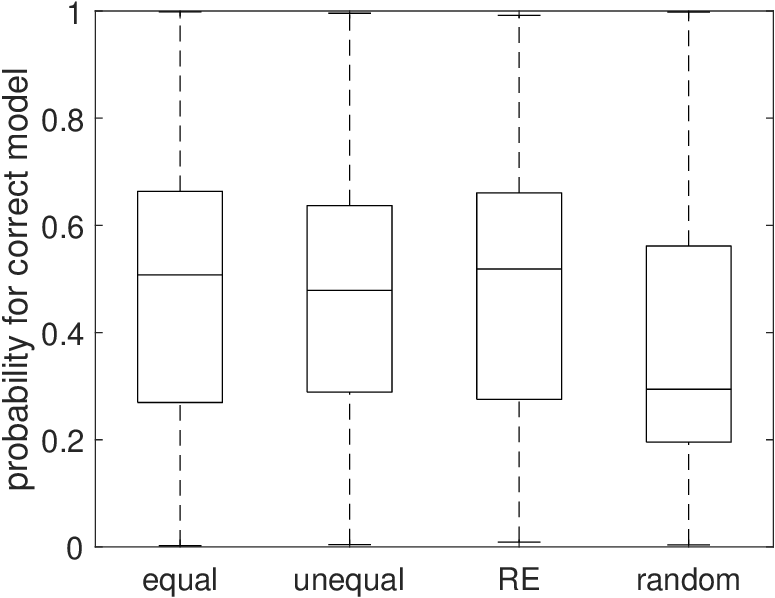}\label{figsub:validation_logreg_equal_n24}} \quad
	\subfigure[$n=48$]{\includegraphics[width=0.45\textwidth]{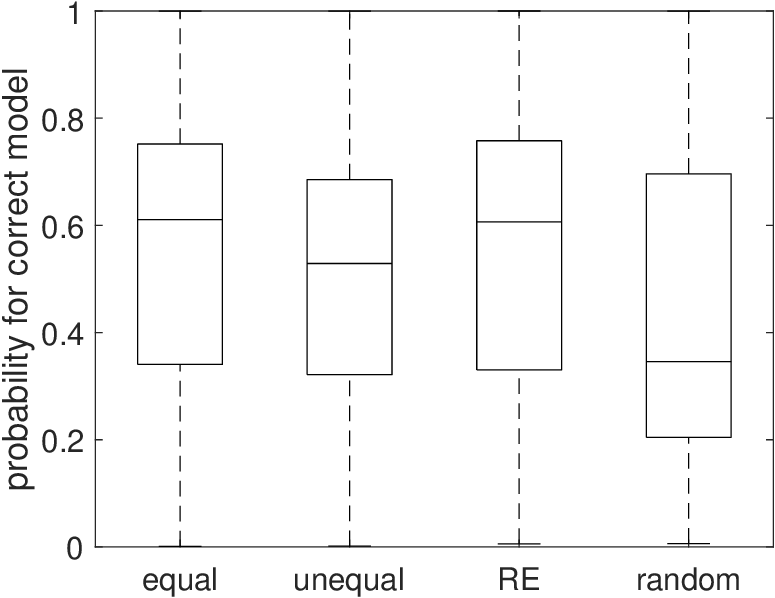}\label{figsub:validation_logreg_equal_n48}}
	\caption{Estimated posterior model probabilities for the correct model by the validation study under the equal prior.  Results based on sample sizes of (a) $n=6$, (b) $n=12$, (c) $n=24$ and (d) $n=48$.  Several designs are considered: optimal design found under the correct (equal) prior, optimal design found under the wrong (unequal) prior, optimal design found under the wrong (RE) structure (no results for $n=6$) and randomly selected designs.}
	\label{fig:logreg_validation_equal}
\end{figure}

\begin{figure}[htbp!]
	\centering
	\subfigure[$n=6$]{\includegraphics[width=0.45\textwidth]{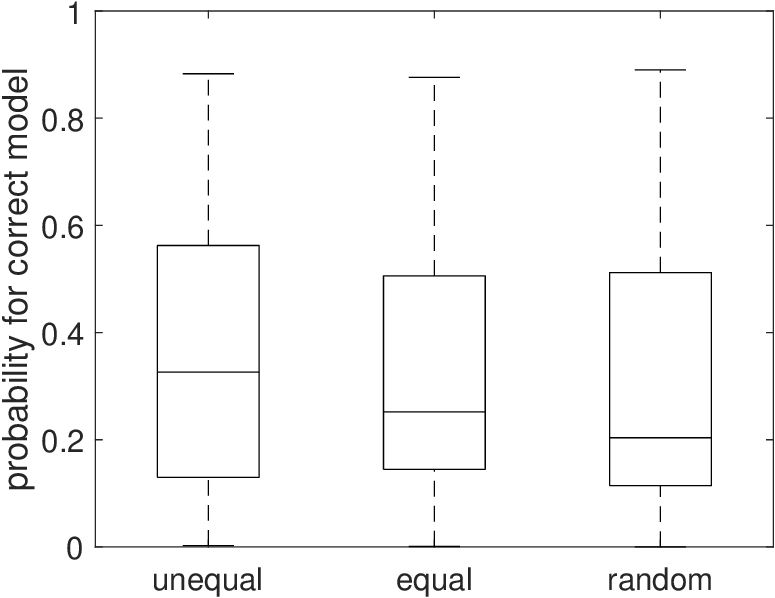}\label{figsub:validation_logreg_multi_n6}} \quad
	\subfigure[$n=12$]{\includegraphics[width=0.45\textwidth]{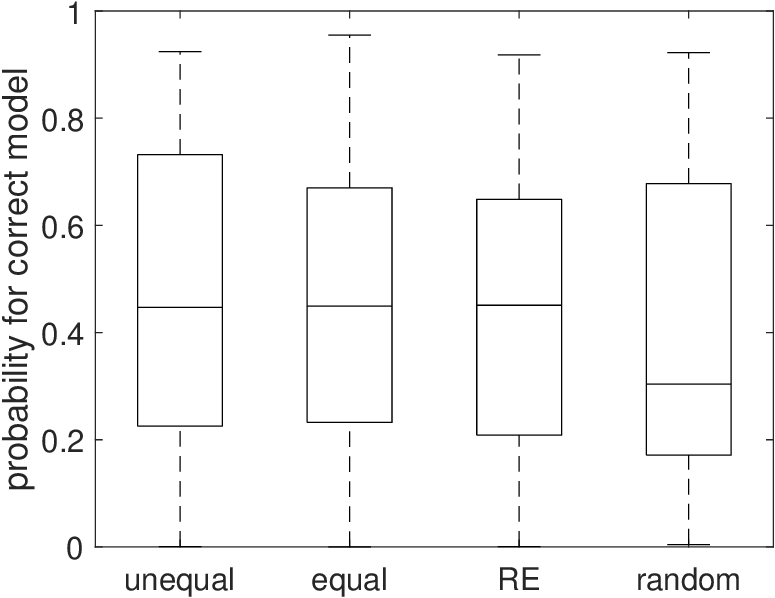}\label{figsub:validation_logreg_multi_n12}}
	\vspace*{1ex}
	\subfigure[$n=24$]{\includegraphics[width=0.45\textwidth]{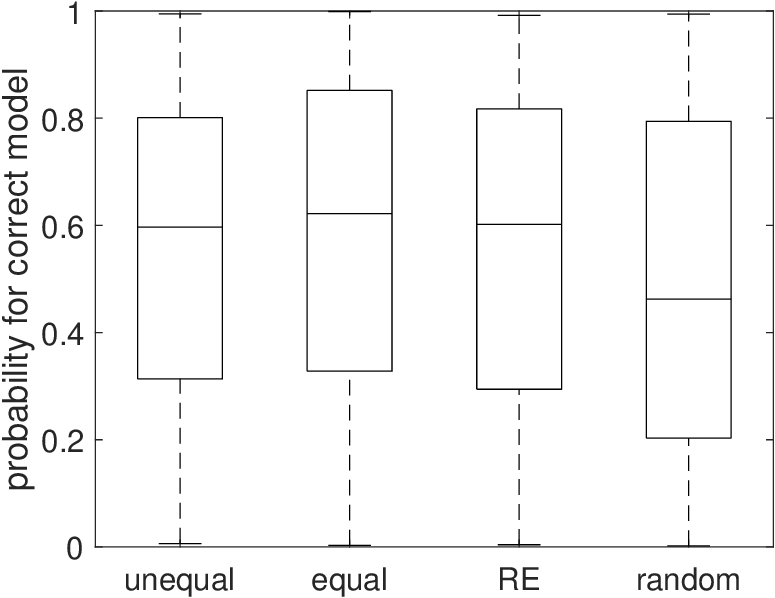}\label{figsub:validation_logreg_multi_n24}} \quad
	\subfigure[$n=48$]{\includegraphics[width=0.45\textwidth]{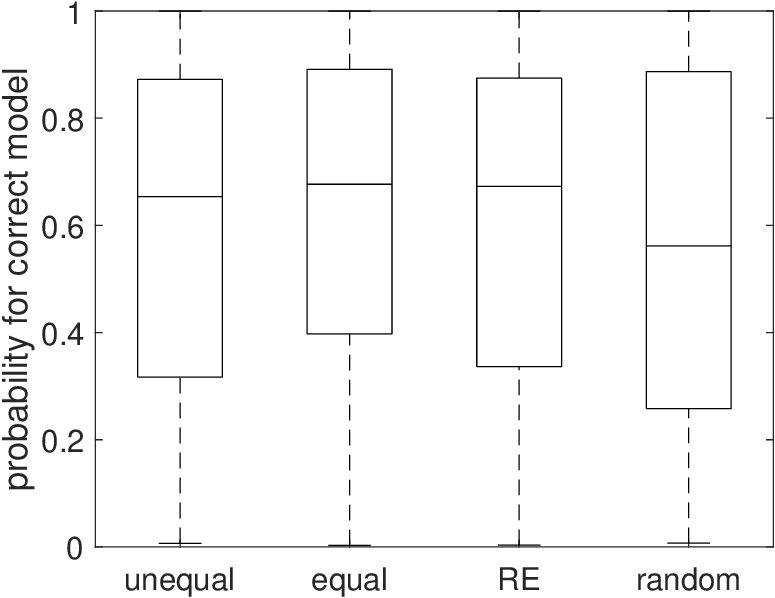}\label{figsub:validation_logreg_multi_n48}}
	\caption{Estimated posterior model probabilities for the correct model by the validation study under the unequal prior.  Results based on sample sizes of (a) $n=6$, (b) $n=12$, (c) $n=24$ and (d) $n=48$.  Several designs are considered: optimal design found under the correct (unequal) prior, optimal design found under the wrong (equal) prior, optimal design found under the wrong (RE) structure (no results for $n=6$) and randomly selected designs.}
	\label{fig:logreg_validation_multi}
\end{figure}

It is also of interest to see how well the optimal designs found under the tree classification approach perform in terms of posterior model probabilities. We conduct a simulation study under the FE structure using either the equal or the unequal prior. For each design we want to assess, we simulate a sample of 800 datasets from the marginal distributions of all the various models, where the proportion of datasets from a particular model in the sample corresponds to that model's prior model probability. For each of the 800 datasets, we approximate the posterior model probability of the model $m$ that generates the dataset $\vecty$ using IS with 100K prior simulations. As for the classification results in Table~\ref{tab:classification_validation_logreg}, we are also interested in the performance of optimal designs found under some wrongly assumed scenarios.  We also consider a `random' setup where we select designs randomly for each of the 800 datasets. Figure~\ref{fig:logreg_validation_equal} shows the boxplots of the estimated posterior model probabilities of the correct model for some of the designs of interest when the true scenario is the FE structure with the equal prior. The resulting boxplots when the true scenario is the FE structure with the unequal prior are shown in Figure~\ref{fig:logreg_validation_multi}. It is again evident that the optimal designs found are robust under the choice of the structure (FE or RE) and the choice of the prior model probabilities (equal or unequal).  We do not perform a simulation study under the RE structure given the increasing difficulty of estimating the posterior model probabilities under this structure. 

It is important to note that the random forest-based validation results in Table~\ref{tab:classification_validation_logreg} were obtained in a small fraction of the time that it took to conduct the simulation study used to produce the results in Figures~\ref{fig:logreg_validation_equal} and \ref{fig:logreg_validation_multi}.   

\begin{figure}[hbtp!]
	\centering
	\subfigure[$n=6$]{\includegraphics[width=0.45\textwidth]{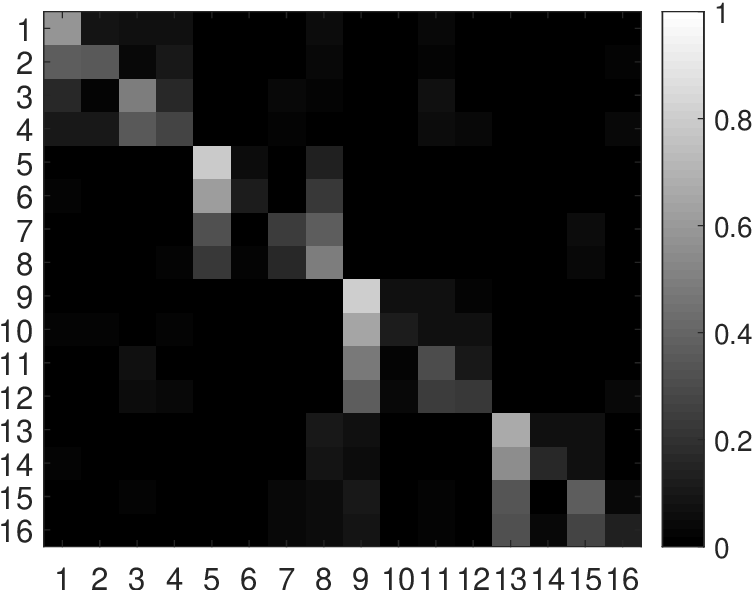}\label{figsub:logreg_class6}} \quad
	\subfigure[$n=12$]{\includegraphics[width=0.45\textwidth]{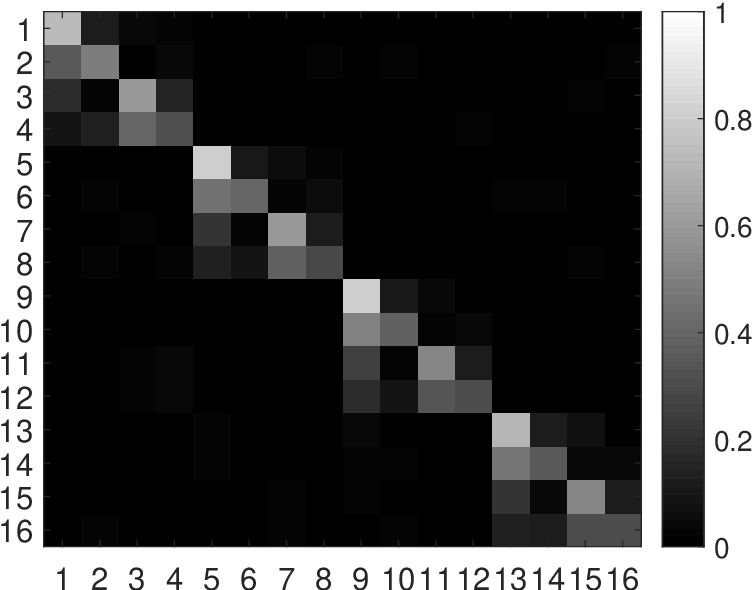}\label{figsub:logreg_class12}}
	\vspace*{1ex}
	\subfigure[$n=24$]{\includegraphics[width=0.45\textwidth]{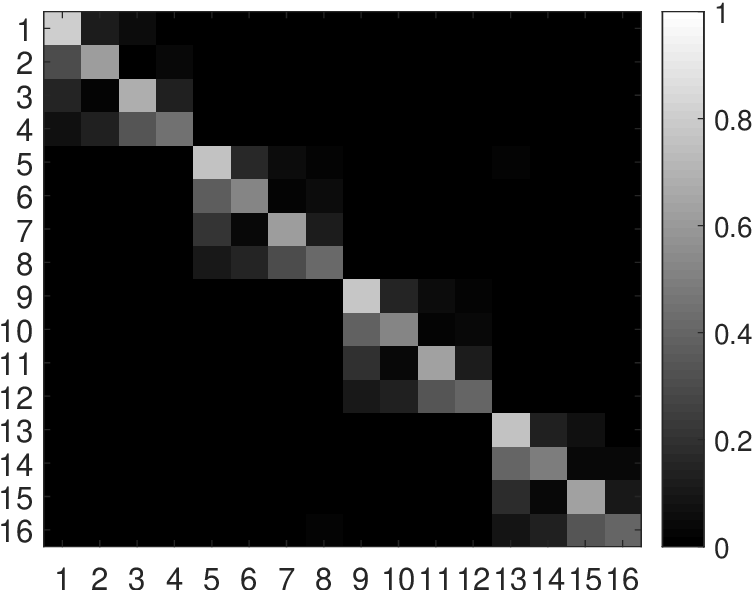}\label{figsub:logreg_class24}} \quad
	\subfigure[$n=48$]{\includegraphics[width=0.45\textwidth]{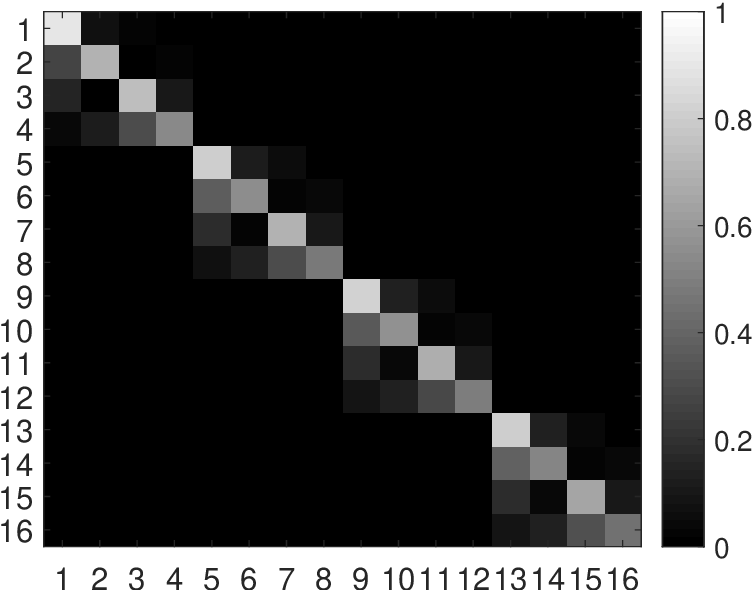}\label{figsub:logreg_class48}}
	\caption{Misclassification matrices obtained for the \emph{FE structures} of the logistic regression example with the \emph{equal prior}.\label{fig:logreg_misclass_equal}}
\end{figure}

\begin{figure}[hbtp!]
	\centering
	\subfigure[$n=6$]{\includegraphics[width=0.45\textwidth]{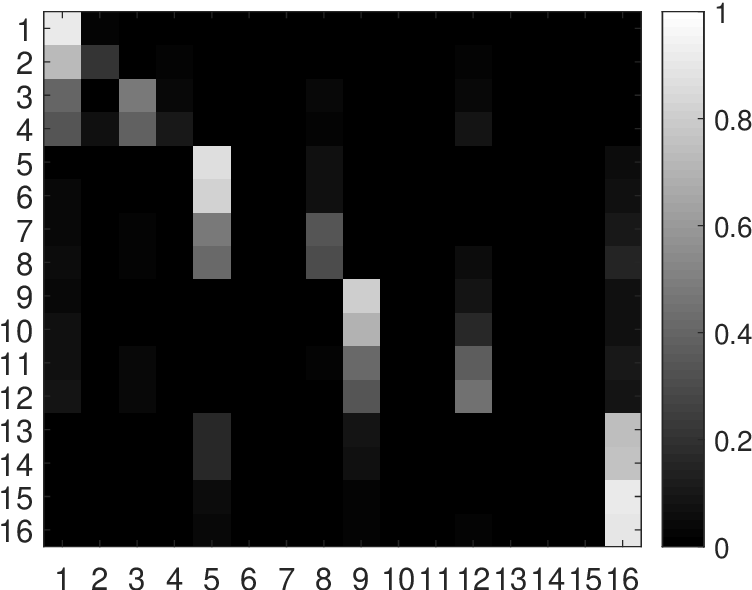}\label{figsub:logreg_class_p6}} \quad
	\subfigure[$n=12$]{\includegraphics[width=0.45\textwidth]{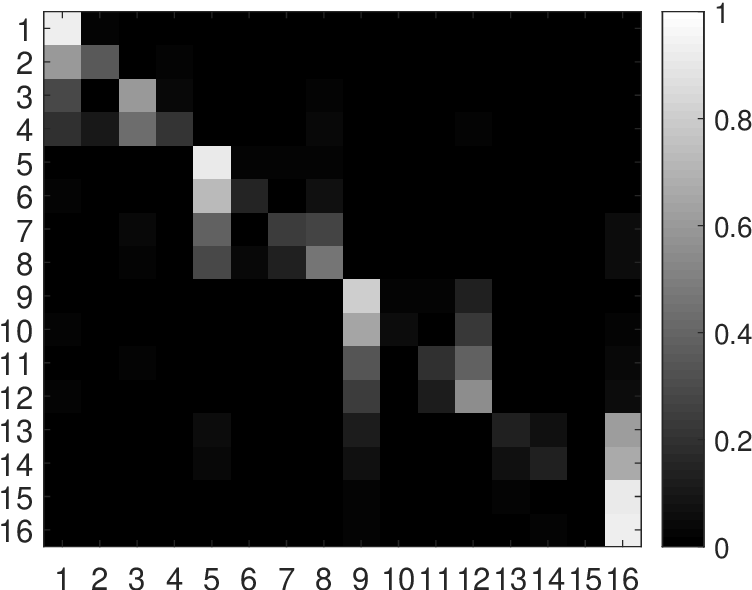}\label{figsub:logreg_class_p12}}
	\vspace*{1ex}
	\subfigure[$n=24$]{\includegraphics[width=0.45\textwidth]{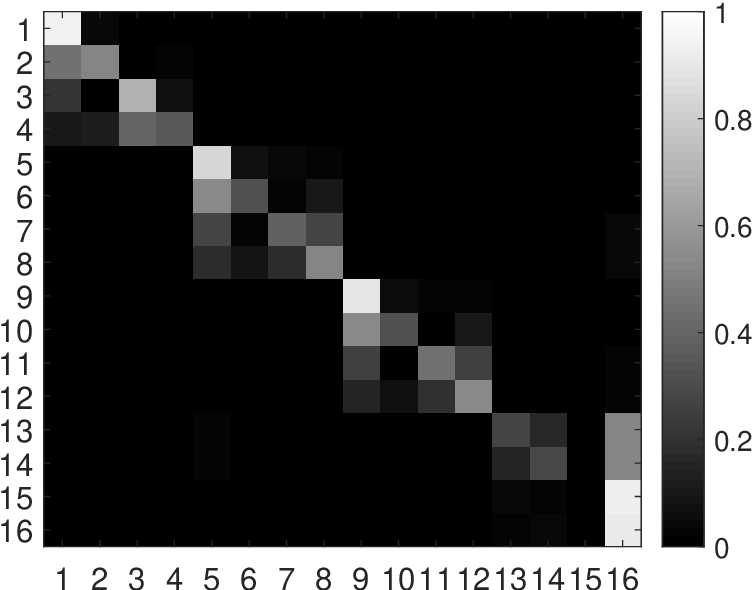}\label{figsub:logreg_class_p24}} \quad
	\subfigure[$n=48$]{\includegraphics[width=0.45\textwidth]{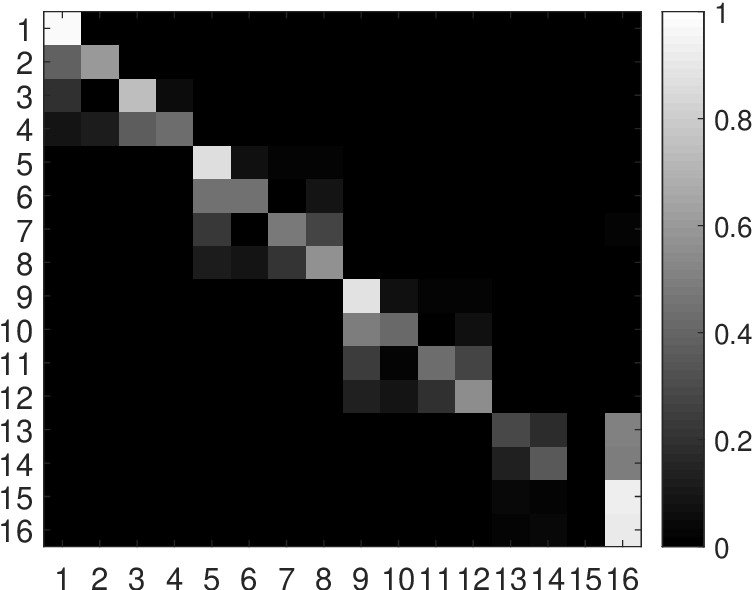}\label{figsub:logreg_class_p48}}
	\caption{Misclassification matrices obtained for the \emph{FE structures} of the logistic regression example with the \emph{unequal prior}.\label{fig:logreg_misclass_multi}}
\end{figure}

Figures \ref{fig:logreg_misclass_equal} and \ref{fig:logreg_misclass_multi} show misclassification matrices for the logistic regression models under the FE structure for the equal and unequal priors, respectively.  To produce the results, 10K simulations from each model are used to train a random forest with 100 trees.  The misclassification matrices are then computed based on a fresh test dataset of size $16 \times 10K = 160K$ with model proportions in the dataset corresponding to the prior model probabilities (under the equal prior, 10K simulations are taken from each model). The improvement in classification accuracy is clear as the sample size is increased.  When the unequal prior is selected, it is evident for small sample sizes that it is easier to classify the models with higher prior probability. The misclassification matrices for the RE structure are omitted because they are very similar.

\clearpage

\section{Spatial Extremes Example}\label{app:spatextr}

In this example, the goal is to place a fixed number of measuring sites in space in order to maximise the ability to discriminate between different spatial models for extreme outcomes (e.g., maximum annual temperatures). There are many spatial models for extreme events, see \citet{davison2012} for an overview. For this example, we consider to discriminate between three isotropic models: two \emph{max-stable} models and one \emph{copula} model.

\subsection{Models}

Max-stable processes are popular for modelling spatial extremes because they are the only possible limits of renormalised pointwise maxima of infinitely many independent copies of a stochastic process \citep{dehaan2006-2}. The advantage of working with the limiting process is that no knowledge about the underlying true process is necessary. Inference for extreme outcomes based on the true underlying process is fraught with high uncertainty and most often not feasible because only the tails of the distribution are observed. If the limiting assumption is (approximately) appropriate, it is much easier to model the extreme data according to a max-stable process.

All the univariate marginal distributions of a max-stable process are members of the family of \emph{generalised extreme value (GEV)} distributions. We assume that all the univariate marginal distributions have a \emph{unit Fr\'{e}chet} distribution ($\Pr\{Y(\vectx) \leq y\} = \exp\{-1/y\}, y > 0$), so the focus is on modelling the dependence structure of the process. The assumption of unit Fr\'{e}chet margins is not too restrictive from a practical perspective since a simple transformation can be applied to the univariate margins to make them unit Fr\'{e}chet distributed, see \citet{davison2012}. The marginal parameters needed for that transformation can be estimated in a separate step. Alternatively, one may estimate the dependence and marginal parameters together.

The \emph{spectral representation} of a max-stable process $\{Y(\vectx), \vectx \in \mathcal{X} \subseteq \mathbb{R}^d\}$ with unit Fr\'{e}chet margins is given by

\begin{align}
Y(\vectx) &= \max_{i\geq1} \varphi_i(\vectx), \hspace{10pt} \vectx \in \mathcal{X}, \label{eq:max_stable_definition}
\end{align}

where the \emph{spectral functions} $\varphi_i(\vectx) = \zeta_i Z_i(\vectx)$ are the products of the realisations $\{\zeta_i\}_{i=1}^{\infty}$ of a Poisson point process on the positive real line with intensity $d\Lambda(\zeta) = \zeta^{-2}d\zeta$ and of the independent realisations $\{Z_i(\vectx), \: \vectx \in \mathcal{X}\}_{i=1}^{\infty}$ of a non-negative stochastic process with continuous sample paths and $\mathrm{E}[Z(\vectx)] = 1 \; \forall \: \vectx \in \mathcal{X}$ (see, e.g., \citet{Ribatet2013}).

Different max-stable processes are obtained by choosing different stochastic processes $Z$. We consider two very popular stationary models, the \emph{extremal-$t$} model \citep{opitz2013} and the \emph{Brown-Resnick} model with power variogram \citep{brown1977,KabluchkoSchlatherdeHaan2009}. The specifications for $Z_i(\vectx)$ for each of the models are 
\begin{alignat*}{3}
&\text{Extremal-}t\text{:}  \qquad		&Z_i(\vectx) &= \sqrt{\pi} \: 2^{-(\nu-2)/2} \: \Gamma \left\{ (\nu+1)/2 \right\}^{-1} \: \max\{0, \epsilon_i(\vectx)\}^{\nu}, \; \nu > 0,\\
&\text{Brown-Resnick:} \qquad	&Z_i(\vectx) &= \exp \left\{\varepsilon_i(\vectx) - \mathrm{Var}[\varepsilon_i(\vectx)]/2\right\},
\end{alignat*}

where $\epsilon_i$ and $\varepsilon_i$ are independent copies of Gaussian processes. 

In the case of the extremal-$t$ model, $\epsilon$ is a stationary Gaussian process defined by the correlation function $\rho(h)$, where $h$ is the Euclidean distance between two points. For our example, we assume the \emph{powered exponential} or \emph{stable} correlation function:
\begin{align}
\rho(h) &= \exp\left[-\left(h/\lambda\right)^\kappa\right], \quad \lambda > 0, \: 0 < \kappa \leq 2. \label{eq:powexp_corfunction}
\end{align}

The Brown-Resnick process is defined by its semi-variogram. If the process $\varepsilon$ is a fractional Brownian motion centred at the origin, the Brown-Resnick process is stationary and the semi-variogram has the form
\begin{align*}
\gamma(h) &= \left( h/\lambda \right)^{\kappa}, \quad \lambda > 0, \: 0 < \kappa \leq 2, 
\end{align*}
where $h$ denotes the distance between two locations.

Both models depend on two parameters governing the dependence between two locations separated by a distance $h$: the \emph{range} parameter $\lambda$ and the \emph{smoothness} parameter $\kappa$. In addition, the extremal-$t$ model has a \emph{degrees of freedom} parameter denoted by $\nu$. We assume there is no discontinuity of the correlation function at $h = 0$ (i.e., no nugget effect).

The third model we consider is a copula model. Similar to the max-stable models, the univariate marginal distributions of the copula model are unit Fr\'{e}chet. However, the extremal dependence between the locations is simply modelled by a standard (non-extremal) copula. For an introduction to copulas see \citet{nelsen2006}. We assume the multivariate Student-$t$ copula in our example. The multivariate cumulative distribution function (CDF) at locations $(\vectx_1,\ldots,\vectx_H)$ implied by the \emph{non-extremal Student-$t$ copula} model \citep{demarta2005} is 
\begin{align*}
\Pr\{Y(\vectx_1) \leq y_1,\ldots,Y(\vectx_H) \leq y_H\} &= T_{H;\nu}\{T_{1;\nu}^{-1}[F(y_1)],\ldots,T_{1;\nu}^{-1}[F(y_H)]; \boldsymbol{\Sigma} \},
\end{align*}
where $F(y) = \exp\{-1/y\}$ is the CDF of the unit Fr\'{e}chet distribution, $T_{1;\nu}^{-1}[\cdot]$ is the quantile function of the univariate Student-$t$ distribution with $\nu$ degrees of freedom, and $T_{H;\nu}\{\cdots; \boldsymbol{\Sigma} \}$ is the CDF of the $H$-variate Student-$t$ distribution with $\nu$ degrees of freedom and dispersion matrix $\boldsymbol{\Sigma}$. The diagonal elements of $\boldsymbol{\Sigma}$ are $1$ and the off-diagonal elements contain the correlations between the locations. Therefore, the entries of $\boldsymbol{\Sigma}$ are given by $\boldsymbol{\Sigma}_{ij} = \rho(h_{ij})$ for $i, j = 1,\ldots,H$, where $h_{ij}$ is the distance between locations $i$ and $j$. As for the extremal-$t$ model, we assume the correlation function to be the powered exponential correlation function \eqref{eq:powexp_corfunction}. This also implies that the non-extremal Student-$t$ copula model has the same set of parameters as the extremal-$t$ model: range ($\lambda$), smoothness ($\kappa$), and degrees of freedom ($\nu$).

\subsection{Summary Statistics}

If a reasonable amount of observations are collected at each location, the data collected quickly becomes very high-dimensional, while each observation is only marginally informative. This diminishes the classification power of the classifiers we use. We therefore aim to reduce the dimension of the data by generating informative summary statistics. Unfortunately, none of the statistics we consider guarantee consistent model choice. This can potentially result in large biases when estimating the posterior model probabilities \citep{robert2011}, which can also affect the estimates of the misclassification error rates. However, trees and random forests work reasonably well with a sizeable amount of moderately informative feature variables. Therefore, we can include a wide variety of summary statistics, where each contains some information about the process. Considering the combined information of all the summary statistics, we expect that only a small loss in information is incurred compared to the full dataset.

First, we include all the \emph{F-madogram} estimates for all the pairs of locations. The F-madogram \citep{cooley2006} is similar to the semi-variogram, but unlike the semi-variogram it also exists if the variances or means of the random variables are not finite. Given $n$ observations $\{y_1(\vectx_1), \ldots, y_n(\vectx_1)\}$ and $\{y_1(\vectx_2), \ldots, y_n(\vectx_2)\}$ collected at locations $\vectx_1$ as well as $\vectx_2$, the pairwise F-madogram between locations $\vectx_1$ and $\vectx_2$ is estimated as
\begin{align*}
\hat{\nu}_F(\vectx_1, \, \vectx_2) &= \frac{1}{2 n} \sum_{i=1}^{n} |F\{y_i(\vectx_1)\} - F\{y_i(\vectx_2)\}|,
\end{align*}
where $F\{y\} = \exp\{-1/y\}$ is the CDF of the unit Fr\'{e}chet distribution.

As a second set of summary statistics, we include estimates for all the pairwise \emph{extremal coefficients} \citep{SchlatherTawn2003}. For a max-stable process, the pairwise extremal coefficient between locations $\vectx_1$ and $\vectx_2$ is defined as the value $\theta(\vectx_1, \vectx_2)$ for which
\begin{align}
\Pr(Y(\vectx_1) \leq y, \: Y(\vectx_2) \leq y) &= \Pr(Y(\vectx_1) \leq y)^{\theta(\vectx_1, \vectx_2)} = \exp\left( -\frac{\theta(\vectx_1, \vectx_2)}{y} \right). \label{eq:def_extremal_coefficient}
\end{align}
The pairwise extremal coefficient can assume values between $1$ and $2$. A value of $\theta(\vectx_1, \vectx_2) = 1$ indicates complete dependence between the two locations. If $\theta(\vectx_1, \vectx_2) = 2$, the two locations are completely independent. We estimate it using the fast estimator of \citet{coles1999},
\begin{align}
\hat{\theta}(\vectx_1, \vectx_2) = \frac{n}{\sum_{i = 1}^{n}{1/\max\{y_i(\vectx_1), y_i(\vectx_2)\}}}. \label{eq:estimate_extremal_coefficient}
\end{align}
The extremal coefficient as defined by \eqref{eq:def_extremal_coefficient} only exists for max-stable processes. In general, the coefficient also depends on the level $y$. However, the quantities computed by Equation~\eqref{eq:estimate_extremal_coefficient} might still provide useful information about the dependence structure. For the $t$ copula model, \citet{lee2018} demonstrate by simulation that the estimates given by \eqref{eq:estimate_extremal_coefficient} are indeed informative about the dependence structure.

The last set of summary statistics we consider is the set of \emph{Kendall's $\tau$} estimates between all pairs of locations. Kendall's $\tau$ between locations $\vectx_1$ and $\vectx_2$ is estimated by
\begin{align*}
\hat{\tau}(\vectx_1, \, \vectx_2) &= \frac{2}{n \, (n-1)} \sum_{1 \leq i < j \leq n} \mathrm{sign}[y_i(\vectx_1)-y_j(\vectx_1)] \: \mathrm{sign}[y_i(\vectx_2)-y_j(\vectx_2)].
\end{align*}
\citet{dombry2018} show that for max-stable processes Kendall's $\tau$ is equal to the probability that the maxima at two locations occur concurrently and are therefore attained for the same extremal function, so
\begin{align*}
\tau(\vectx_1, \, \vectx_2) &= \Pr\left(\underset{i \geq 1}{\arg \max} \: \varphi_i(\vectx_1) = \underset{i \geq 1}{\arg \max} \: \varphi_i(\vectx_2)\right).
\end{align*}

All of the summary statistics we incorporate are also considered by \citet{lee2018}, who perform ABC model selection using the summary statistic projection method of \citet{prangle2014-2} for a very similar set of models as in this example. Therefore, a more detailed discussion of the summary statistics can be found in \citet{lee2018}.

\subsection{Bayesian Inference for Spatial Extremes Models}

The likelihood functions of max-stable models are practically intractable for most models for dimensions greater than two or three. Composite likelihood methods have been the most popular way to conduct classical inference for max-stable models, so model discrimination is usually based on the composite likelihood information criterion (CLIC) \citep{padoan2010}.

The observed extrema at several locations might occur at the same time, which means that the extrema at these locations arise from the same extremal function $\varphi_i(\vectx)$ in Equation~\eqref{eq:max_stable_definition}. The locations can then be partitioned according to which extremal functions $\varphi_i(\vectx)$ produce the extreme observations at the different locations. \citet{stephenson2005} show that the joint likelihood of the extreme observations and the partitions is substantially simpler than the likelihood of the extreme observations without knowledge of the partitions. \citet{thibaud2016} and \citet{dombry2017-1} use this property to devise a Gibbs sampler with the partitions as auxiliary variables to conduct Bayesian inference for max-stable models. However, even the augmented likelihoods are expensive to evaluate for the Brown-Resnick and extremal-$t$ model because they include multivariate Gaussian (Brown-Resnick) and Student-$t$ (extremal-$t$) CDFs.

Due to the intractability of the likelihoods, ABC has also been a popular method for Bayesian inference of max-stable models, see, e.g., \citet{ErhardtSmith2012} or the overview in \linebreak \citet{erhardt2015review}. \citet{lee2018} present an ABC application with the joint goal of model selection and parameter estimation for the same set of models we consider. \citet{hainy2016} seek to find optimal designs for parameter estimation for the extremal-$t$ model with $\nu = 1$ (called the `Schlather model'). They use ABC to estimate the posterior variances, which they use as design criterion. Their design algorithm is confined to very low-dimensional design spaces in order to be able to store the reference table for all possible designs. They sequentially select the best single location among a small set of possible locations. With our classification approach, we are able to overcome these limitations.

\subsection{Settings and Results}

In our example, we want to select $H$ ($H = 3, \ldots,8$) locations on a regular grid such that the ability to discriminate between the three models as measured by the misclassification error rate is optimised. We search the $H$ optimal design points over a regular $6 \times 6$ grid laid over a square with edge length $10$. The data consist of $n = 10$ independent realisations of the process collected at all the locations. Due to the isotropic nature of the processes, there are potentially many equivalent optimal solutions. With our modification of the coordinate exchange algorithm using 20 random starts, we seek to find one of these designs or at least a nearly optimal design.

We assume the following prior distributions:
\begin{align*}
\log\left(\lambda\right) &\sim \mathcal{N}(1, \, 4), \\
\kappa &\sim \mathcal{U}(0, \, 2), \\
\log(\nu) &\sim \mathcal{N}(0, \, 1) \text{ truncated on } [-2.5, \, 2.5].
\end{align*}
Furthermore, we assume equal prior model probabilities (= 1/3) for all models.

Simulating from the $t$ copula model is straightforward. It only requires simulating from a multivariate $t$ distribution and then transforming the margins with respect to the univariate $t$ CDF followed by the inverse unit Fr\'{e}chet CDF. For simulating from the max-stable models, we use the exact simulation algorithm via extremal functions of \citet{dombry2016}.

\begin{figure}[hbtp!]
	\centering
	\subfigure[3 design points]{\includegraphics[width=0.35\textwidth]{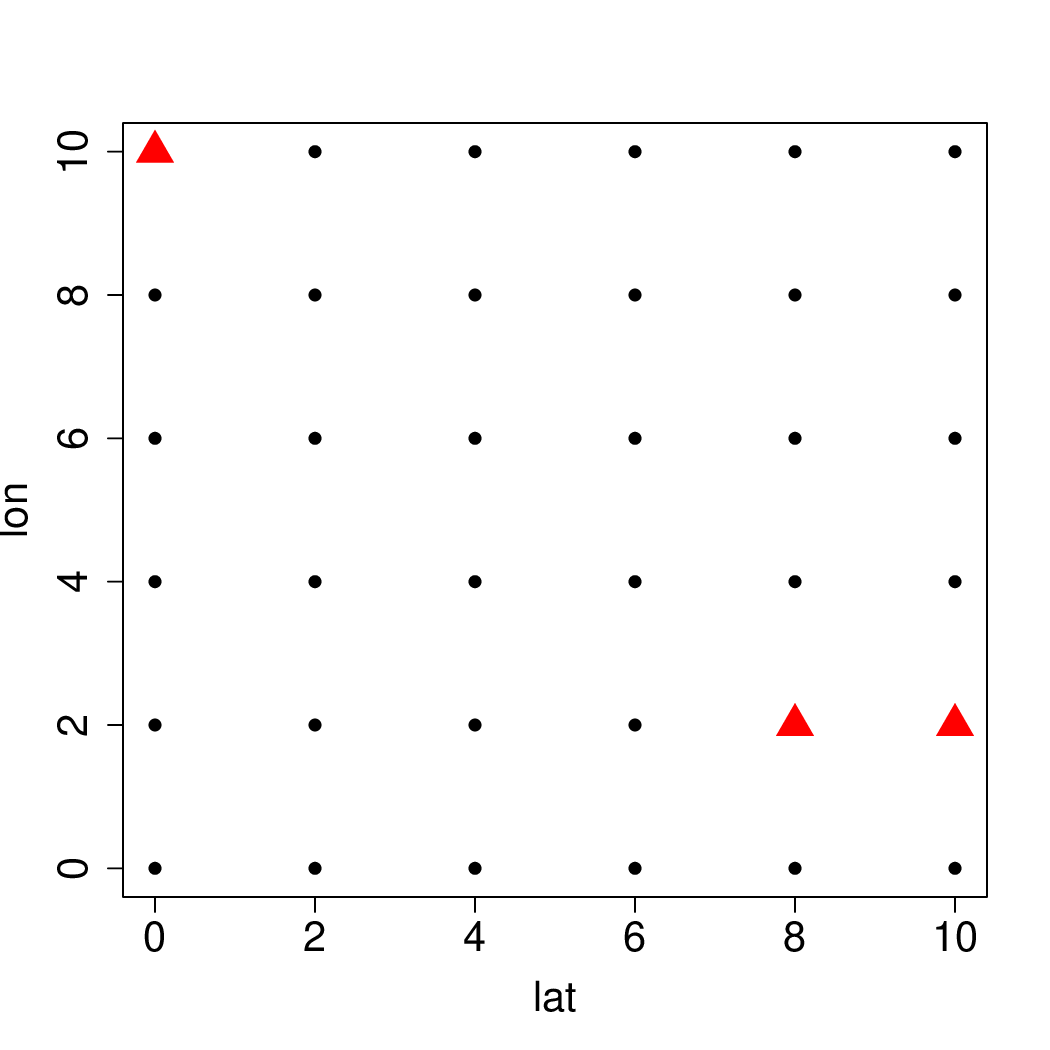}}
	\subfigure[4 design points]{\includegraphics[width=0.35\textwidth]{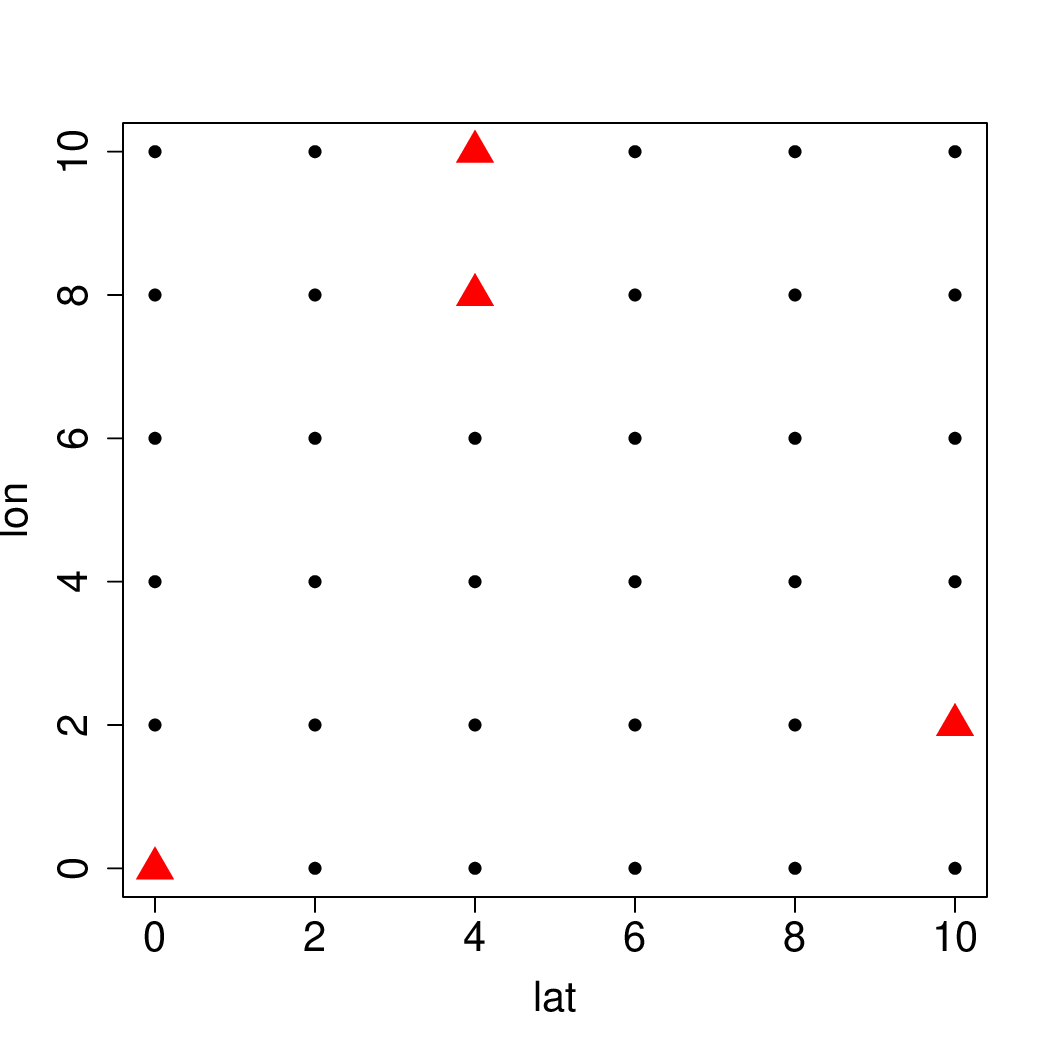}}
	\subfigure[5 design points]{\includegraphics[width=0.35\textwidth]{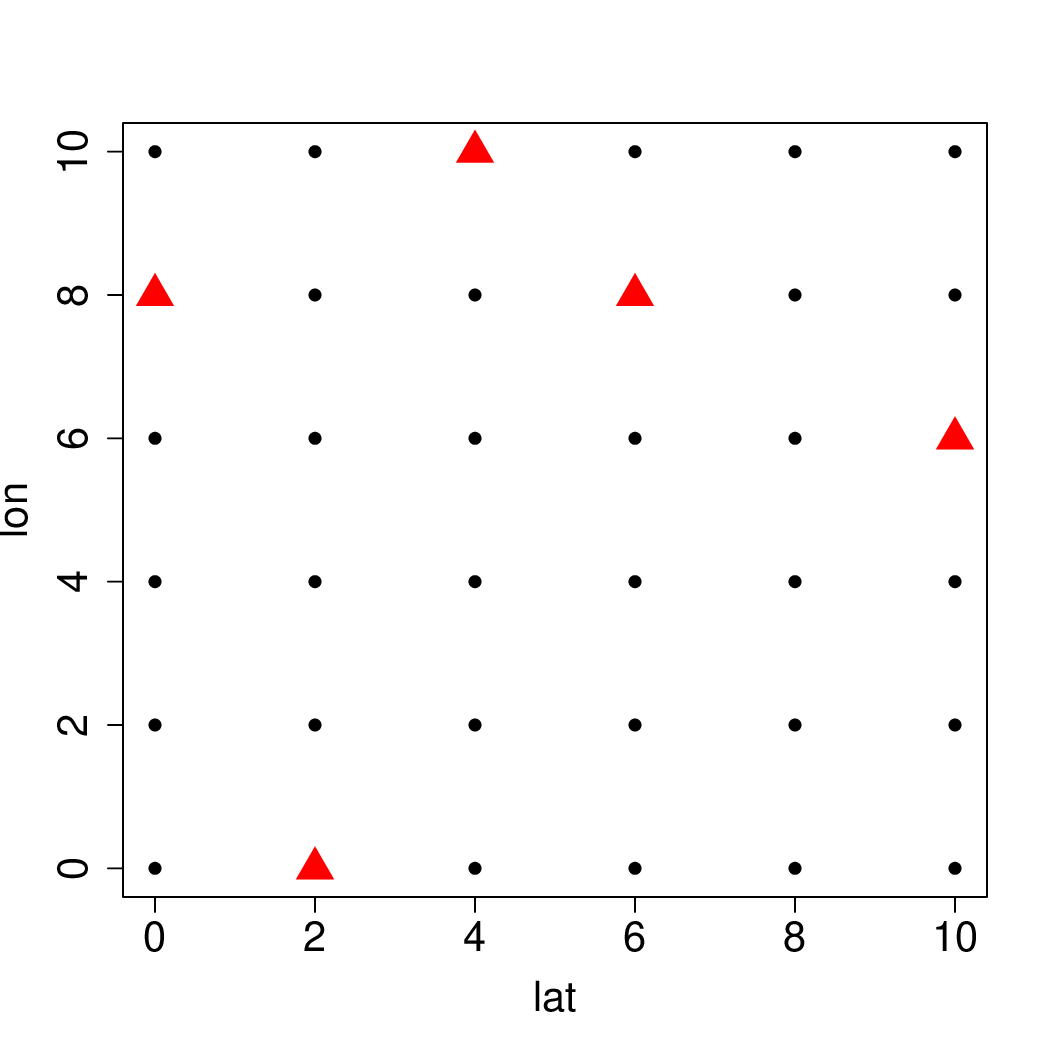}}
	\subfigure[6 design points]{\includegraphics[width=0.35\textwidth]{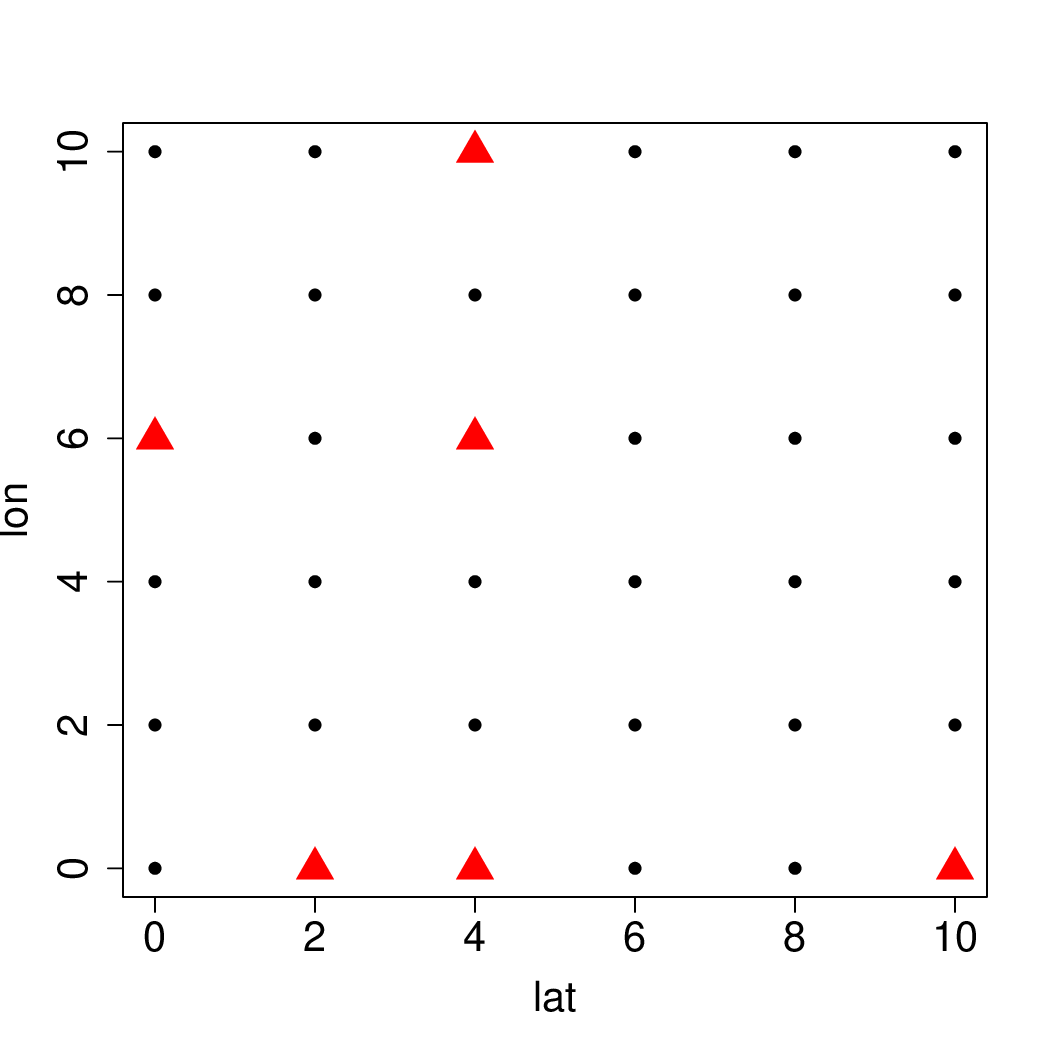}}
	\subfigure[7 design points]{\includegraphics[width=0.35\textwidth]{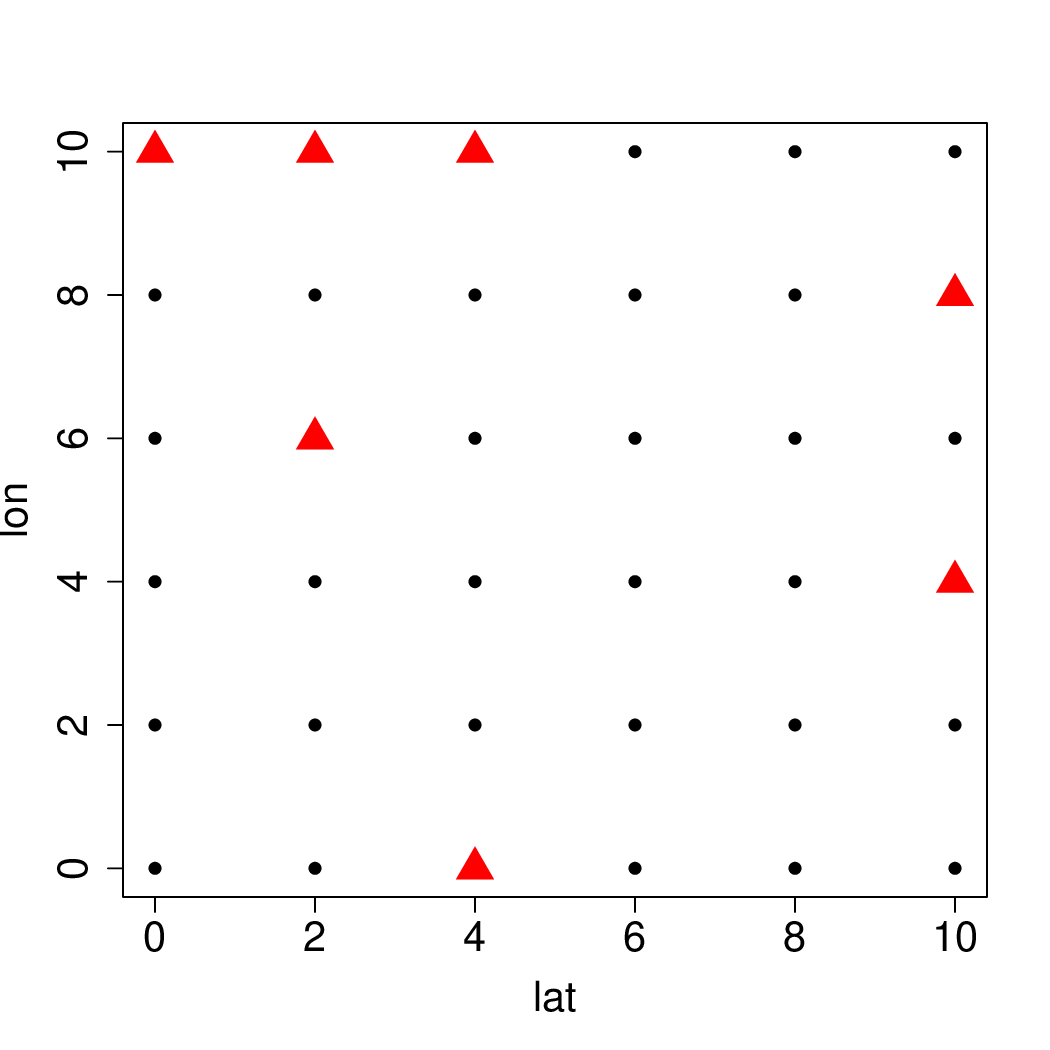}}
	\subfigure[8 design points]{\includegraphics[width=0.35\textwidth]{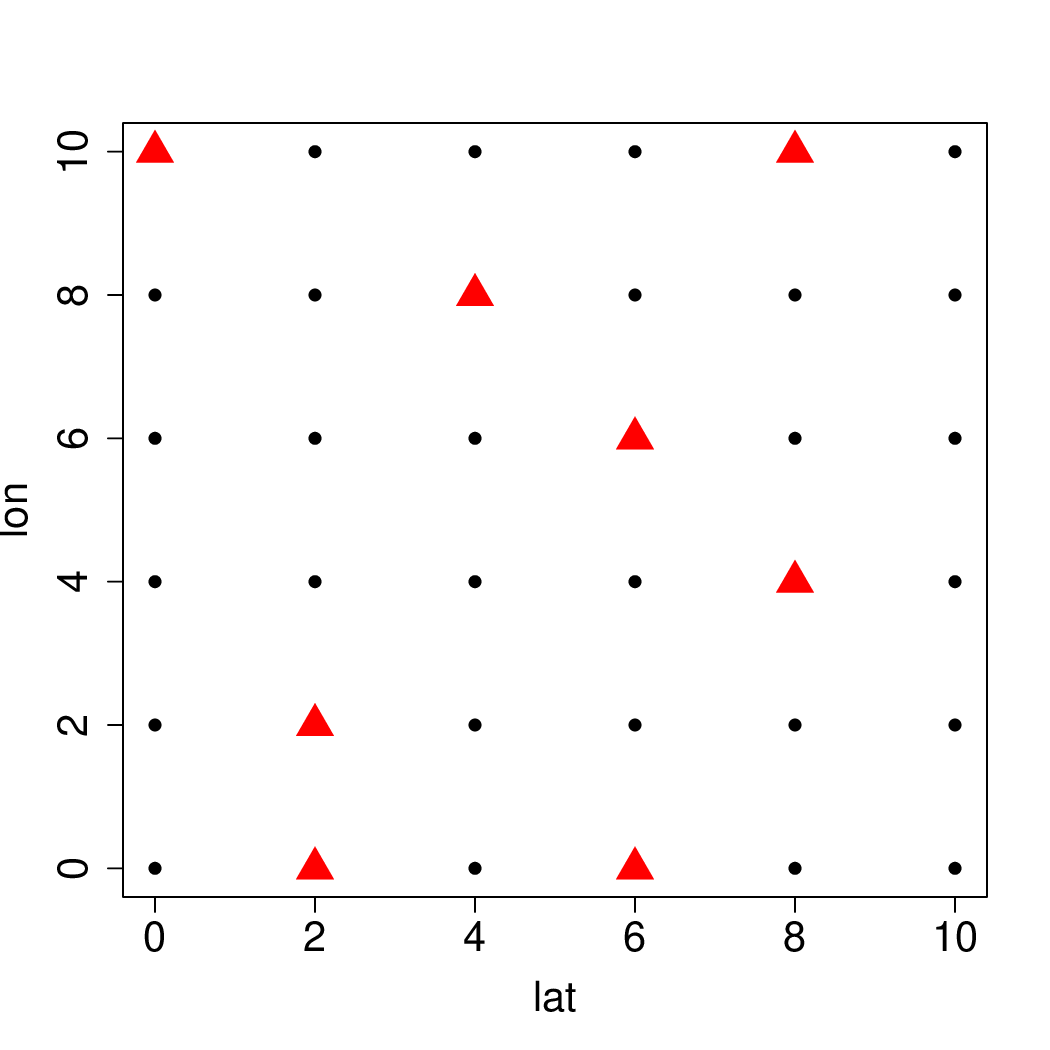}}
	\caption{Optimal classification designs found using \emph{trees} for design sizes from three to eight for the spatial extremes example. Selected design points are marked by red triangles.\label{fig:designs_spatextr_tree}}
\end{figure}

\begin{figure}[hbtp!]
	\centering
	\subfigure[3 design points]{\includegraphics[width=0.35\textwidth]{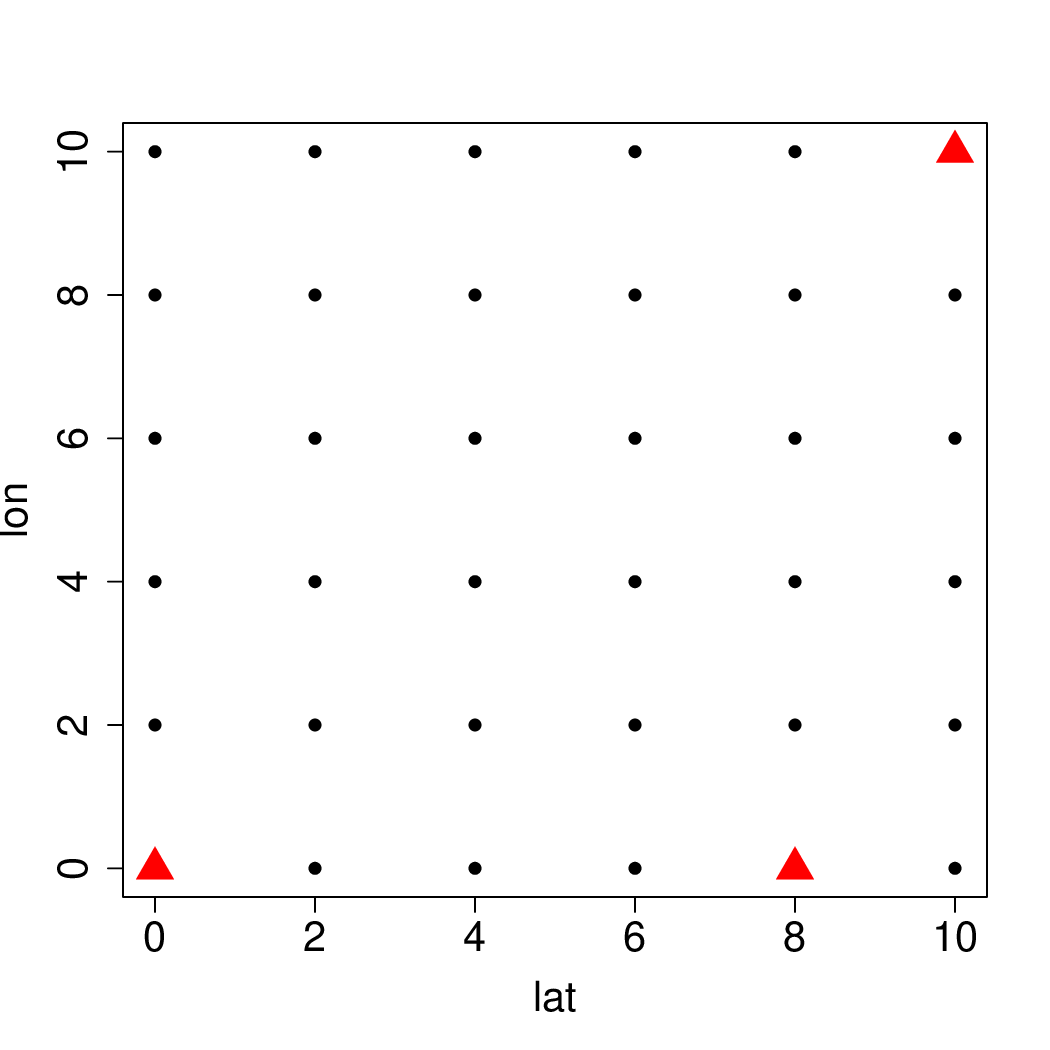}}
	\subfigure[4 design points]{\includegraphics[width=0.35\textwidth]{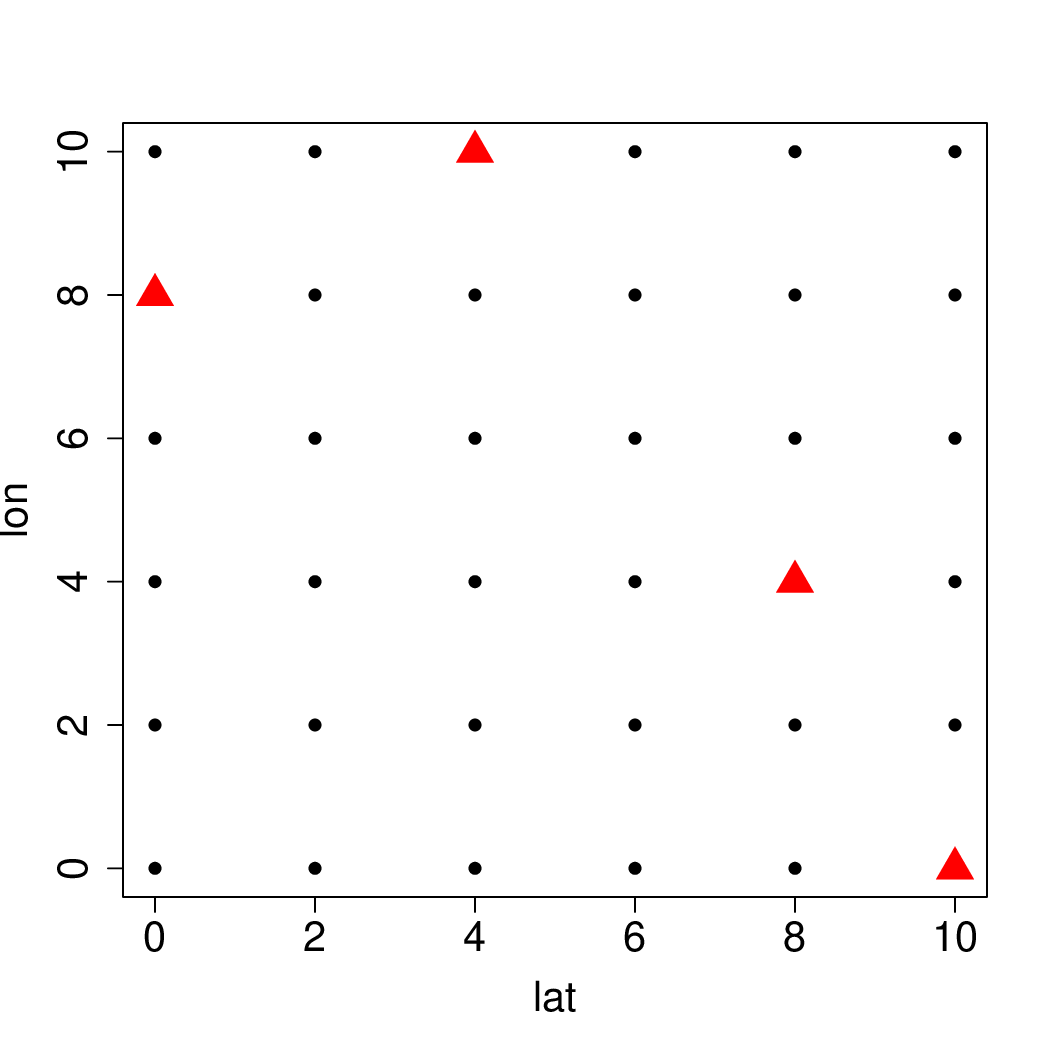}}
	\subfigure[5 design points]{\includegraphics[width=0.35\textwidth]{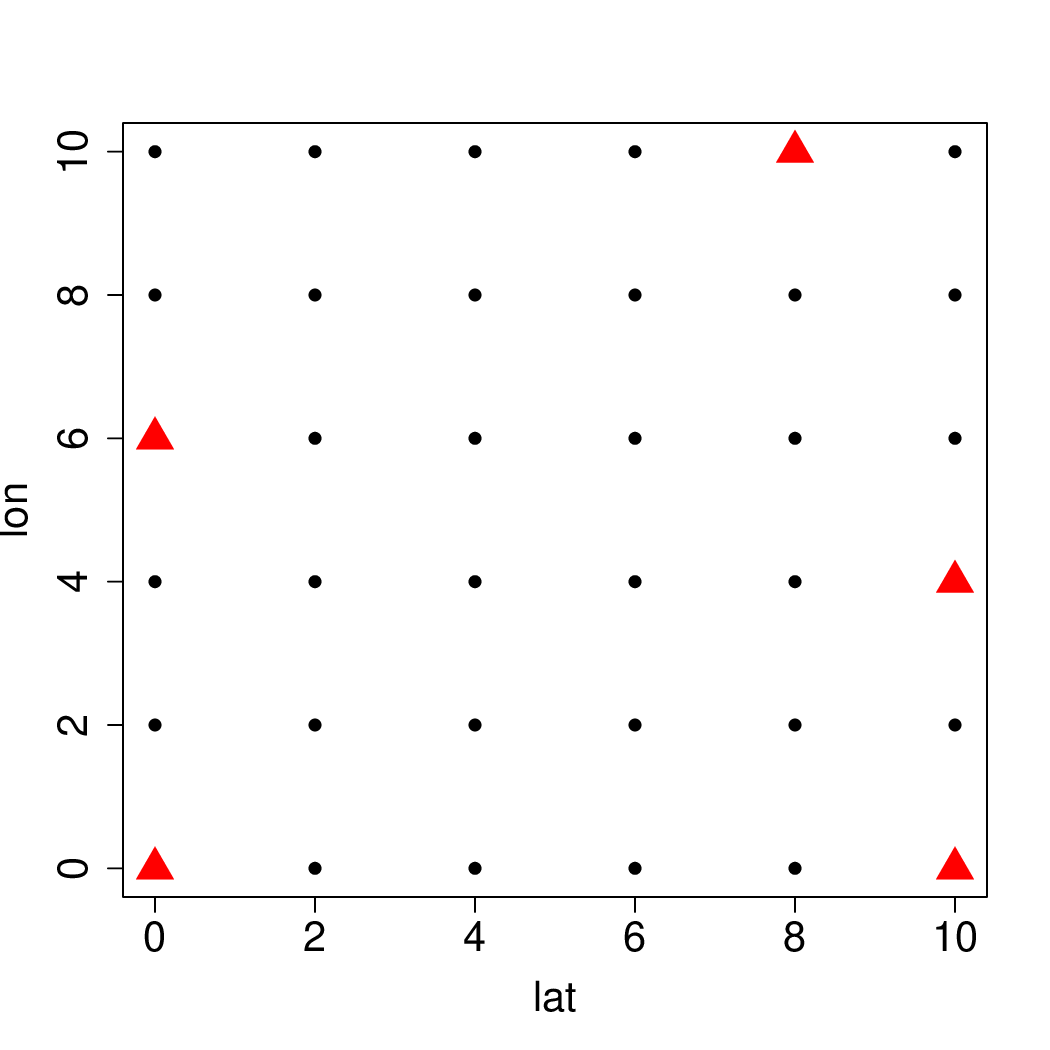}}
	\subfigure[6 design points]{\includegraphics[width=0.35\textwidth]{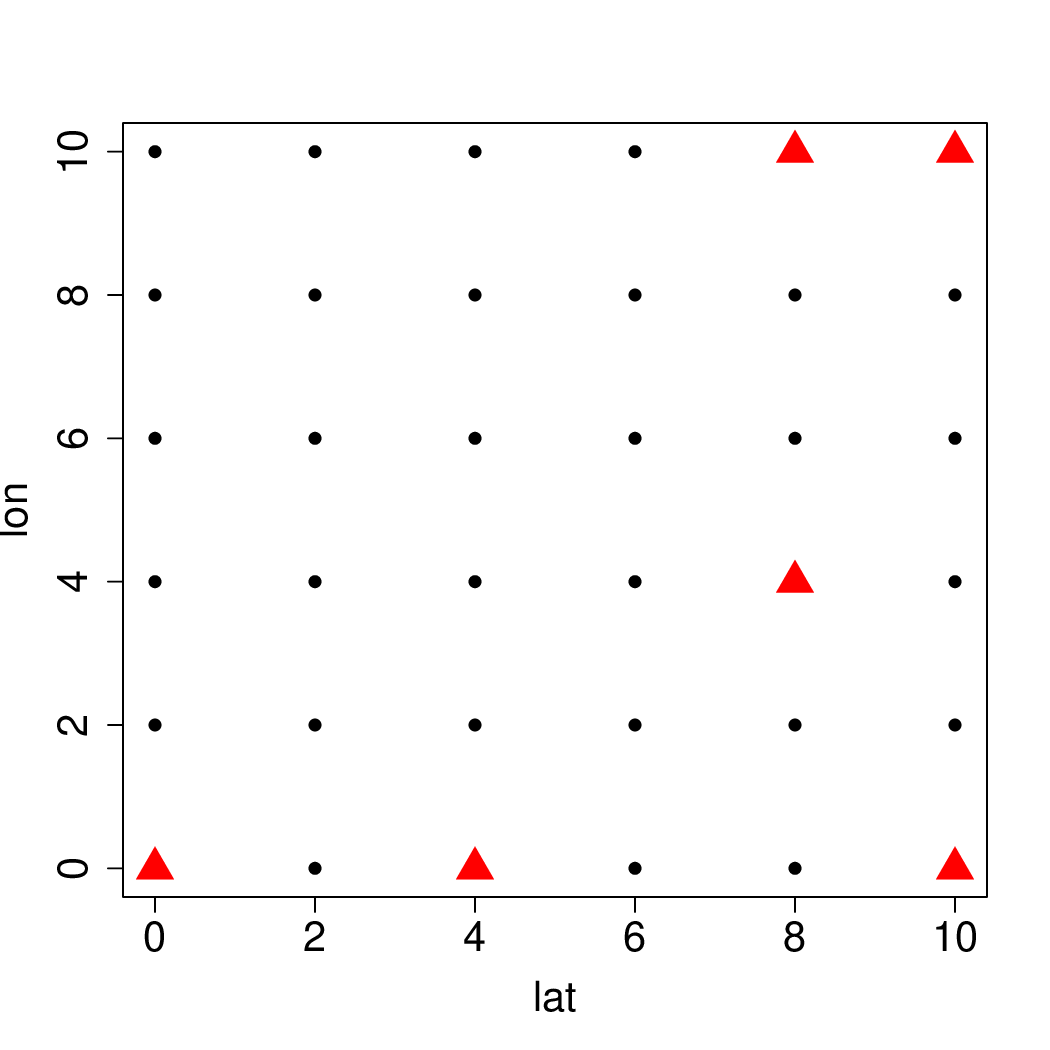}}
	\subfigure[7 design points]{\includegraphics[width=0.35\textwidth]{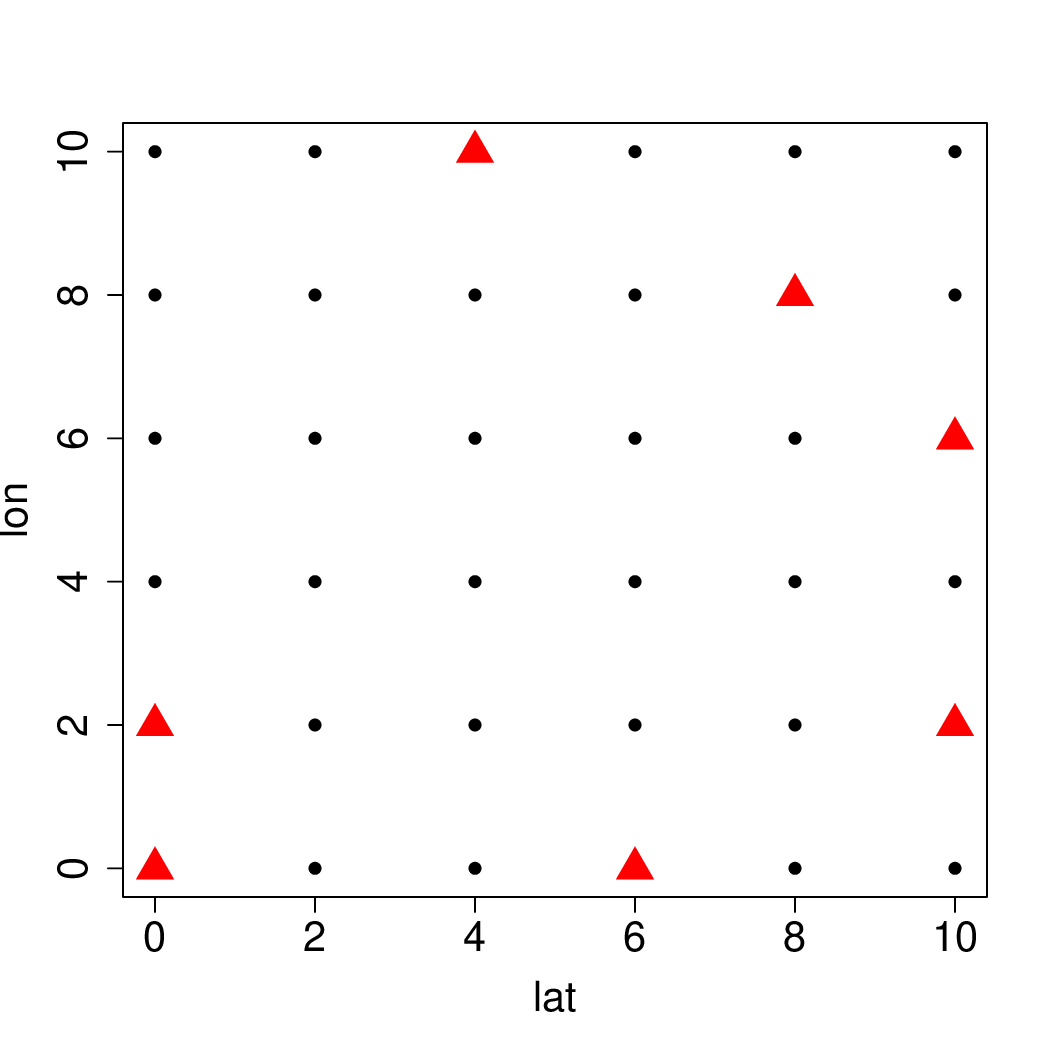}}
	\subfigure[8 design points]{\includegraphics[width=0.35\textwidth]{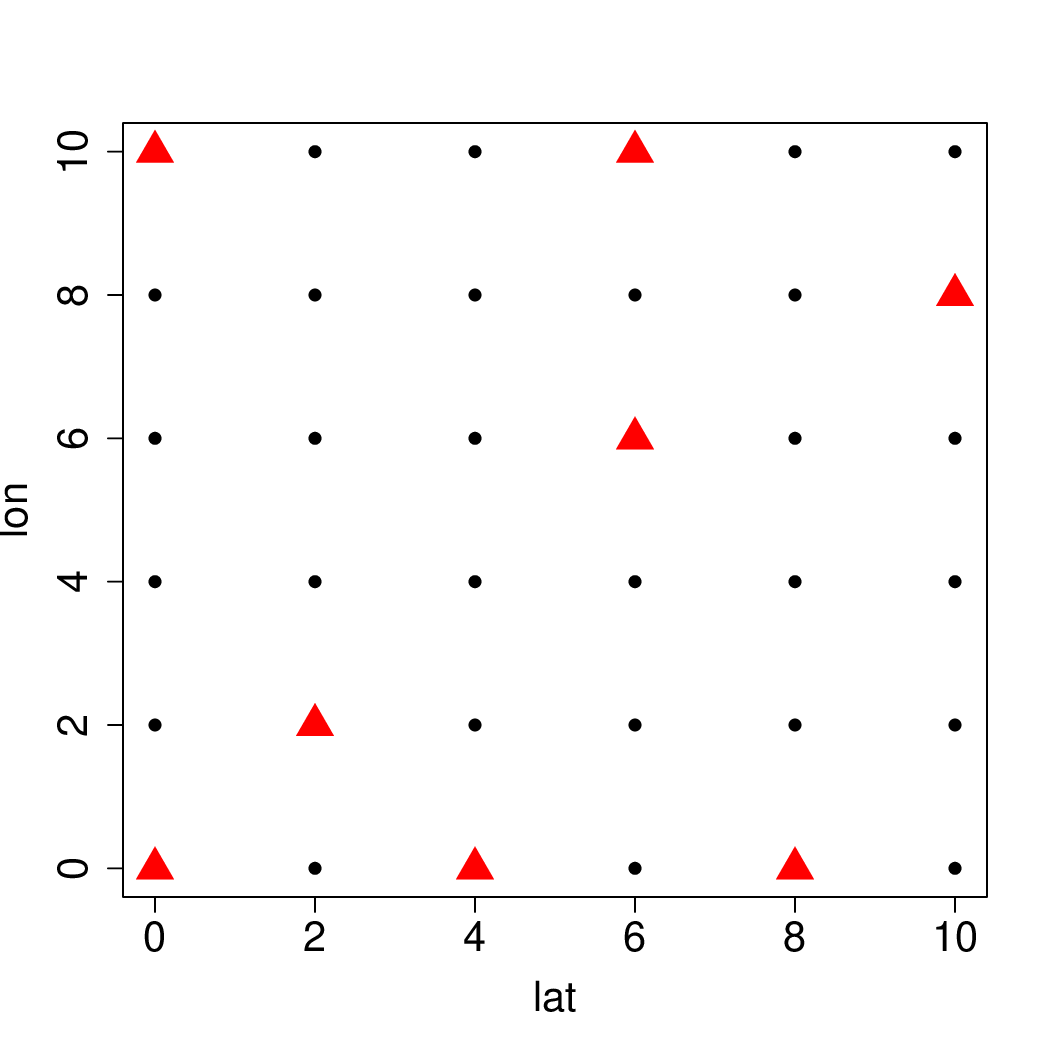}}
	\caption{Optimal classification designs found using \emph{random forests} for design sizes from three to eight for the spatial extremes example. Selected design points are marked by red triangles.\label{fig:designs_spatextr_rf}}
\end{figure}

During the design phase, we use cross-validated classification trees as well as random forests with $500$ trees using out-of-bag class predictions to estimate the misclassification error rates. We had implemented the simulator functions for this example in R, therefore we use the R function \texttt{rpart} for classification trees, for which we keep all the default settings except for not considering any surrogate splits to speed up computing time. For random forests, we employ the function \texttt{randomForest} from the R \citep{RSoftware} package of the same name \citep{liaw2002}. The simulated sets for both methods contain 5K simulations per model. The optimal designs obtained for these two methods are shown in Figures~\ref{fig:designs_spatextr_tree} (trees) and \ref{fig:designs_spatextr_rf} (random forests). 

\begin{figure}[htbp!]
	\centering
	\includegraphics[width=0.8\textwidth]{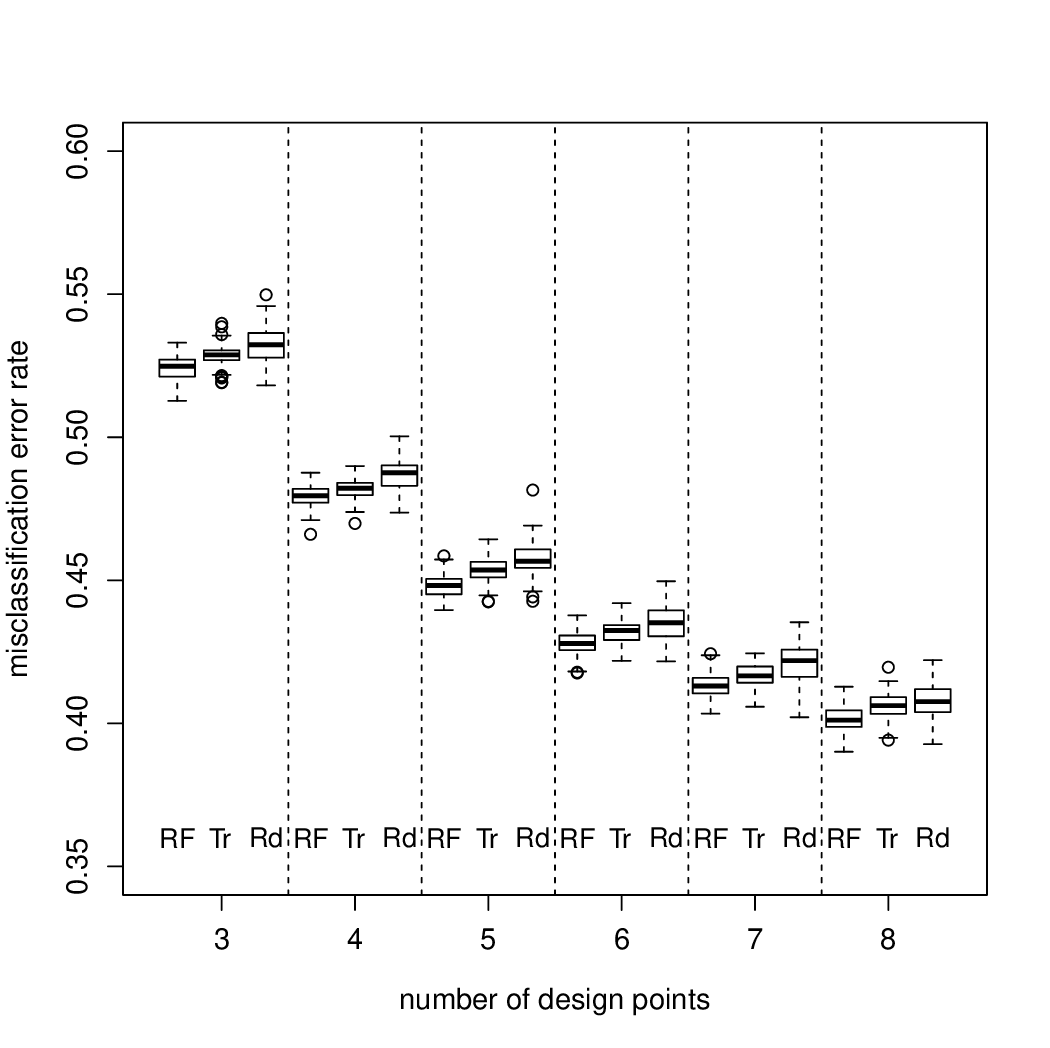}
	\caption{Spatial extremes example: distributions of the random forest-estimated misclassification error rates over 100 random samples of size 15K generated from the prior predictive distribution at the optimal classification designs found using random forests (rf) or trees (tr) for design sizes from three to eight. The distributions of the misclassification error rates over 100 random samples of size 15K generated from the prior predictive distribution at 100 random designs (rd) are also shown for the same design sizes.} \label{fig:loss_distr_spatextr}
\end{figure}

To evaluate the designs found by our classification approach, we repeat estimating the misclassification error rate via random forests with $500$ trees using out-of-bag class predictions on $100$ different simulated samples of size 15K (5K simulations per model) from the prior predictive distribution. The distributions of the estimated misclassification error rates are plotted in Figure~\ref{fig:loss_distr_spatextr}. We also include the distributions of the estimated misclassification error rates for $100$ simulated samples from the prior predictive distribution generated on $100$ randomly selected designs. The optimal classification designs found using random forests clearly perform best for all design sizes. Using classification trees with cross-validation instead of random forests leads to designs which are a bit worse. However, the average misclassification error rate of the classification tree designs is still smaller than the average error rate of the random designs up until $7$ design points.

\begin{figure}[hbtp!]
	\centering
	\subfigure[3 design points]{\includegraphics[width=0.45\textwidth]{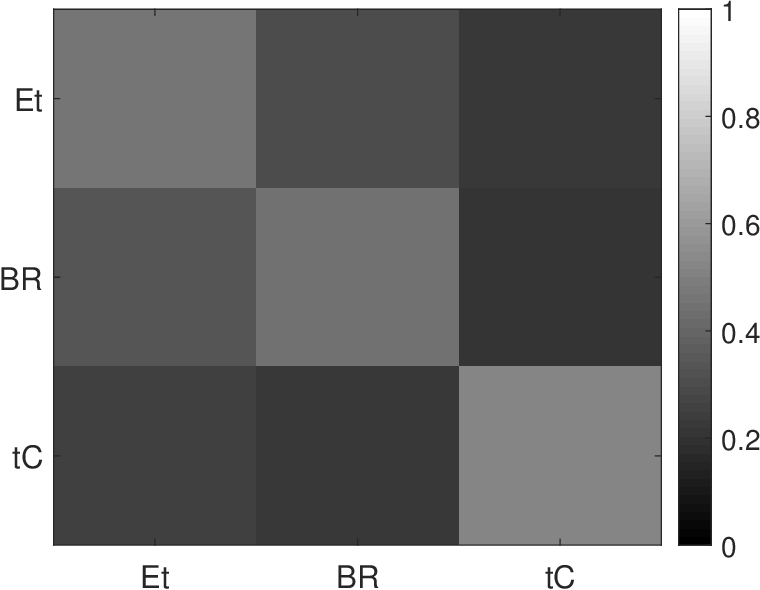}} \quad
	\subfigure[4 design points]{\includegraphics[width=0.45\textwidth]{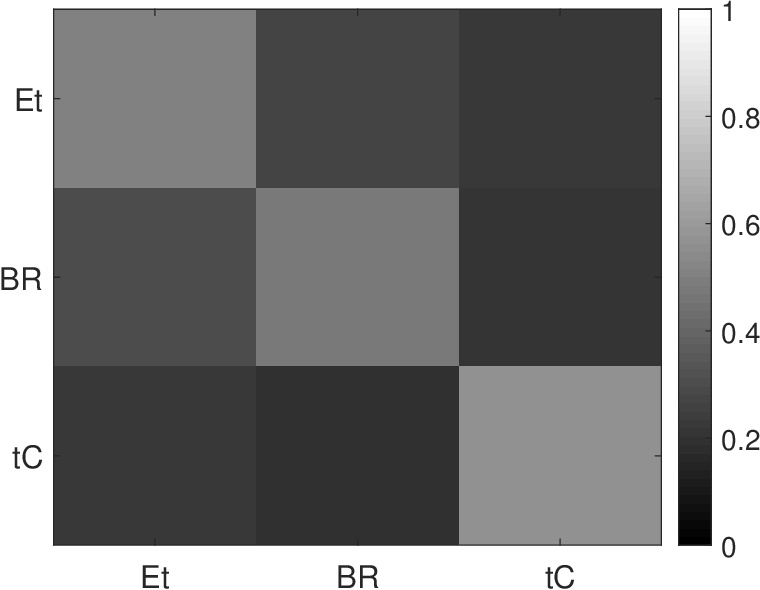}}
	\vspace*{1ex}
	\subfigure[5 design points]{\includegraphics[width=0.45\textwidth]{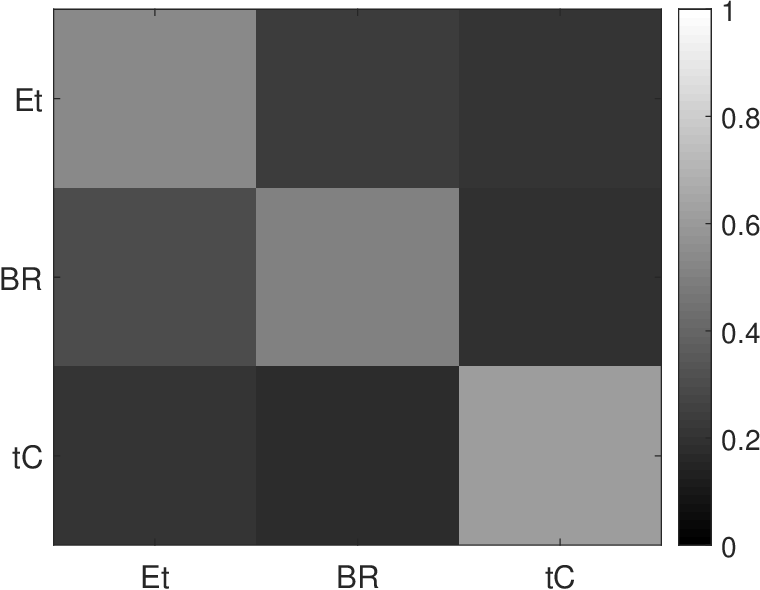}} \quad
	\subfigure[6 design points]{\includegraphics[width=0.45\textwidth]{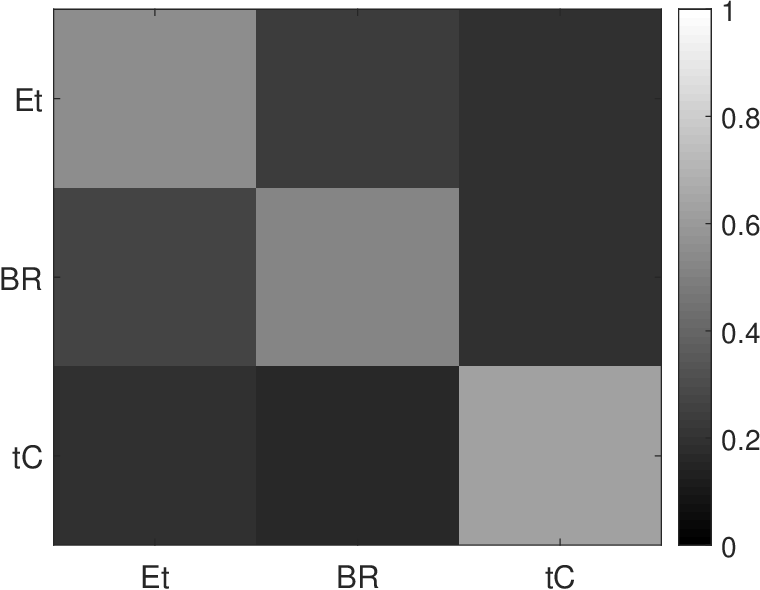}}
	\vspace*{1ex}
	\subfigure[7 design points]{\includegraphics[width=0.45\textwidth]{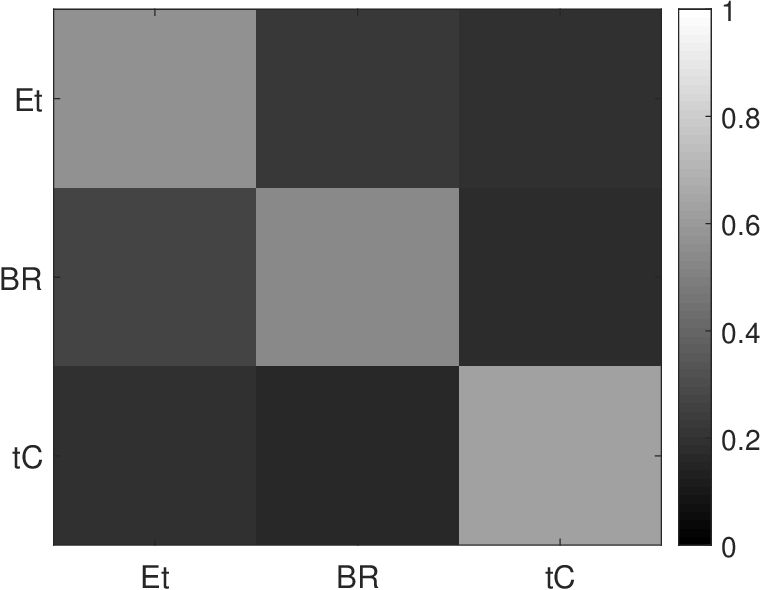}} \quad
	\subfigure[8 design points]{\includegraphics[width=0.45\textwidth]{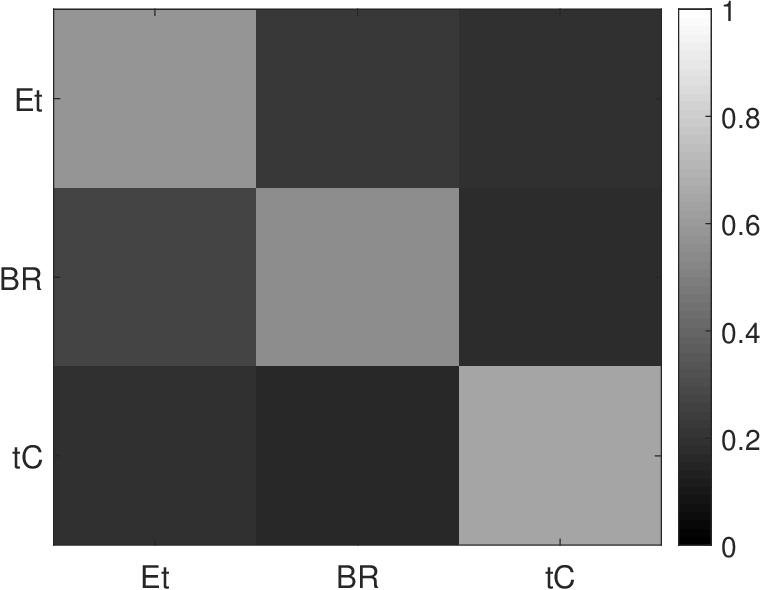}}
	\caption{Average misclassification matrices over 100 simulated prior predictive samples obtained for the \emph{random forest classification designs} for the spatial extremes example. Design sizes from 3 -- 8 design points are considered.\label{fig:spatextr_misclassmatrix}}
\end{figure}

In addition to the misclassification error rate, we also compute the misclassification matrix yielded by the random forest classifier for each of the $100$ simulated samples for each evaluated design. The average misclassification matrices over the $100$ samples are depicted in Figure~\ref{fig:spatextr_misclassmatrix} for the optimal designs obtained by the random forest classification approach. They show that discriminating between the two max-stable models is more difficult than discriminating between the $t$ copula model and either of the max-stable models.

\end{appendices}

\end{document}